\documentclass[a4paper,fleqn]{cas-dc}

\usepackage[numbers]{natbib}

\usepackage{multirow}
\newcommand{\tc}{\textdegree\xspace}
\newcolumntype{P}[1]{>{\centering\arraybackslash}p{#1}}
\newcolumntype{M}[1]{>{\centering\arraybackslash}m{#1}}

\def\tsc#1{\csdef{#1}{\textsc{\lowercase{#1}}\xspace}}
\tsc{WGM}
\tsc{QE}

\newdefinition{CAH}{Contact Angle Hysteresis}

\begin{document}
\let\WriteBookmarks\relax
\def\floatpagepagefraction{1}
\def\textpagefraction{.001}

\shorttitle{Contact Angle Hysteresis on Pillared Surfaces via Energy Dissipation}    

\shortauthors{\textit{Kumar et al.}}  

\title [mode = title]{Predicting Contact Angle Hysteresis on Surfaces with Randomly and Periodically Distributed Cylindrical Pillars via Energy Dissipation}  



%

\author[1]{Pawan Kumar} [type=editor,
        orcid=0000-0002-8654-8514,
]

 \ead{kumar.p@unimelb.edu.au}



\affiliation[1]{organization={Department of Chemical Engineering },
            addressline={University of Melbourne, Parkville}, 
            city={Melbourne},
            postcode={3010}, 
            state={Victoria},
            country={Australia}}


\author[2]{Paul Mulvaney} [type=editor,
        orcid=0000-0002-8007-3247,
]

 \ead{mulvaney@unimelb.edu.au}



\affiliation[2]{organization={School of Chemistry},
            addressline={University of Melbourne, Parkville}, 
            city={Melbourne},
            postcode={3010}, 
            state={Victoria},
            country={Australia}}

\author[1]{Dalton J. E. Harvie} [type=editor,
        orcid=0000-0002-8501-1344,
]
\cormark[3]
\ead{daltonh@unimelb.edu.au}
\cortext[3]{Corresponding author}

\fntext[1]{}


\begin{abstract}
Hypothesis: Understanding contact angle hysteresis on rough surfaces is important as most industrially relevant and naturally occurring surfaces possess some form of random or structured roughness. We hypothesise that hysteresis originates from the energy dissipation during the \textit{stick-slip} motion of the contact line and that this energy dissipation is key to developing a predictive equation for hysteresis.\\
Experiments: We measured hysteresis on surfaces with randomly distributed and periodically arranged microscopic cylindrical pillars for a variety of different liquids in air. The inherent (flat surface) contact angles tested range from lyophilic ($\theta_{\rm{e}} = 33.8$\textdegree) to lyophobic ($\theta_{\rm{e}} = 112.0$\textdegree).\\ 
Findings: A new methodology for calculating the average advancing and receding contact angles on random surfaces is presented. Also, the correlations for roughness-induced energy dissipation were derived, and a predictive equation for the advancing and receding contact angles during homogeneous (Wenzel) wetting on random surfaces is presented. Significantly, equations that predict the onset of the alternate wetting conditions of hemiwicking, split-advancing, split-receding and heterogeneous (Cassie) wetting are also derived, thus defining the range of validity for the derived homogeneous wetting equation. A novel feature `cluster' concept is introduced which explains the measurably higher hysteresis exhibited by structured surfaces compared to random surfaces observed experimentally.
\end{abstract}



\maketitle

\section{Introduction}\label{sec:exp_intro}

Surface wetting refers to the spreading of a liquid on a solid in the presence of another immiscible liquid or gas. Self-cleaning surfaces \cite{sun2005bioinspired,xu2016biomimetic}, efficient microfluidic devices \cite{jia2019effect}, metal-ore separation processes \cite{coleman2009flotation}, and enhanced oil recovery \cite{bonn2001complex} are just a few applications where surface wetting plays an important role. The wettability of a surface can be characterised by measuring the contact angle of a liquid when deposited on the surface; which is the angle between the average solid surface and the fluid/fluid interface measured from the side of the reference fluid \cite{Marmur2006}. The contact angle thus measured (experimentally) is actually the macroscopic contact angle ($\theta_{\rm{m}}$), which is the angle between the fluid/fluid interface and a plane parallel to the average solid surface but at a certain distance away from it (usually a few tens of microns). 

On an ideal surface (S, which is chemically homogeneous, perfectly flat and non-reactive) the contact angle of a droplet (fluid-1) in the presence of another immiscible fluid (fluid-2 or air) is given by Young's law \cite{young1805iii}, that is
\begin{equation}
    \cos \theta_{\rm{e}} = \left( \frac{\sigma_{2\rm{S}}-\sigma_{1\rm{S}}}{\sigma_{12}}\right),
    \label{eqn:exp_young}
\end{equation}
where $\sigma_{12}, \sigma_{\rm{1S}}$ and $\sigma_{\rm{2S}}$ are the interfacial tension of the fluid-1/fluid-2, fluid-1/solid and fluid-2/solid interfaces respectively. $\theta_{\rm{e}}$ in equation (\ref{eqn:exp_young}) is known as Young's angle (which is also the macroscopic angle ($\theta_{\rm{m}}$) on an ideal surface). The magnitude of Young's angle dictates the wetting nature of a surface. A high contact angle ($\theta_{\rm{e}}>90$\textdegree) defines a lyophobic surface while a small contact angle ($\theta_{\rm{e}}<90$\textdegree) defines a lyophilic surface.

In reality, however, it is not possible to satisfy all the requirements of an ideal surface, and the contact angle on real surfaces deviates from that determined by Young's law. Real surfaces possess some form of roughness \cite{quere2008wetting} and can be chemically heterogeneous as well, however, in this work we only focus on the effect of surface roughness. The wetting on rough surfaces can either be described by the Wenzel \cite{wenzel1936resistance} or the Cassie-Baxter \cite{cassie1944wettability} models, depending upon the wetting state exhibited by the liquid, is homogeneous (the liquid completely fills up all the surface crevices) or heterogeneous state (the surrounding fluid (fluid-2 or air) is trapped between the liquid and the top of the surface undulations).

For homogeneous (or Wenzel) wetting states, Wenzel \cite{wenzel1936resistance} proposed an equation for the contact angle on a rough surface ($\theta_{\rm{m}}=\theta_{\rm{W}}$), that is
\begin{equation}
    \cos\theta_{\rm{W}} = r \cos\theta_{\rm{e}},
    \label{eqn:exp_wenzel}
\end{equation}
where $r$ is the ratio of true to apparent surface areas, known as the roughness ratio. According to equation (\ref{eqn:exp_wenzel}), surface roughness amplifies the wetting nature of the parent surface: A lyophilic surface ($\theta_{\rm{e}}<90$\textdegree) becomes more lyophilic in the presence of roughness, while a lyophobic surface ($\theta_{\rm{e}}>90$\textdegree) becomes more lyophobic due to roughness. However, as shown by the experimental studies \cite{forsberg2010contact}, a lyophilic surface can exhibit contact angles greater than 90\tc due to roughness. Also, as increased roughness may affect the state of the wetting, the notion of an amplified wetting nature due to roughness may not be always true. For a heterogeneous (or Cassie/composite) wetting state, the Cassie-Baxter equation \cite{cassie1944wettability} relates the contact angle ($\theta_{\rm{m}}=\theta_{\rm{C}}$) to Young's angle ($\theta_{\rm{e}}$) and the area fractions $f_1$ and $f_2$ defined as the area of the liquid which is in contact with the roughness and the liquid area which is not contacting the roughness per unit projected area of the surface, respectively:
\begin{equation}
    \cos\theta_{\rm{C}} = f_1 \cos \theta_{\rm{e}} - f_2.
    \label{eqn:exp_cassie}
\end{equation}
Like the Wenzel equation, the Cassie-Baxter equation also predicts that roughness affects the contact angle on a surface. However, note that the wetted and non-wetted area fractions $f_1$ and $f_2$ are not known \textit{a-priori} for an arbitrary surface. 

While useful concepts, both the Cassie-Baxter and Wenzel equations only predict a single contact angle instead of a range of contact angles as observed experimentally \cite{huh1977effects,oliver1977resistance,ramiasa2014influence}. On a real surface, the contact angle depends not only upon the surface roughness and its chemical nature but also upon the history of the wetting process. When a particular state is obtained by increasing the volume of the liquid, the contact angle (known as the advancing contact angle) is greater than the contact angle obtained when the volume of the liquid is reduced (known as the receding contact angle). The difference between the advancing ($\theta_{\rm{a}}$) and receding ($\theta_{\rm{r}}$) contact angle is known as contact angle hysteresis ($\Delta \theta_{\rm{cah}}=\theta_{\rm{a}}-\theta_{\rm{r}}$). Contact angle hysteresis (CAH) is an important parameter which significantly characterizes the nature of wetting processes on real surfaces as it determines the mobility of contact lines. For example, a super-hydrophobic surface requires a high contact angle ($\theta_{\rm{m}} > 150$\textdegree) and very low hysteresis ($\Delta \theta_{\rm{cah}}<10$\tc or less) \cite{kim2018superhydrophobicity}. Note that liquids wetting in the heterogeneous state generally exhibit a high contact angle and low hysteresis, while liquids wetting in the homogeneous state generally exhibit large CAH, with values in excess of 100\tc being reported \cite{reyssat2009contact}.

Due to this very high contact angle hysteresis, the study of liquids wetting in a homogeneous state is important as it may be detrimental to the performance of systems where liquid mobility on the surface is important - example, liquid-repellent surfaces. Since the discovery of the self-cleaning property of a lotus leaf \cite{barthlott1997purity}, many studies have been conducted to  understand the behaviour of liquids wetting in the Cassie or heterogeneous state \cite{choi2009modified,li2022mimicking,shirtcliffe2010introduction,marmur2004lotus,quere2008wetting,bonn2009wetting}. Conversely, with a few exceptions \cite{swain1998contact,wolansky1998actual,wolansky1999apparent,dorrer2008drops,yeh2008contact,forsberg2010contact}, the study of wetting in the homogeneous state has received far less attention. Forsberg et al. \cite{forsberg2010contact} measured advancing and receding contact angles on structured surfaces with square cross-section pillars arranged in a square array. Even though the inherent surface was hydrophilic (inherent advancing angle, that is the advancing angle on a chemically similar but flat surface ($\theta_{\rm{A}}$) was 72\textdegree) they observed advancing angles as high as 140\textdegree, clearly demonstrating the inapplicability of Wenzel's equation (which predicts $\theta_{\rm{m}}<72$\textdegree). With their numerical and experimental results, Dorrer et al. \cite{dorrer2008drops} also showed the limitations of the Wenzel equation in relating contact angles to surface roughness. Other researchers \cite{bartell1953effect,bartell1953surface,pease1945significance,erbil2014debate,gao2007wenzel} have noted the inability of Wenzel's equation to predict the wetting behaviour of rough surfaces.

With the deficiencies of Wenzel and Cassie-Baxter models recognised, some research has focused on relating CAH and surface roughness (and Young's angle). Cox \cite{cox1983spreading} proposed an analytical model for CAH during a homogeneous wetting state on smoothly varying rough surfaces with low roughness gradients. Iliev et al. \cite{iliev2018contact} and Promraksa et al. \cite{promraksa2012modeling} numerically obtained the advancing and receding contact angles on smoothly varying rough surfaces, but did not extend this to the wetting behaviour on surfaces with mesa defects (with step irregularities). In our previous work \cite{kumar23}, we proposed an equation for predicting advancing and receding contact angles on surfaces with mesa defects, but that analysis only considered defects arranged in a regular pattern. Semprebon et al. \cite{semprebon2012advancing} also investigated homogeneous wetting numerically, but on structured surfaces. With the current state-of-the-art micro-fabrication techniques, the wetting on surfaces with a structured array of pillars has been studied to some extent \cite{reyssat2009contact, gauthier2014finite, dorrer2008drops, forsberg2010contact, yeh2008contact, xu2012sticky, papadopoulos2012wetting, barbieri2007water, callies2005microfabricated}. However, wetting on surfaces with randomly distributed pillars is scantly studied. Random surfaces are important to study as most of the industrially relevant and naturally occurring surfaces possess some type of random roughness, be it randomness of defect geometry, defect dimensions or defect distribution. Most of the work on random surfaces is either done for chemically heterogeneous surfaces \cite{Iliev2013, David2010a,  Woodward2000, Decker1997} or surfaces with a random distribution of asperities of random geometries \cite{Iliev2018a, Ramos2003, Xiao2018, Yuan2017, David2013}. Wetting on surfaces with roughness in the form of identical pillars (both in geometry and chemical nature) distributed randomly has not been studied. 

To develop a fully predictive model for real surfaces, it is important to understand the effect of each parameter which defines the surface roughness, i.e. roughness geometry, chemical nature and distribution. Through careful experiments, in this work, we understand the effect of the roughness distribution, while keeping the other parameters unchanged. We experimentally measured the advancing and receding contact angles on surfaces with randomly distributed cylindrical pillars of unit aspect ratio using different liquids; specifically, DI water, dimethyl sulfoxide (DMSO), dimethylformamide (DMF), acetonitrile (ACN) and heptanol. Based on the framework of mechanical energy balance (MEB) \cite{dhcontact09} and our measurements, we propose a predictive equation for contact angle hysteresis on random surfaces with cylindrical pillars of unit aspect ratio over a range of Young's angles. We also predict the range of area fractions for a given inherent advancing ($\theta_{\rm{A}}$) or receding ($\theta_{\rm{R}}$) contact angle, inside which the proposed equation is valid based on a mechanical energy analysis of alternative wetting states. Finally, we discuss the difference in the wetting behaviour of random and structured (hexagonally arranged pillars) surfaces in the Wenzel state by analysing the contact angles measured on them.

\section{Experimental section}
\label{sec:experimental}

\subsection{Surface fabrication}

\begin{figure}
    \centering
    \includegraphics[width=0.48\textwidth]{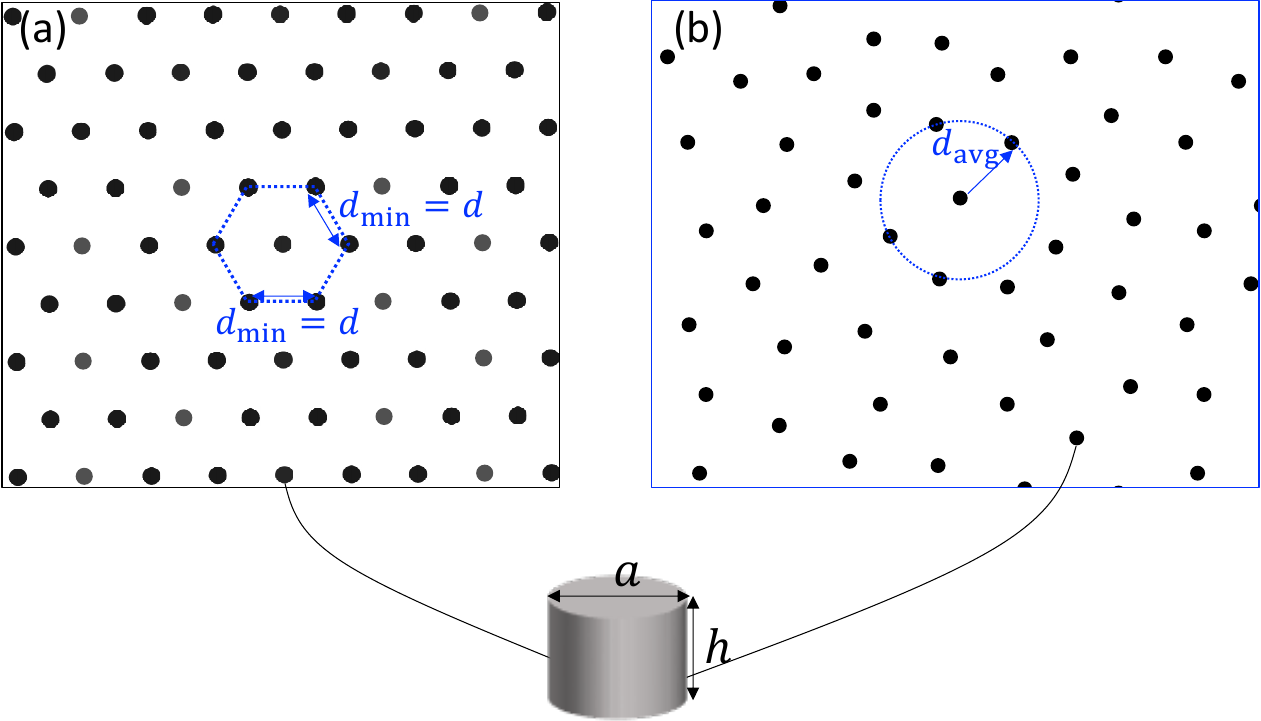}
    \caption{Schematic representation of (a) hexagonal and (b) random surface designs. Prepared surfaces have a unit aspect ratio, i.e. $a=h=10$ $\mu$m. The average distance and minimum distance between pillar centres are represented by $d_{\rm{avg}}$ and $d_{\rm{min}}$ respectively and the distance between neighbouring pillar centres on the hexagonal surface is represented by $d$.}
    \label{fig:exp_surface_design}
\end{figure}

Surfaces with random and hexagonally arranged cylindrical pillars (figure \ref{fig:exp_surface_design}) were prepared by U.V. photolithography on silicon wafers (refer to supplementary material S1 for more information on photolithography technique).
The silicon wafers were then diced into $1\times1$ cm$^2$ squares to facilitate measurements on the goniometer (contact angle measurement). To generate random surfaces, we used an in-house code to create a Gerber format file (.gbr) which was then used to write the photomask. The random surface designs were produced by randomly placing surface defects until a specified area fraction was achieved, subject to a minimum separation distance ($d_{\rm{min}}$) between pillar centres. On a random surface, the pillar centres are constrained to be at least $d_{\rm{min}}$ apart, whereas 
on a structured surface $d_{\rm{min}}$ is constant and depends upon the structure geometry. For example, for hexagonally structured pillars of circular cross section (of diameter $a$), $d_{\rm{min}}$ (which is the same as the inter-pillar distance $d$) is related to the pillar area fraction ($\phi$) as
\begin{equation}
    \frac{d_{\rm{\rm{min}}}}{a} = \sqrt{\frac{\pi}{2\sqrt{3}\phi}}.
    \label{eqn:exp_d_min}
\end{equation}
Apart from the minimum distance between pillar centres, we can also define an average distance between the pillar centres ($d_{\rm{avg}}$), as
\begin{equation}
    \frac{d_{\rm{avg}}}{a} = \frac{1}{\sqrt{\phi}}.
    \label{eqn:exp_d_avg}
\end{equation}
For a given pillar diameter ($a$), the average pillar separation ($d_{\rm{avg}}$) is only a function of the pillar area fraction ($\phi$) and is independent of their distribution. 

In this study, cylindrical pillars of height ($h=10$ $\mu$m) and diameter ($a=10$ $\mu$m) were used. 
Table \ref{tab:exp_surfaces_liquids} lists the surfaces and different liquids used in this study. 
\begin{table*}[t]
  \centering
  \caption{\label{tab:exp_surfaces_liquids}Details of the different surfaces and liquids used in the study. $d_{\rm{avg}}$ and $d_{\rm{min}}$ are the average and minimum distance between the pillar centres as defined in equations (\ref{eqn:exp_d_min}) and (\ref{eqn:exp_d_avg}) respectively.}
  \begin{tabular}{M{1 cm}M{2.5 cm}M{1.0 cm}M{1.0 cm}M{0.5 cm}M{0.5 cm}M{0.5 cm}M{0.5 cm}M{0.5 cm}}
  \toprule
    $\phi$ & design & \makecell{$d_{\rm{avg}}$\\($\mu$m)} & \makecell{$d_{\rm{min}}$\\($\mu$m)} & \rotatebox{90}{DI water} & \rotatebox{90}{DMSO} & \rotatebox{90}{DMF} & \rotatebox{90}{ACN} & \rotatebox{90}{Heptanol}\\
  \midrule
            0.02 & random & 70.7 & 13.0 & \checkmark & \checkmark & \checkmark & \checkmark & \checkmark\\
              0.02 & hexagonal & 70.7 & 67.3 & \checkmark & \checkmark &  &  & \\
                      
             0.03 & random & 57.7 & 12.0 & \checkmark & \checkmark & \checkmark & \checkmark & \checkmark\\
               
             0.03 & hexagonal & 57.7 & 55.0 & \checkmark & \checkmark &  &  & \\ 
             
             0.04 & random & 50.0 & 11.5 & \checkmark & \checkmark & \checkmark & \checkmark & \checkmark\\
               
             0.04 & hexagonal & 50.0 & 47.6 & \checkmark & \checkmark &  &  & \\ 
               
             0.05 & random & 44.7 & 11.5 & \checkmark & \checkmark & \checkmark & \checkmark & \checkmark\\
                 
             0.05 & hexagonal & 44.7 & 42.6 & \checkmark & \checkmark &  &  & \\ 
               
             0.08 & random & 35.4 & 11.0 & \checkmark & \checkmark & \checkmark & \checkmark & \checkmark\\
              
             0.08 & hexagonal & 35.4 & 33.7 & \checkmark & \checkmark &  &  & \\ 
                
             0.10 & random & 31.6 & 11.0 & \checkmark & \checkmark & \checkmark & \checkmark & \checkmark\\
                  
             0.10 & hexagonal & 31.6 & 30.1 & \checkmark & \checkmark &  &  & \\ 
            
             0.15 & random & 25.8 & 10.8 & \checkmark & \checkmark & \checkmark & \checkmark & \checkmark\\
                  
             0.15 & hexagonal & 25.8 & 24.6 & \checkmark & \checkmark &  &  & \\ 
             
             0.20 & random & 22.4 & 10.5 & \checkmark & \checkmark & \checkmark & \checkmark & \checkmark\\
                
             0.20 & hexagonal & 22.4 & 21.3 & \checkmark & \checkmark &  &  & \\ 
           
             0.25 & random & 20.0 & 10.2 & \checkmark & \checkmark & \checkmark & \checkmark & \checkmark\\
               
             0.25 & hexagonal & 20.0 & 19.0 & \checkmark & \checkmark &  &  & \\ 
            
             0.30 & random & 18.3 & 10.2 & \checkmark & \checkmark & \checkmark & \checkmark & \checkmark\\            
             0.30 & hexagonal & 18.3 & 17.4 & \checkmark & \checkmark &  &  & \\ 
    \bottomrule
  \end{tabular}
\end{table*}

\subsection{Surface cleaning}

Prior to coating the samples with a suitable monolayer, the samples were cleaned by dipping in acetone for 2 minutes followed by isopropanol for 1 minute and then rinsing with DI water. The samples were then dried in a nitrogen gas stream and exposed to oxygen plasma (under a vacuum of 350-400 mTorr) for 10 minutes. 
More details on the cleaning methods are given in supplementary material S2.

\subsection{Surface coating}

The gold-coated surfaces (see supplementary material S1 for details about the gold coating of the samples) were hydrophobized by dipping in a 1 mM solution of octadecanethiol (ODT) in toluene for 30 minutes, which forms a uniform monolayer on the surface. Samples were then washed thoroughly with ethanol, rinsed in DI water and then dried under a stream of nitrogen gas (refer to supplementary material S2 for details). Octadecanethiol was used to make the surfaces hydrophobic and chemically homogeneous as it forms a stable monolayer on gold surfaces which does not undergo any deterioration for prolonged periods \cite{bain1989formation} and has a low inherent CAH.

\subsection{Contact angle measurements}
\label{sec:exp_ca_measurement}

The advancing and receding contact angles were measured by using the sessile drop method on a Dataphysics OCA 15Pro goniometer\footnote{A typical goniometer setup is accurate within $\pm 2$\tc \cite{law2016surface}.}. To measure the advancing and receding contact angles, a small droplet of 2 $\mu \rm{L}$ was deposited on the sample using a gas-tight syringe (Hamilton, 1750 TLL 500 $\mu \rm{L}$) and a 30 gauge (0.312 mm outer diameter and 0.159 mm inner diameter) needle (Hamilton N730). After depositing the droplet, the needle was kept in place and 4 $\mu \rm{L}$ of more liquid was pumped into the droplet at a rate of 0.06 $\mu \rm{L}/s$ for DI water and 0.1 $\mu \rm{L}/s$ for dimethyl sulfoxide (DMSO), dimethylformamide (DMF), acetonitrile (ACN) and heptanol. A higher flow rate was used for these liquids due to their hygroscopic nature and high volatility. In previous studies, different flow rates have been used for the measurement of static advancing and receding contact angles: For example, Dorrer et al. \cite{dorrer2008drops} and Di et al. \cite{di2021exploring} used 0.1 $\mu \rm{L}/s$ while Forsberg et al. \cite{forsberg2010contact} and Fetzer at al. \cite{fetzer2011exploring} used 0.06 $\mu \rm{L}/s$. The flow rate is usually kept below 0.2 $\mu \rm{L}/s$ \cite{law2016surface} so that static, rather than dynamic contact angles are measured. Videos of the droplet motion were recorded using the SCA 20 software at a frame rate of 50 fps. Once the drop volume reached 6 $\mu \rm{L}$, the device setting was moved to \textit{reverse displacement} mode and the liquid was pumped out of the droplet at a rate of 0.1 (or 0.06 for DI water) $\mu \rm{L}/s$. The entire video was then analysed to measure the advancing and receding contact angles and droplet base diameter by using the \textit{ellipse fit method} provided by the SCA 20 software.

\begin{figure*}
    \centering
    \includegraphics[width=0.98\textwidth]{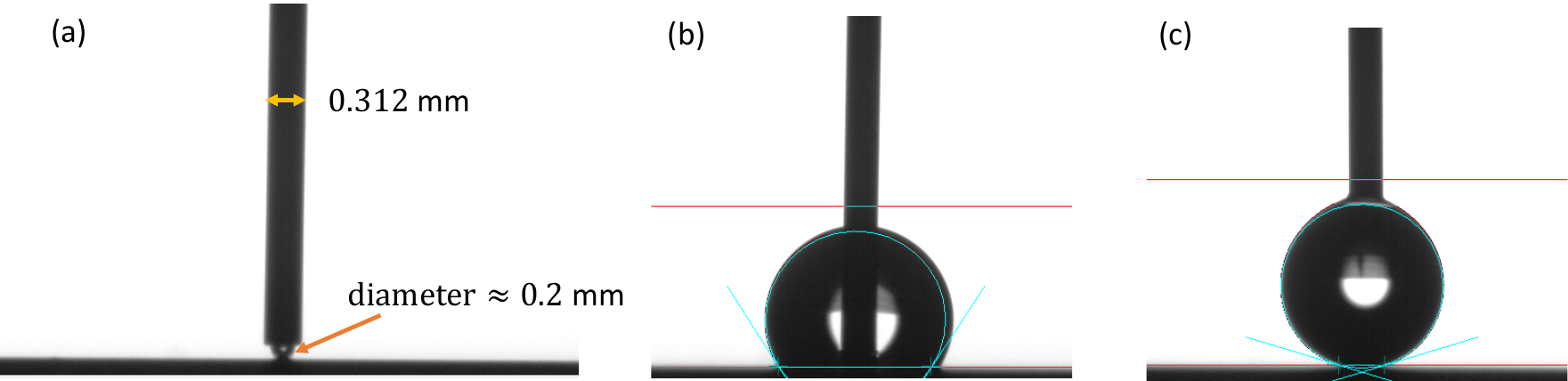}
    \caption{Shows a 2 $\mu \rm{L}$ DI water droplet being deposited on a surface coated with octadecanethiol (ODT) ($\phi=0.05$) demonstrating the influence of needle height on achieved wetting state. (a) and (b): The drop is deposited with a needle approximately 0.2 mm away from the surface. The resulting droplet is in the Wenzel state. (c) The drop is deposited with a needle approximately 1.5 - 2 mm away from the surface. The resulting droplet is in Cassie's state.  In (b) and (c), the straight lines parallel to the solid surface set the region of interest (ROI) and an elliptical profile (shown in cyan) is fitted to the portion of the droplet which is inside the ROI. In (b), the fitted profile shows the interface tracking from a previous time point.}
    \label{fig:cassie_or_wenzel}
\end{figure*}

On a rough surface, a droplet can exist in either a Wenzel or a Cassie wetting state\footnote{Assuming that $\theta_{\rm{A}}$ is large enough to avoid any hemiwicking behaviour.}, depending upon the pillar area fraction and the chemical nature of the surface. We observed that in addition, the way a droplet is deposited can also influence the type of wetting regime. For example, the height of the needle above the surface during the initial drop deposition plays an important role in deciding the wetting state of the droplet. Figure \ref{fig:cassie_or_wenzel}(a) shows a drop being deposited on a surface. If the needle is held too close to the surface, the initial droplet has a very small radius and a high Laplace pressure when it touches the surface. In the case shown in figure \ref{fig:cassie_or_wenzel}(a), the needle (30G needle with an inner diameter of 0.159 mm) is held at a distance of approximately 0.2 mm from the surface. The drop generated has a diameter of 0.2 mm (approximately). For this diameter, the DI water droplet would have an excess Laplace pressure of $2 \sigma_{12} / R \approx 1.46$ $\rm{kPa}$ (where, $\sigma_{12}$ is the surface tension of the water-air interface, 0.0728 N/m at room temperature and $R$ is the radius of the droplet which is approximately 0.1 mm). The high Laplace pressure pushes the liquid inside the rough structure and the droplet can exist in a Wenzel wetting state even for moderately high values of $\phi$ (figure \ref{fig:cassie_or_wenzel}(b)). Conversely, holding the needle further from the substrate can result in a heterogeneous wetting state for the same surface (figure \ref{fig:cassie_or_wenzel}(c)). With DI water droplets and a close needle position, we were able to achieve a Wenzel wetting state up to a pillar area fraction of 0.08 (see figures \ref{fig:cassie_or_wenzel}(a) and \ref{fig:cassie_or_wenzel}(b)) by keeping the needle tip close to the rough surface during deposition. For our measurements, we have used the close needle approach for generating droplets as we are primarily interested in Wenzel wetting states in this study.

\section{Results and discussion}
\label{sec:result_discussion_ch7}

\subsection{Inherent (flat) advancing and receding contact angle: Definition}
\label{sec:angle_definition}
\begin{figure}
    \centering
        \includegraphics[width=0.42\textwidth]{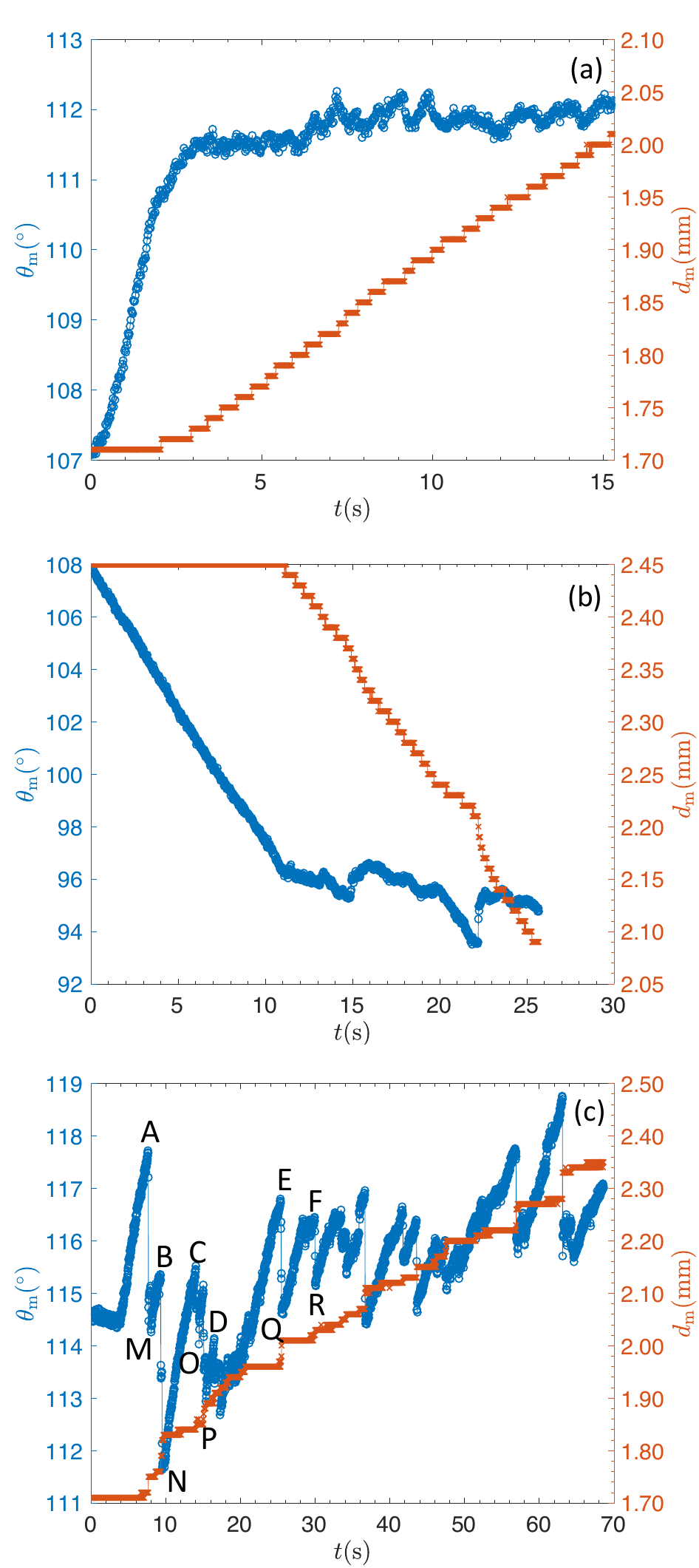}
        \caption{Variation in contact angle ($\theta_{\rm{m}}$, in degrees) and base diameter ($d_{\rm{m}}$, in mm, resolution 0.01 mm) with time ($t$, in seconds) during the advancing (a) and receding (b) motion of the interface on a flat surface. (c) Typical variation in contact angle ($\theta_{\rm{m}}$, in degrees) and base diameter ($d_{\rm{m}}$, in mm, resolution 0.01 mm) with time ($t$, in seconds) during the advance of an interface on a random surface. Points A, B, C, D, E and F represent some of the peaks and points M, N, O, P, Q and R represent some of the troughs in the contact angle during the interface advancement. In all of the figures, contact angles are shown by blue circles and base diameters by red crosses. }
        \label{fig:adv_rec_flat}
\end{figure}

Figures \ref{fig:adv_rec_flat}(a) and \ref{fig:adv_rec_flat}(b) show a typical variation in the droplet contact angle and base diameter on a flat surface\footnote{A surface is called flat here if it is free of microscopic pillars. However, the surface may still have roughness at length scales below a few microns. In fact, the surface does have a roughness at the nanometric length scales (see supplementary material S4).} as the liquid volume is added and then removed from the droplet. Initially, as the droplet volume is increased the contact angle increases while the base diameter ($d_{\rm{m}}$) remains constant. This continues until the contact angle crosses a threshold, after which the base diameter starts increasing and the triple-phase contact line (TPCL) starts advancing. During this advance, the contact angle is at the inherent advancing contact angle ($\theta_{\rm{A}}$) \cite{de2004capillarity} and within the resolution of the data, the base diameter increases continuously. The initial increase in the contact angle occurs while the TPCL is pinned, as evidenced by the constant droplet diameter during this period. Similarly, when the droplet volume is decreased, the contact angle initially decreases while the base diameter remains constant until the TPCL becomes depinned. The contact angle measured while the TPCL recedes is the inherent receding contact angle ($\theta_{\rm{R}}$) \cite{de2004capillarity}. 

Before proceeding further, we address the presence of inherent hysteresis on surfaces - that is, $\theta_{\rm{A}} \neq \theta_{\rm{R}}$. This may be due to the presence of surface roughness at lengthscales below a few microns, chemical heterogeneity and/or irreversibility in the creation/destruction of surfaces at a molecular scale. We used AFM to scan the pillar tops and base surfaces (see supplementary material S4) and observed a nanometric scale roughness on both, which is consistent with the inherent CAH being caused by this small-scale roughness. Also, $\theta_{\rm{A}}$ and $\theta_{\rm{R}}$ were measured on both the tops and base surfaces and consistent with the AFM roughness measurements, found to be the same.
Additionally, from figures \ref{fig:adv_rec_flat}(a) and \ref{fig:adv_rec_flat}(b), the average velocity of the TPCL during advancing and receding motion turns out to be approximately 0.010 mm/s and 0.013 mm/s respectively which is small enough to consider the contact angles as static advancing and receding contact angles \cite{dhcontact09}.

\subsection{Wetting on random surfaces}
\label{sec:random_wetting}

Wetting on a surface with random roughness at micron scales is, however, very different. Figure \ref{fig:adv_rec_flat}(c) shows a typical contact angle and base diameter variation for the wetting on a randomly structured surface. During the entire advance of the interface, the TPCL exhibits a \textit{stick-slip} behaviour which is evident from the appearance of much larger peaks and troughs in the contact angle data. Referring back to figure \ref{fig:adv_rec_flat}(c), as the droplet expands the TPCL remains pinned until point A is reached (the base diameter does not change). After this, the TPCL jumps and again pins at a new position (point M), evidenced by an increase in base diameter and a drop in the contact angle. Thus point A represents a critical point at which the TPCL depins. Similarly, points B, C, D, E, F etc. are also critical points at which the TPCL depins, while points, M, N, O, P, Q, R etc. are the points where the TPCL sticks again after depinning from a previous location. We can see that the magnitude of the contact angle is different at all of these points, rather than being constant as per the conventional advancing contact angle definition. 

To better quantify this behaviour, a methodology is presented to find an appropriate average advancing (or receding) contact angle that captures the wetting state from this data. The idea of an average advancing/receding contact angle has been used previously in the literature \cite{david2013contact,david2010computation,iliev2003wetting, iliev2014contact, brandon2003partial, iliev2013contact}, but not in the context of homogeneous wetting of surfaces with identical pillars distributed randomly. For example, David et al. \cite{david2013contact} numerically studied the variation in contact angle as a solid plate with random self-affine roughness is lowered into and then withdrawn from a pool of liquid. The average advancing and receding contact angles were obtained by averaging the contact angles as the plate was virtually moved. However, it was not clearly specified when the averaging of the contact angle, especially the receding contact angle, should be started, as the plate moves a certain distance before the sawtooth variation in the receding contact angle can be observed. A similar averaging methodology was used in a computational study of the wetting of surfaces with a random chemical heterogeneity by David and Neumann \cite{david2010computation}. Most of the averaging methods in the literature are for averaging the contact angle over the entire length of the TPCL for a stationary interface \cite{iliev2003wetting, iliev2014contact, brandon2003partial, iliev2013contact} (i.e., over space), while here we present an approach for averaging the contact angles during the interface movement over a surface (i.e., over time).

To explain our averaging method, we first carefully summarise our experimental procedure. First, a small droplet (2 $\mu \rm{L}$) is generated in contact with the surface. This droplet serves as the starting point for measuring the average advancing angle. The contact angle exhibited by this initial droplet is neither necessarily the advancing nor receding angle. There are two main reasons for this: Firstly, the initial droplet has some inertia when deposited which may or may not be sufficient for the depinning of the TPCL, meaning that a particular point along the TPCL could be anywhere within the CAH range after the initial deposition. The second reason is that due to evaporation the contact angle of the deposited droplet generally decreases with time. Evaporation becomes even more important for liquids which are highly volatile. Therefore, starting the averaging of the contact angles at the moment when the droplet is deposited on a surface may result in an under-prediction of the average advancing contact angle. Hence, after deposition and before movement of the local interface around the TPCL the droplet could have a measured contact angle ($\theta_{\rm{m}}$) anywhere between the receding ($\theta_{\rm{r}}$) and advancing ($\theta_{\rm{a}}$) contact angle and is not until all surface is advancing over the roughness structure that $\theta_{\rm{m}}$ approaches $\theta_{\rm{a}}$. The essence of this averaging method is that not until the interface has moved by a significant multiple ($\zeta$) of $d_{\rm{avg}}$ can we be sure that we are measuring $\theta_{\rm{a}}$ (or alternatively, $\theta_{\rm{r}}$). Note that $d_{\rm{avg}}$ is the average distance between centres of the neighbouring pillars as given by equation (\ref{eqn:exp_d_avg}). Similarly at the end of an advance (or recede) motion of the TPCL, the measured angle may not be representative of the true advancing (or receding ) angle within $\zeta d_{\rm{avg}}$ of the stopping position of the interface. Hence, our averaging method neglects angles measured within $\zeta d_{\rm{avg}}$ of both the start and end of interface movement. The specific algorithm we use to accomplish this is detailed in supplementary material S6. We also calculate the standard deviation in advancing and receding contact angles ($\Delta \theta_{\rm{sd,a}}$, $\Delta \theta_{\rm{sd,r}}$) within the same starting and stopping time limits, as additional parameters characterising the wetting states.

In addition to the average value of the advancing angle, another important characteristic of random surface wetting is the contact angle peaks corresponding to the points where the TPCL slips and troughs corresponding to the point where the TPCL sticks after slipping, based on the changes in the droplet base diameter. The difference between the contact angle peaks and the troughs represents the magnitude of contact angle jumps occurring during the advancing motion of the interface and is a function of the pillar area fraction ($\phi$). For example, if the pillars are spaced sparingly, the TPCL covers a larger distance before sticking to the next set of pillars and hence the peak-to-trough angle difference is high. The average peak-to-trough angle difference, therefore, should approximately scale with the average distance between pillar centres ($d_{\rm{avg}}$). The method for calculating the average peak-to-trough difference is detailed in supplementary material S6. 

In summary, we find the following measures that characterise the advancing wetting process
\begin{itemize}
    \item average advancing contact angle ($\theta_{\rm{a}}$) and its standard 
    deviation ($\Delta \theta_{\rm{sd,a}}$).
    \item average peak-to-trough angle difference ($\Delta \theta_{\rm{pt,a}}$) 
    and its standard deviation ($\Delta \theta_{\rm{sd,pt,a}}$).
\end{itemize}

The above parameters are all calculated over the same data range that excludes $\zeta d_{\rm{avg}}$ worth of data from the start and end of each advance or recede movement. 

What remains is to chose a suitable $\zeta$ parameter such that starting/ending effects are neglected, but minimal data loss occurs. We chose $\zeta=2.0$ because it represents an interface advancing $d_{\rm{avg}}$ everywhere around the circumference of a droplet, that is the TPCL has advanced by a distance of $d_{\rm{avg}}$ in every direction prior to commencing angle averaging. Another reason for choosing $\zeta=2.0$ is to minimize the loss in data points while still capturing the averaged angle based on the physical interpretation of the distance moved by the contact line. The particular choice of $\zeta$ is discussed in detail in supplementary material S6.

\subsection{Energy dissipation during advancing motion of an interface in a homogeneous wetting state}
\label{sec:exp_dissipation}

In \S\ref{sec:random_wetting} we presented a typical contact angle and base diameter variation for a droplet during the advancing (or receding) motion of an interface on a solid with random roughness. We observed that the contact angle exhibits discontinuities and the base diameter exhibits step-like behaviour with time, even though the drop volume is being increased (or decreased) continuously. This behaviour of the contact angle and base diameter is a result of the \textit{stick-slip} motion of the TPCL, meaning that the TPCL moves in rapid and discrete jumps while remaining pinned to the pillars for most of the time. Energy is dissipated every time the TPCL executes a jump \cite{huh1977effects}. Recently, Harvie \cite{dhcontact09} showed that this dissipation in energy is related to the macroscopically measured advancing and receding contact angles ($\theta_{\rm{a}}, \theta_{\rm{r}})$, and can be calculated from the change in the interfacial energies within in a control volume (CV) that is moving along with the TPCL at a constant but slow speed (see \cite{dhcontact09} for details). In this section, we develop a relationship between energy dissipation and surface roughness (i.e., area fraction and Young's angle for a fixed pillar geometry) based on our experimental results which relate to an interface advancing over a rough surface in a Wenzel (or homogeneous) wetting state.

The advancing/receding contact angle is related to Young's angle ($\theta_{\rm{e}}$), surface roughness ($r=(1+4\phi)$, for unit aspect ratio pillars) and energy dissipation as \cite{dhcontact09}
\begin{equation}
\begin{split}
 \cos\theta_{\rm{a}} &= r\cos\theta_{\rm{e}} - D_{\rm{a}}/ \sigma_{12},\\
 \cos\theta_{\rm{r}} &= r\cos\theta_{\rm{e}} + D_{\rm{r}}/ \sigma_{12},
 \end{split}
 \label{eqn:exp_meb}
\end{equation}
where $D_{\rm{a}}/D_{\rm{r}}$ represents the total energy dissipation caused by the \textit{stick-slip} motion during the advancing/receding motion of the interface divided by the average area traversed by the TPCL (denoted as $A_{\rm{CV}}$). The total energy dissipation ($D$) is related to the changes in interfacial areas (fluid-1/fluid-2, solid/fluid-1 and solid/fluid-2 interfaces) during TPCL jumping events and the TPCL area, $A_{\rm{CV}}$, that is \cite{dhcontact09}
\begin{equation}
    D = -\sum_{k=1}^N \sum_{i<j} \sigma_{ij} \frac{\widehat{\Delta A_{ij}}_k}{A_{\rm{CV}}},
    \label{eqn:exp_diss_dalton}
\end{equation}
where $\widehat{\Delta A_{ij}}_k$ is the change in interfacial areas during the $k$th TPCL jump and $N$ is the total number of TPCL jumps. Equations (\ref{eqn:exp_meb}) and (\ref{eqn:exp_diss_dalton}) are valid under certain constraints, that is
\begin{equation}
    \begin{split}
        h_{\rm{mol}} &\ll h_{\rm{rough}} \ll r_{\rm{CV}} \ll h_{\rm{surround}}, r_{\rm{CV,grav}},\\
        &v_{\rm{CV}} \ll \text{min}(v_{\rm{CV,ke}}, v_{\rm{CV,vis}}, v_{\rm{CV,cap}}).
    \end{split}
    \label{eqn:exp_lengthscales}
\end{equation}
Here $h_{\rm{mol}},h_{\rm{rough}},h_{\rm{surround}}$ represent molecular, roughness and surrounding fluid flow lengthscales, $r_{\rm{CV}}$ is the control volume size, $r_{\rm{CV},grav}$ is the capillary lengthscale, $v_{\rm{CV}}$ is the control volume velocity and $v_{\rm{CV,ke}},v_{\rm{CV,vis}},v_{\rm{CV,cap}}$ are the limiting control volume velocities based on the transport of the kinetic energy through the control surface, viscous dissipation at the TPCL and the capillary velocity (ensuring that TPCL jumping time is very small compared to the timescale of the macroscopic flow). Out of the three limiting CV velocities, the capillary velocity limit is more restricting (refer to \cite{dhcontact09} for details) and therefore the control volume velocity (or the TPCL velocity during experiments) should be less than $v_{\rm{CV,cap}}$. The roughness lengthscale ($h_{\rm{rough}}$) in the present study is 10 $\mu$m, therefore the control volume size based on $h_{\rm{rough}}$ can be approximated as, $r_{\rm{CV}} \approx 10 h_{\rm{rough}}=100 \mu$m. The capillary lengthscale for the water/air system is $r_{\rm{CV,grav}}= \sqrt{\sigma_{12}/ \rho g} \approx 2700 \mu$m, where $\sigma_{12}=0.0728$ N/m, $\rho=1000$ kg/m$^3$  are the surface tension (in the air) and density of water at room temperature and $g=9.81$ m/s$^2$ is the magnitude of the acceleration due to gravity. The control volume size ($r_{\rm{CV}}=100\mu$m) is therefore much smaller than the capillary length scale ($r_{\rm{CV,grav}}=2700\mu$m) and much larger than the surface roughness ($h_{\rm{rough}}$). Also, the molecular lengthscale ($h_{\rm{mol}}$) is of the order of a few nanometers, therefore, all the lengthscale requirements mentioned in equation (\ref{eqn:exp_lengthscales}) are satisfied. Before using equations (\ref{eqn:exp_meb}) and (\ref{eqn:exp_diss_dalton}) we also need to check that the control volume (or TPCL velocity, $v_{\rm{CV}}$) is much smaller than the control volume velocity based on the capillary speed, that is $v_{\rm{CV}} \ll v_{\rm{CV,cap}}$. The capillary velocity (for water/air system) is, 
\begin{equation}
 v_{\rm{cap}} = \text{min} \left(\frac{\sigma_{12}}{\mu},\sqrt{\frac{\sigma_{12}}{\rho h_{\rm{rough}}}} \right) \approx 1.6 \:\text{m/s},
\end{equation}
where $\mu$ is the dynamic viscosity of water at room temperature. The capillary velocity based limiting control volume velocity ($v_{\rm{CV,cap}}$) can be calculated as, $v_{\rm{CV,cap}}=v_{\rm{cap}} h_{\rm{rough}}/r_{\rm{CV}} = v_{\rm{cap}}/10 \approx 0.16$ m/s. As already discussed in \S\ref{sec:angle_definition} the experimentally measured TPCL velocity in this study is of the order of a fraction of a few mm/s, therefore the condition $v_{\rm{CV}} \ll v_{\rm{CV,cap}}$ is always satisfied. Hence, all the conditions as mentioned in equation (\ref{eqn:exp_lengthscales}) are satisfied in the present study and we can use equations (\ref{eqn:exp_meb}) and (\ref{eqn:exp_diss_dalton}) (which are based on a mechanical energy balance \cite{dhcontact09}) for analysing the experimental data.

We use the experimentally measured advancing/receding contact angles along with equation (\ref{eqn:exp_meb}) to calculate the dissipation in energy during advancing ($D_{\rm{a}}$) and receding ($D_{\rm{r}}$) motion of an interface for different Young's angles ($\theta_{\rm{e}}$). To choose a form for the dissipation functions, we assume that the energy dissipated per pillar transversed, that is $D_{1}$, varies with the pillar area fraction as suggested by Joanny \& de Gennes \cite{joanny1984model} and others \cite{reyssat2009contact, kumar23} as
\begin{equation}
 D_{1} = A^{'}\ln\phi + B^{'},  
 \label{eqn:exp_D1}
\end{equation}
where $A^{'}$ and $B^{'}$ are constants depending on the surface roughness and Young's angle. If the TPCL traverses an area $A_{\rm{CV}}$ and interacts with $N$ number of pillars during the travel, then the energy dissipation (total) per unit area traversed by the TPCL, that is $D$ can be related to $D_1$ as
\begin{equation}
D = \frac{ND_1}{A_{\rm{CV}}}.
\label{eqn:exp_D_N}
\end{equation}
Noting that the total number of pillars ($N$) are related to the pillar area fraction ($\phi$) as $N = \phi A_{\rm{Cv}}/A_{\rm{top}}$, where $A_{\rm{top}}$ is the area of a pillar top, equation (\ref{eqn:exp_D_N}) can also be written as
\begin{equation}
    \overline{D} = A \phi \ln \phi + B \phi,
    \label{eqn:exp_D_bar}
\end{equation}
where $A=A^{'}/(\sigma_{12} A_{\rm{top}})$ and $B=B^{'}/(\sigma_{12} A_{\rm{top}})$ are constants and $\overline{D}/ \sigma_{12}$ is the total non-dimensional energy dissipation. With this dissipation form (equation (\ref{eqn:exp_D_bar})), the expressions for advancing/receding contact angles become
\begin{equation}
\begin{split}
 \cos\theta_{\rm{a}} &= r\cos\theta_{\rm{e}} - \overline{D}_{\rm{a}},\\
 \cos\theta_{\rm{r}} &= r\cos\theta_{\rm{e}} + \overline{D}_{\rm{r}},
 \end{split}
 \label{eqn:exp_meb_non_dim}
\end{equation}
respectively, where $\overline{D}_{\rm{a}}$ and $\overline{D}_{\rm{r}}$ are the total non-dimensional energy dissipation during the advancing and receding motion of an interface respectively.

\subsection{A predictive equation for CAH}
\label{sec:exp_equation_development}

In this section, 
we use the advancing contact angles measured with DMSO and DMF droplets and the receding contact angles measured with the DMF droplets
to calculate the non-dimensional energy dissipation using equation (\ref{eqn:exp_meb_non_dim}). We then use the form of non-dimensional energy dissipation depicted in equation (\ref{eqn:exp_D_bar}) to calculate the value of the constants $A$ and $B$. Out of all the liquids tested in the present study, we selected DMSO (advancing contact angles only) and DMF (both advancing and receding contact angles) to develop the equation as they cover a large range of inherent advancing angles ($\theta_{\rm{A}}$). Specifically, for DMSO, $\theta_{\rm{A}}=79.0$\tc and for DMF $\theta_{\rm{A}}=72.3$\tc and $\theta_{\rm{R}}=58.2$\textdegree, providing a range of equivalent inherent advancing angles ($\theta_{\rm{A}}$) from 72.3\tc to 121.8\tc ($=180$\textdegree$-\theta_{\rm{R}}$ for the receding DMF).

\begin{figure}
    \centering
    \includegraphics[width=0.48\textwidth]{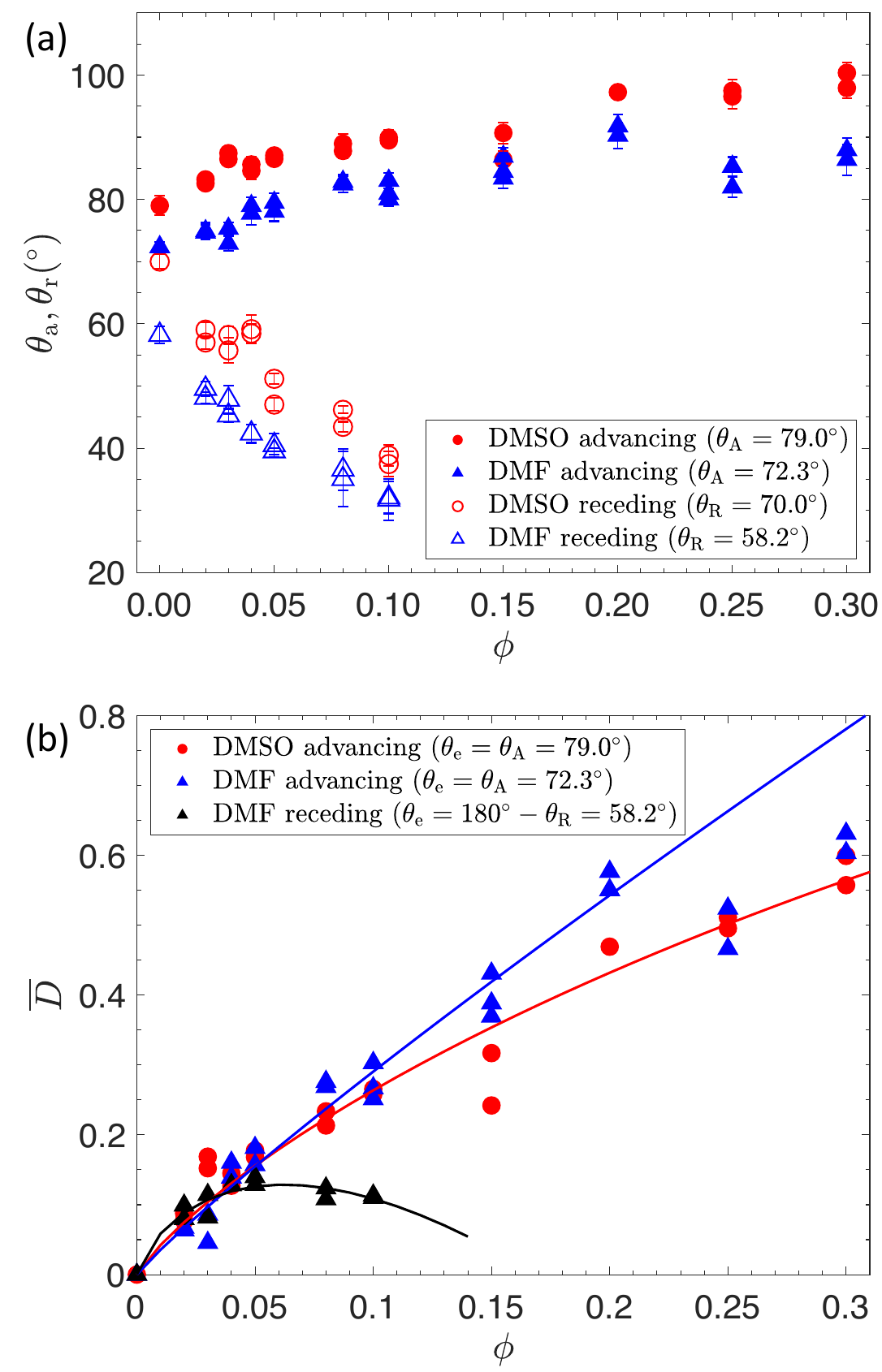}
    \caption{(a) Advancing and receding contact angles (in degrees) measured on random surfaces with DMSO and DMF droplets. The advancing and receding contact angles are shown by filled and empty symbols respectively. All the results are for the Wenzel wetting state. (b) Variation in the total non-dimensional energy dissipation ($\overline{D}$) with the pillar area fraction ($\phi$) calculated from the experimentally measured advancing (DMSO and DMF) and receding (DMF) contact angles on random surfaces. The fitting equations of the form as given in equation (\ref{eqn:exp_D_bar}) are also shown as solid lines.}
    \label{fig:dmso_dmf_exp}
\end{figure}

Figure \ref{fig:dmso_dmf_exp}(a) shows the variation in the average advancing and receding contact angles plotted against the pillar area fraction ($\phi$). Results are shown for dimethyl sulfoxide (DMSO) and dimethylformamide (DMF) droplets on randomly structured surfaces. 
With the increase in pillar area fraction ($\phi$), we observed an increase in advancing contact angle for the DMSO and DMF droplets. This is expected as the number of effective pinning sites increases with the area fraction. Similarly, we observed a decrease in the receding contact angle with the pillar area fraction for the same reason. The receding contact angles are shown for area fractions only up to 0.1 due to difficulty in accurately resolving the droplet profile at low contact angles as the area fraction is further increased. Also, for the DMF droplets for $\phi \geq 0.25$, we observed hemiwicking (discussed in \S\ref{sec:exp_hemiwicking}) and therefore, we neglect the advancing data for DMF droplets for $\phi \geq 0.25$ in calculating the energy dissipation. Another observation from figure \ref{fig:dmso_dmf_exp}(a) is that for both DMSO and DMF droplets, the inherent advancing angle ($\theta_{\rm{A}}$) is less than 90\textdegree. Therefore, according to Wenzel's equation (\ref{eqn:exp_wenzel}), the advancing and receding contact angles should decrease with an increase in the area fraction. However, on the contrary, we observe an increase in the advancing contact angle with area fraction. This is due to the presence of a non-zero dissipation in energy during the advancement of the interface \cite{dhcontact09}.

We now develop equations for the total non-dimensional energy dissipation ($\overline{D}$) for each of the three systems following equation (\ref{eqn:exp_D_bar}). 
Since we observed a non-zero CAH on flat surfaces with all the liquids tested in the present study, we modify the equation (\ref{eqn:exp_meb_non_dim}) to incorporate the inherent hysteresis present in the system by replacing $\theta_{\rm{e}}$ with $\theta_{\rm{A}}$ and $\theta_{\rm{R}}$ for the advancing and receding motion of the interface respectively. In figure \ref{fig:dmso_dmf_exp}(b) we show the variation in the total non-dimensional energy dissipation ($\overline{D}$) with the pillar area fraction ($\phi$) for DMSO and DMF droplets. The fitting equations of the form as proposed in equation (\ref{eqn:exp_D_bar}) are also shown. 
%
%
We observe good agreement between equation (\ref{eqn:exp_D_bar}) and the experimentally obtained values of $\overline{D}$ giving confidence that we are capturing the physical mechanisms responsible for CAH. The $A$ and $B$ parameters used for this plot are given in table \ref{tab:diss_dmso_dmf}.
\begin{table}[]
\caption{Fitting parameters and the corresponding $R^2$ values in the total non-dimensional energy dissipation equation (\ref{eqn:exp_D_bar}) calculated using equation (\ref{eqn:exp_meb_non_dim}) and the advancing (DMSO and DMF) and receding (DMF) contact angles on random surfaces measured experimentally.}\label{tab:diss_dmso_dmf}
\begin{tabular*}{\tblwidth}{p{3.0cm}p{1.5cm}p{1.5cm}p{1.0cm}}
\toprule
 System & $A$ & $B$ & $R^2$\\ 
\midrule
 DMSO advancing & -0.69 & 1.05 & 0.951\\
 DMF advancing & -0.28 & 2.27 & 0.967\\
 DMF receding & -2.07 & -3.68 & 0.602\\
\bottomrule
\end{tabular*}
\end{table}

The parameters $A$ and $B$ developed above are specific to the inherent equilibrium angle (that is $\theta_{\rm{e}}=\theta_{\rm{A}}$ or $\theta_{\rm{e}}=\theta_{\rm{R}}$ for an advancing or receding interface respectively) for each liquid advance/recede combination. To generalize this model to variations in inherent angle, for the considered pillar geometry, the parameters $A$ and $B$ can be represented by a power series function of $\theta_{\rm{e}}$, or alternatively, $\cos\theta_{\rm{e}}$. A simple form which has two degrees of freedom (to fit the three data points) is a linear function of $\cos\theta_{\rm{e}}$. Hence, we use the following equations to represent $A$ and $B$,
\begin{equation}
    \begin{split}
    &A = a_0\cos\theta_{\rm{e}} + a_1,\\
    &B = b_0\cos\theta_{\rm{e}} + b_1,
    \end{split}
    \label{eqn:exp_A_and_B}
\end{equation}
where $a_0,a_1$ and $b_0,b_1$ are the parameters depending upon the pillar geometry only. As previously, here $\theta_{\rm{e}}=\theta_{\rm{A}}$ or 180\textdegree$-\theta_{\rm{R}}$ depending on whether an advancing or receding interface is being considered. 
The values of parameters $a_0=2.08,a_1=-0.99,b_0=6.97$ and $b_1=-0.04$ are obtained by fitting equations (\ref{eqn:exp_A_and_B}) and (\ref{eqn:exp_D_bar}) to the total non-dimensional energy dissipation for DMSO and DMF droplets (see figure \ref{fig:dmso_dmf_exp}(b)). Good agreement is found across the considered data range, lending confidence that equations (\ref{eqn:exp_D_bar}) and (\ref{eqn:exp_A_and_B}) are capturing the variation in dissipation with area fraction and inherent contact angle, respectively. Based on equations (\ref{eqn:exp_D_bar}) and (\ref{eqn:exp_A_and_B}) and the value of the parameters $a_0,a_1,b_0$ and $b_1$ the dissipation terms ($\overline{D}_{\rm{a}}$ and $\overline{D}_{\rm{r}}$) in equation (\ref{eqn:exp_meb_non_dim}) can be written as 
\begin{equation}
    \overline{D}_{\rm{a}} = (2.08\cos\theta_{\rm{A}}-0.99)\phi\ln\phi + (6.97\cos\theta_{\rm{A}}-0.04)\phi,
\label{eqn:exp_diss_adv}
\end{equation}
and
\begin{equation}
    \overline{D}_{\rm{r}} = (-2.08\cos\theta_{\rm{R}}-0.99)\phi\ln\phi + (-6.97\cos\theta_{\rm{R}}-0.04)\phi.
\label{eqn:exp_diss_rec}
\end{equation}
%
Since $\overline{D}$ is a function of $\cos\theta_{\rm{A}}/ \cos\theta_{\rm{R}}$ for an advancing/receding interface, it is asymmetric in nature - i.e., the dissipation is different during the advancing and receding modes \cite{kumar23}. We denote the energy dissipation coefficients (equation (\ref{eqn:exp_A_and_B})) with subscripts `a' and `b' to represent the advancing and receding modes respectively, that is as $A_{\rm{a}},B_{\rm{a}}$ and $A_{\rm{r}},B_{\rm{r}}$.
This asymmetry in energy dissipation is also reflected in figure \ref{fig:dmso_dmf_exp}(a) where the rate of increase in the advancing contact angle ($\theta_{\rm{a}}$) with the pillar area fraction is different from the rate of decrease in the receding contact angle ($\theta_{\rm{r}}$, with pillar area fraction). 

\begin{figure}
    \centering
    \includegraphics[width=0.48\textwidth]{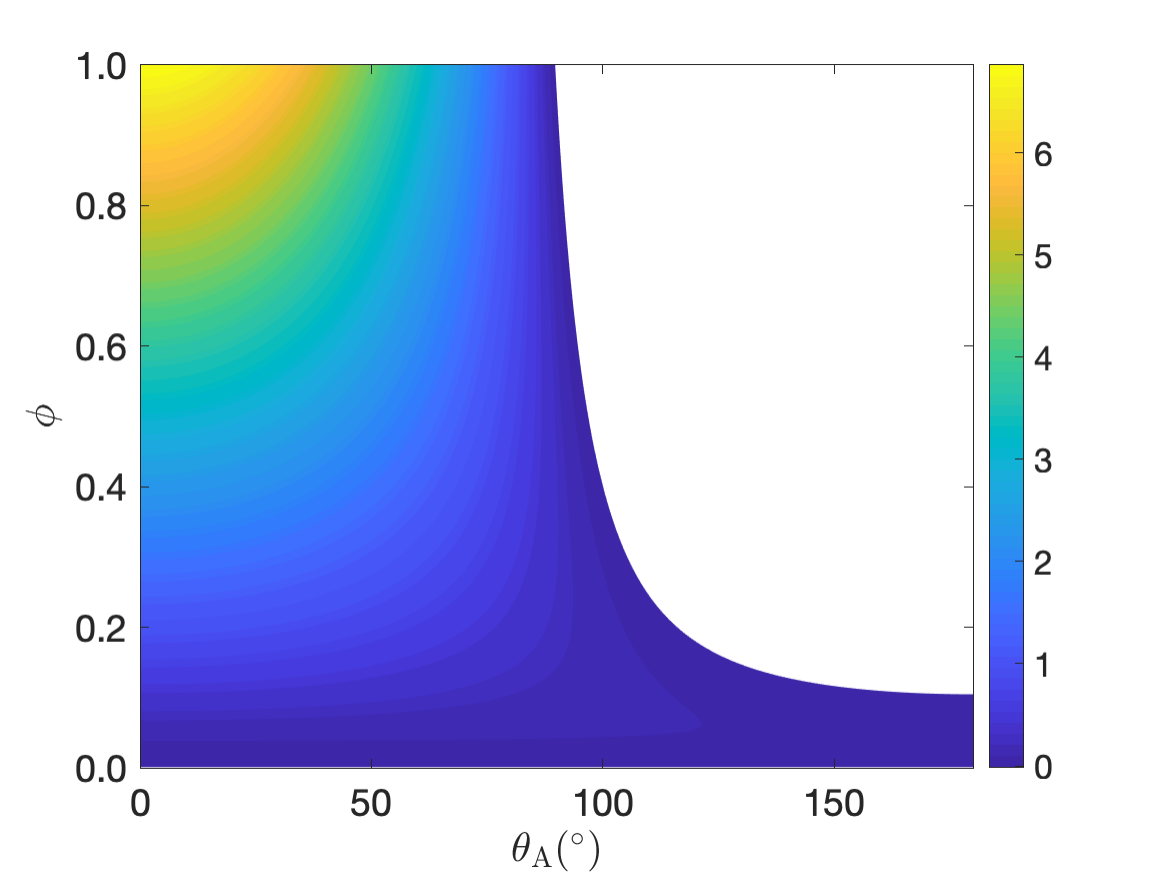}
    \caption{Variation in the total non-dimensional energy dissipation for an advancing interface on a random surface exhibiting a homogeneous wetting state ($\overline{D}_{\rm{a}}$) with the inherent advancing angle ($\theta_{\rm{A}}$, in degrees) and the pillar area fraction ($\phi$) calculated using equation (\ref{eqn:exp_diss_adv}). Only $\overline{D}_{\rm{a}}>0$ regions are shown.}
    \label{fig:exp_dissipation_3D}
\end{figure}

In figure \ref{fig:exp_dissipation_3D} we plot the variation in total non-dimensional energy dissipation for an advancing interface ($\overline{D}_{\rm{a}}$, equation (\ref{eqn:exp_diss_adv})) against the inherent advancing angle ($\theta_{\rm{A}}$) and pillar area fraction ($\phi$). We observe that $\overline{D}_{\rm{a}}$ is large at smaller advancing angles ($\theta_{\rm{A}}$) and high area fractions ($\phi$). Note that since the dissipation equation is developed based on limited experimental data (ranging from $\theta_{\rm{A}}=72.3$\tc to $\theta_{\rm{A}}=121.8$\tc and area fractions up to $\phi=0.30$), the correlation for $\overline{D}_{\rm{a}}$ is not valid for all $\theta_{\rm{A}}$ and $\phi$. In particular, for some $\theta_{\rm{A}}$ and $\phi$ combinations the predicted $\overline{D}_{\rm{a}}$ is negative, and hence non-physical. For this reason, only those values for which $\overline{D}_{\rm{a}}>0$ are plotted in figure \ref{fig:exp_dissipation_3D}.

Combining the dissipation equations (\ref{eqn:exp_diss_adv}) and (\ref{eqn:exp_diss_rec}) with the mechanical energy balance equation (\ref{eqn:exp_meb_non_dim}) gives
a predictive equation for the advancing ($\theta_{\rm{a}}$) and receding contact angles ($\theta_{\rm{r}}$) over a range of system parameters ($\theta_{\rm{A}}$, $\theta_{\rm{R}}$ and $\phi$). Specifically,
\begin{equation}
\begin{split}
    \cos\theta_{\rm{a}} = r \cos\theta_{\rm{A}} - & \left[(2.08\cos\theta_{\rm{A}}-0.99)\phi\ln\phi \right. \\
    & \left. + (6.97 \cos\theta_{\rm{A}} - 0.04)\phi \right], \\
    \cos\theta_{\rm{r}} = r \cos\theta_{\rm{R}} - & \left[(2.08\cos\theta_{\rm{R}}+0.99)\phi\ln\phi \right. \\
    & \left.+ (6.97 \cos\theta_{\rm{R}} + 0.04)\phi \right].
    \end{split}
    \label{eqn:exp_proposed_eqn}
\end{equation}

\subsection{Alternative wetting states}
\label{sec:exp_validity}

\begin{figure}
    \centering
    \includegraphics[width=0.48\textwidth]{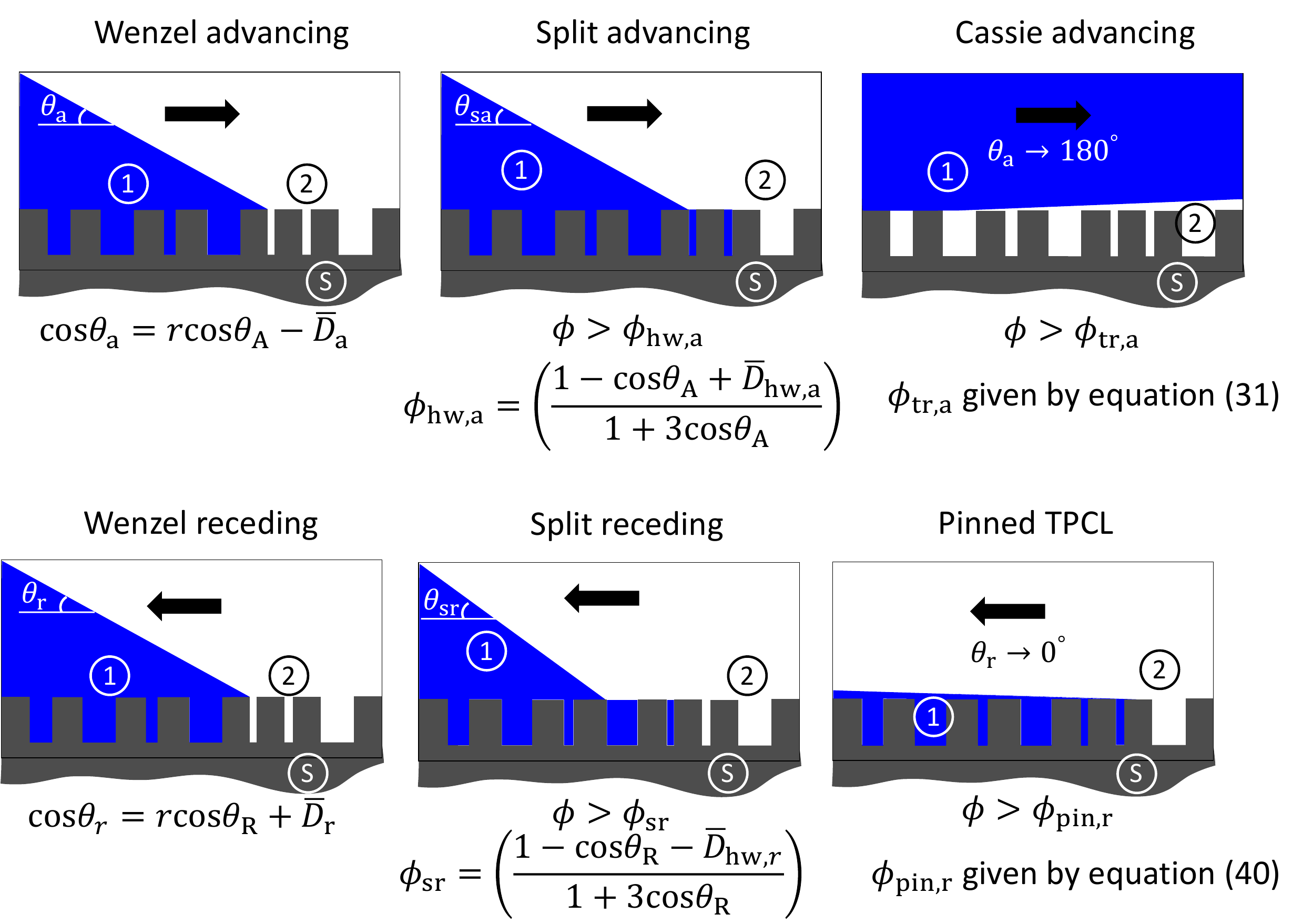}
    \caption{Schematic representation of different wetting regimes. The proposed model (equation(\ref{eqn:exp_proposed_eqn})) is valid only for Wenzel advancing and Wenzel receding modes. The limiting area fractions for split-advancing ($\phi_{\rm{hw,a}}$), Cassie advancing ($\phi_{\rm{tr,a}}$), split-receding ($\phi_{\rm{sr}}$) and pinned TPCL ($\phi_{\rm{pin,r}}$) are based on equations (\ref{eqn:phi_hw}), (\ref{eqn:phi_tr_combined}), (\ref{eqn:phi_split_receding}) and (\ref{eqn:phi_pin_r_combined}) respectively. }
    \label{fig:exp_wetting_regimes}
\end{figure}

Equation (\ref{eqn:exp_proposed_eqn}) is valid for a range of inherent advancing/receding angles ($\theta_{\rm{A}}, \theta_{\rm{R}}$) on surfaces with randomly distributed, micron-sized, cylindrical pillars of unit aspect ratio as long as the wetting state remains homogeneous. However, at a certain low advancing angle ($\theta_{\rm{A}}$) and/or high area fraction in the advancing case, we may observe hemiwicking in which a thin liquid film spontaneously precedes the bulk of the droplet. Alternatively, above a certain advancing angle ($\theta_{\rm{A}}$) and/or area fraction the droplet may not exist in a homogeneous wetting state and again in these cases, the proposed equation (\ref{eqn:exp_proposed_eqn}) is not valid. Similarly, for a receding interface under certain conditions a thin film of liquid may be left behind, pinned on the pillars as the TPCL recedes. We may also observe a permanently pinned TPCL during the receding motion. In all of these scenarios, equation (\ref{eqn:exp_proposed_eqn}) is not valid. In figure \ref{fig:exp_wetting_regimes} we illustrate various wetting states that can exist during the advancing/receding motion of the interface. In the next sections, we use the mechanical energy balance framework to analyse all of these cases and discuss the range of area fraction and inherent advancing/receding angles ($\theta_{\rm{A}}, \theta_{\rm{R}}$) over which the proposed equation (\ref{eqn:exp_proposed_eqn}) is valid.

\subsubsection{Hemiwicking and split-advancing}
\label{sec:exp_hemiwicking}

\begin{figure*}
\centering
  \includegraphics[width=0.94\textwidth]{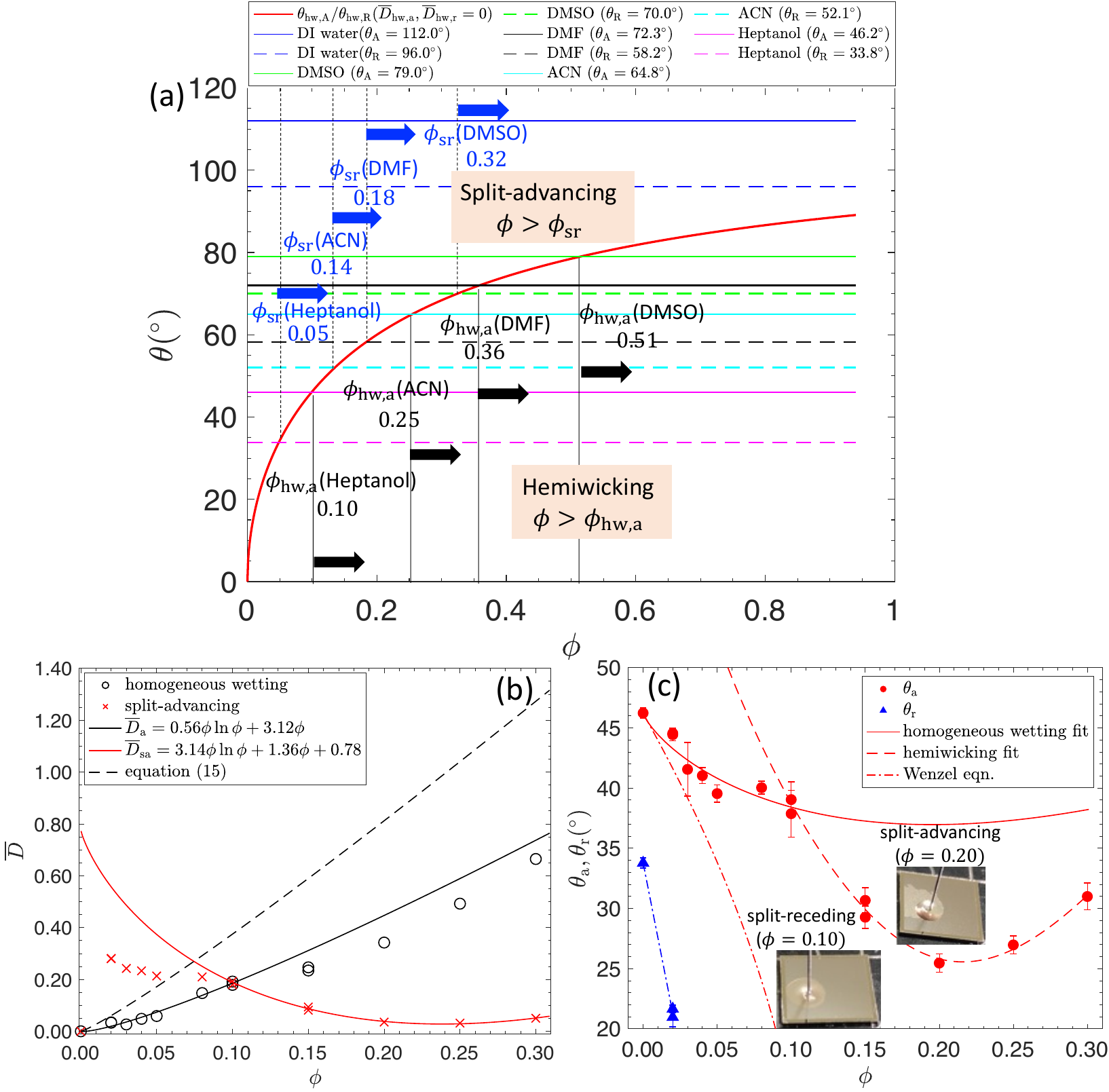}
  \caption{(a) Variation in the critical angle for hemiwicking ($\theta_{\rm{hw,A}}$) and split-receding ($\theta_{\rm{hw,R}}$) with the pillar area fraction ($\phi$) according to equations (\ref{eqn:exp_hemiwicking}) and (\ref{eqn:exp_sr_critical_angle}) and assuming $\overline{D}_{\rm{hw,a}}=0$ and $\overline{D}_{\rm{hw,r}}=0$. The minimum pillar area fraction required by DMSO, DMF, ACN and heptanol droplets for exhibiting hemiwicking and split-receding are also shown. (b) Variation in the total non-dimensional energy dissipation ($\overline{D}$) with the pillar area fraction ($\phi$) calculated from the experimentally measured advancing contact angles on random surfaces with heptanol droplets using equation (\ref{eqn:exp_meb_non_dim}) shown by black circles and equation (\ref{eqn:exp_meb_sa}) shown by red crosses. The solid black and red lines represent the total non-dimensional energy dissipation for a homogeneous wetting state ($\overline{D}_{\rm{a}}$) and split-advancing ($\overline{D}_{\rm{sa}}$) respectively and the black dashed lines represent the total non-dimensional energy dissipation based on equation (\ref{eqn:exp_D_bar}). (c) Variation in the experimentally measured advancing and receding contact angles ($\theta_{\rm{a}},\theta_{\rm{r}}$) with the pillar area fraction ($\phi$) for heptanol droplets on randomly structured surfaces. Inset shows the split-advancing and split-receding behaviour exhibited by the heptanol droplets. Due to the low inherent advancing angle of heptanol droplets, the onset of hemiwicking makes the measurement of contact angles difficult and prone to measurement errors. The solid and dashed lines represent the predictive equations (\ref{eqn:exp_meb_non_dim}) and (\ref{eqn:exp_meb_sa}) and the dashed-dotted line represents the Wenzel equation (\ref{eqn:exp_wenzel}). The dashed blue line is an aid for visualizing the receding contact angles.}
    \label{fig:exp_critical_angle_hemiwicking}
\end{figure*}

Hemiwicking refers to the formation of a spontaneously-spreading thin film of liquid, which fills in the spaces between the pillars \cite{Wemp2017,Ishino2004,Ishino2007,Quere2008a,Semprebon2014b}. A rough surface is likely to undergo hemiwicking if $\theta_{\rm{A}}$ is less than a certain critical value (see supplementary material S7 for a detailed analysis), that is
\begin{equation}
    \theta_{\rm{A}} < \theta_{\rm{hw,A}} = \cos ^{-1} \left( \frac{1-\phi + \overline{D}_{\rm{hw,a}}}{r-\phi} \right),
    \label{eqn:exp_hemiwicking}
\end{equation}
where $\overline{D}_{\rm{hw,a}}$ is the energy dissipation within the spreading film. The term on the right-hand side of the equation (\ref{eqn:exp_hemiwicking}) is the critical angle ($\theta_{\rm{hw,A}}$), and a system can only exhibit a homogeneous wetting state (without hemiwicking) if $\theta_{\rm{A}} > \theta_{\rm{hw,A}}$. Therefore, for a given $\theta_{\rm{A}}$, the critical angle ($\theta_{\rm{hw,A}}$) sets an upper limit on the area fraction up to which a droplet can be in a homogeneous wetting state with no hemiwicking. We denote this maximum area fraction above which the system can undergo hemiwicking as $\phi_{\rm{hw,a}}$. For a surface with pillars of unit aspect ratio (that is $r=1+4\phi$), $\phi_{\rm{hw,a}}$ can be obtained from equation (\ref{eqn:exp_hemiwicking}) as
\begin{equation}
    \phi_{\rm{hw,a}} = \left( \frac{1 - \cos\theta_{\rm{A}} + \overline{D}_{\rm{hw,a}}}{1 + 3 \cos\theta_{\rm{A}}}  \right).
    \label{eqn:phi_hw}
\end{equation}
In figure \ref{fig:exp_critical_angle_hemiwicking}(a) we plot the critical angle ($\theta_{\rm{hw,A}}$) and the inherent advancing angle ($\theta_{\rm{A}}$) for different liquids (DI water, DMSO, DMF, ACN and heptanol) with the pillar area fraction ($\phi$). Given that little is known about dissipation occurring during hemiwicking, we assume that $\overline{D}_{\rm{hw,a}}=0$ in equation (\ref{eqn:exp_hemiwicking}), which gives an upper estimate of $\theta_{\rm{hw,A}}$. We observe that for DI-water, the $\theta_{\rm{A}}$ line never intersects the $\theta_{\rm{hw,A}}$ curve, therefore, DI-water will not exhibit hemiwicking at any value of the pillar area fraction. On the other hand, the $\theta_{\rm{A}}$ lines for DMSO, DMF, ACN and heptanol intersect the $\theta_{\rm{hw,A}}$ curve at area fractions 0.51, 0.36, 0.25 and 0.10 respectively, which sets a minimum area fraction upper limit up to which the system can exist in a homogeneous wetting state without undergoing hemiwicking (minimum as $\overline{D}_{\rm{hw,a}}=0$ has been assumed). 
%
%

During a typical interface advancement, if the inherent advancing angle ($\theta_{\rm{A}}$) and the area fraction ($\phi$) are such that $\phi > \phi_{\rm{hw,a}}$, hemiwicking will result. Consequently, when more liquid is slowly added to the droplet, the bulk of the droplet will advance over a thin film which surrounds it \cite{quere2002rough}. We refer to this phenomenon as `split-advancing'. In figure \ref{fig:exp_critical_angle_hemiwicking}(c), split-advancing is shown in the inset when a heptanol droplet is placed on a surface with $\phi=0.25$ ($\phi_{\rm{hw,a}}$ for heptanol is 0.10). Also, when such a droplet is made to recede by removing the liquid from it, a thin film is left behind while the bulk of the droplet has receded. We refer to this phenomenon as `split-receding'. Note however that, hemiwicking is not essential for observing split-receding, as discussed further in \S\ref{sec:split_receding}. The split-advancing and split-receding behaviours are analysed in detail in supplementary material S9.

Before proceeding further, we discuss the nature of energy dissipation during split-advancing. Applying the general mechanical energy balance equation \cite{dhcontact09} to this case, the macroscopic advancing angle ($\theta_{\rm{sa}}$) can be related to the total non-dimensional energy dissipation during the interface advancement ($\overline{D}_{\rm{sa}}$) and pillar area fraction ($\phi$) as (refer to supplementary material S9 for a detailed analysis)
\begin{equation}
   \cos\theta_{\rm{sa}} = \phi(\cos\theta_{\rm{A}}-1) + 1 - \overline{D}_{\rm{sa}}.
   \label{eqn:exp_meb_sa}
\end{equation}
For a heptanol droplet ($\theta_{\rm{A}}=46.2$\textdegree) the critical area fraction for the onset of hemiwicking ($\phi_{\rm{hw,a}}$) is 0.10 (neglecting energy dissipation within the spreading film) - that is up to an area fraction of 0.10 heptanol droplets are in a homogeneous wetting state. Hence for $\phi>0.10$, heptanol droplets exhibit the `split-advancing' behaviour. We plot the total non-dimensional energy dissipation against the pillar area fraction for the advancing motion of heptanol droplets on random surfaces in figure \ref{fig:exp_critical_angle_hemiwicking}(b). We plot two sets of energy dissipation data: $\overline{D}_{\rm{a}}$ calculated using equation (\ref{eqn:exp_meb_non_dim}) and corresponding to conventional homogeneous wetting which are shown as black circles, and $\overline{D}_{\rm{sa}}$ calculated using equation (\ref{eqn:exp_meb_sa}) (corresponding to split-advancing) which are shown as red crosses. We observe that the total non-dimensional energy dissipation during split-advancing ($\overline{D}_{\rm{sa}}$) decreases with the pillar area fraction.
To understand why, we use equation (\ref{eqn:exp_meb_sa}) to explicitly express the total non-dimensional energy dissipation during split-advancing ($\overline{D}_{\rm{sa}}$) as a function of advancing contact angle ($\theta_{\rm{sa}}$), inherent advancing angle ($\theta_{\rm{A}}$) and the pillar area fraction ($\phi$), that is
\begin{equation}
    \overline{D}_{\rm{sa}} = \phi(\cos\theta_{\rm{A}}-1) + 1 - \cos\theta_{\rm{sa}}. 
    \label{eqn:exp_D_sa_explicit}
\end{equation}
From equation (\ref{eqn:exp_D_sa_explicit}) it follows that $d\overline{D}_{\rm{sa}}/d\phi=(\cos\theta_{\rm{A}}-1) \leq 1$. Since, for heptanol droplets $\phi_{\rm{hw,a}}$ is 0.10, we use equation (\ref{eqn:exp_meb_non_dim}) up to $\phi=0.10$ to obtain a fitting equation for $\overline{D}_{\rm{a}}$ and equation (\ref{eqn:exp_meb_sa}) for $\phi>0.10$ to obtain a fitting equation for $\overline{D}_{\rm{sa}}$ which are shown as black and red lines respectively in figure \ref{fig:exp_critical_angle_hemiwicking}(b). We also plot the total non-dimensional energy dissipation based on the proposed equation, that is equation (\ref{eqn:exp_diss_adv}) as a dashed black line. Overall the combination of the two dissipation equations ($\overline{D}_{\rm{a}}$ and $\overline{D}_{\rm{sa}}$) and hemiwicking area fraction constraint ($\phi_{\rm{hw,a}}$) represents the data well, giving confidence that our analyses are capturing the physics of these processes. We do observe that equation (\ref{eqn:exp_diss_adv}) slightly overpredicts the energy dissipation, which may be due to the fact that the coefficients were determined with $\theta_{\rm{A}}$ between 72.3\tc and 121.8\textdegree, whereas $\theta_{\rm{A}}$ for heptanol is 46.2\tc which is outside this range. 

In figure \ref{fig:exp_critical_angle_hemiwicking}(c) we plot the variation in advancing/receding contact angles against the pillar area fraction for heptanol droplets on random surfaces. We also plot the predictive equations (\ref{eqn:exp_meb_non_dim}) and (\ref{eqn:exp_meb_sa}) based on the MEB framework; using the non-dimensional energy dissipations $\overline{D}_{\rm{a}}$ and $\overline{D}_{\rm{sa}}$ as shown in figure \ref{fig:exp_critical_angle_hemiwicking}(b) for the advancing interface as solid and dashed red lines respectively. We observe excellent agreement between the experimentally obtained $\theta_{\rm{a}}$ and those predicted using equation (\ref{eqn:exp_meb_non_dim}) (up to $\phi=0.10$) and equation (\ref{eqn:exp_meb_sa}) (for $\phi>0.10$). We also plot the Wenzel equation (\ref{eqn:exp_wenzel}) as a dashed-dotted red line. 
Both the Wenzel equation and the experimentally measured $\theta_{\rm{a}}$ values show a decreasing trend; however, the Wenzel equation overpredicts the decrease in $\theta_{\rm{a}}$ with $\phi$ as it neglects the dissipation in energy. In figure \ref{fig:exp_critical_angle_hemiwicking}(c) we observe a comparatively sharp drop in $\theta_{\rm{a}}$ at $\phi=0.10$, which we expect is due to the onset of hemiwicking. 
%
%

\subsubsection{Wenzel to Cassie transition during advancing motion}
\label{sec:wenzel-to-cassie}

At higher pillar area fractions ($\phi$) or inherent advancing angles ($\theta_{\rm{A}}$), a heterogeneous wetting state can be favourable as compared to the homogeneous wetting state for which equation (\ref{eqn:exp_proposed_eqn}) is developed. These combinations ($\phi$ and $\theta_{\rm{A}}$) are important for identifying the validity limit for the proposed equation. To determine which wetting state is preferred, we note that due to the kinematics of liquid advance, the particular wetting state that is realised will be the one with the smallest $\theta_{\rm{a}}$, or alternatively, the largest $\cos\theta_{\rm{a}}$. Our previous analysis for the heterogeneous wetting state (supplementary material S10) has already shown that for this state $\theta_{\rm{a}} \approx 180$\textdegree, or $\cos\theta_{\rm{a}} \approx -1$. Hence, a condition for heterogeneous wetting to be preferred over homogeneous wetting is (following equation (\ref{eqn:exp_meb_non_dim}))
\begin{equation}
    \cos \theta_{\rm{a}} = r \cos\theta_{\rm{A}} - \overline{D}_{\rm{a}} > -1,
\end{equation}
or assuming unit aspect ratio pillars,
\begin{equation}
    (1+4 \phi) \cos\theta_{\rm{A}} - \overline{D}_{\rm{a}} > -1,
\end{equation}
or
\begin{equation}
    \overline{D}_{\rm{a}} < 1 + (1+4 \phi) \cos\theta_{\rm{A}}.
 \label{eqn:D_min_phi_tr}    
\end{equation}
Hence the area fraction that represents the transition from homogeneous (Wenzel) to heterogeneous (Cassie) wetting state is given by
\begin{equation}
    \phi_{\rm{tr,a}} = \frac{\overline{D}_{\rm{a}} - (1 + \cos\theta_{\rm{A}})}{4 \cos\theta_{\rm{A}}}.
    \label{eqn:phi_tr_general}    
\end{equation}
If we assume $\overline{D}_{\rm{a}}$ to be zero, the area fraction above which a heterogeneous wetting state is favoured over a homogeneous wetting is
\begin{equation}
    \phi_{\rm{tr,a}} = \frac{- (1 + \cos\theta_{\rm{A}})}{4 \cos\theta_{\rm{A}}}.
    \label{eqn:phi_tr_D_r_zero}
\end{equation}
As discussed previously, the energy dissipated during the advancing motion of an interface ($\overline{D}_{\rm{a}}$) in a homogeneous wetting state can be obtained from equation (\ref{eqn:exp_D_bar}).
Substituting this into equation (\ref{eqn:D_min_phi_tr}) yields
\begin{equation}\begin{split}
 &A_{\rm{a}}\phi \ln \phi + B_{\rm{a}}\phi < (1+4 \phi)\cos\theta_{\rm{A}} + 1, \quad \text{OR} \\
 &\phi(A_{\rm{a}}\ln\phi - 4\cos\theta_{\rm{A}}+B_{\rm{a}}) - (1+\cos\theta_{\rm{A}}) < 0,
 \end{split}
 \label{eqn:full_curve}
\end{equation}
where $A_{\rm{a}}=2.08\cos\theta_{\rm{A}}-0.99$ and $B_{\rm{a}}=6.97\cos\theta_{\rm{A}}-0.04$. Equation (\ref{eqn:full_curve}) can be solved analytically as
\begin{equation}
    \phi_{\rm{tr,a}} = \frac{1 + \cos\theta_{\rm{A}}}{A_{\rm{a}} W \left( \frac{(1+\cos\theta_{\rm{A}})e^{\left({\frac{-4\cos\theta_{\rm{A}}}{A_{\rm{a}}} + \frac{B_{\rm{a}}}{A_{\rm{a}}}} \right)} } {A_{\rm{a}}} \right)},
    \label{eqn:exp_phi_tr_ana}
\end{equation}
where $W$ is the product log function (Lambert $W$-function) \cite{Mathematica}. We solve equation (\ref{eqn:full_curve}) numerically to obtain the area fraction ($\phi_{\rm{tr,a}}$) above which a heterogeneous wetting state is favourable as compared to a homogeneous wetting state. However, at certain higher area fractions equation (\ref{eqn:exp_diss_adv}) may not be able to correctly predict the dissipation ($\overline{D}_{\rm{a}}$), which can result in equation (\ref{eqn:full_curve}) having no solution. For such cases, as an approximation, we use the maximum value of total non-dimensional energy dissipation at a particular $\theta_{\rm{A}}$ based on equation (\ref{eqn:exp_diss_adv}) to solve equation (\ref{eqn:full_curve}). The maximum total non-dimensional energy dissipation ($\overline{D}_{\rm{max,a}}$) is calculated at the area fraction at which the first derivative of $\overline{D}_{\rm{a}}$ with respect to $\phi$ becomes zero, that is
\begin{equation}
 \overline{D}_{\rm{max,a}} = -A_{\rm{a}} e^{-(1+B_{\rm{a}}/A_{\rm{a}})}.
 \label{eqn:D_c}
\end{equation}
Substituting $\overline{D}_{\rm{max,a}}$ from equation (\ref{eqn:D_c}) as $\overline{D}_{\rm{a}}$ into equation (\ref{eqn:D_min_phi_tr}) yields
\begin{equation}
     \phi_{\rm{tr,a}} = \left( \frac{-A_{\rm{a}} e^{-(1 + B_{\rm{a}}/A_{\rm{a}})} - (1+\cos\theta_{\rm{A}})}{4 \cos\theta_{\rm{A}}} \right).
      \label{eqn:phi_tr}
\end{equation}
In a more general way, we can write the condition for the area fraction above which a heterogeneous wetting state is preferred over a homogeneous wetting state as
\begin{equation}
\begin{split}
    \phi_{\rm{tr,a}} = \text{min} \Biggl[ \frac{1 + \cos\theta_{\rm{A}}}{A_{\rm{a}} W \left( \frac{(1+\cos\theta_{\rm{A}})e^{\left({\frac{-4\cos\theta_{\rm{A}}}{A_{\rm{a}}} + \frac{B_{\rm{a}}}{A_{\rm{a}}}} \right)} } {A_{\rm{a}}} \right)}, \\
     \left( \frac{-A_{\rm{a}} e^{-(1 + B_{\rm{a}}/A_{\rm{a}})} - (1+\cos\theta_{\rm{A}})}{4 \cos\theta_{\rm{A}}} \right) \Biggr].
    \end{split}
    \label{eqn:phi_tr_combined}
\end{equation}
Hence, $\phi_{\rm{tr,a}}$, which is the area fraction representing the transition from a homogeneous to a heterogeneous wetting state is determined as the minimum values of equations (\ref{eqn:exp_phi_tr_ana}) and (\ref{eqn:phi_tr}). 

\subsubsection{Split-receding motion}
\label{sec:split_receding}

In \S\ref{sec:exp_hemiwicking} we discussed the hemiwicking phenomena and the interface's resulting split-advancing and split-receding motion. However, it is possible for the interface to exhibit split-receding without exhibiting hemiwicking during the advancing motion. This behaviour is discussed in detail in supplementary material S8. Here, we only report the necessary condition for the onset of split-receding ($\phi_{\rm{sr}}$), that is
\begin{equation}
 \phi_{\rm{sr}} > \frac{1 - \cos\theta_{\rm{R}} - \overline{D}_{\rm{hw,r}}}{1 + 3 \cos\theta_{\rm{R}}},
 \label{eqn:phi_split_receding}
\end{equation}
where $\overline{D}_{\rm{hw,r}}$ the energy dissipation within the receding film. From equation (\ref{eqn:phi_split_receding}) we observe that the presence of a non-zero dissipation within the film ($\overline{D}_{\rm{hw,r}}$) promotes split-receding behaviour. Equation (\ref{eqn:phi_split_receding}) can be rearranged to give a critical angle for the onset of split-receding ($\theta_{\rm{hw,R}}$) such that when $\theta_{\rm{R}}<\theta_{\rm{hw,R}}$ split-receding is observed, as
\begin{equation}
   \cos\theta_{\rm{hw,R}} = \frac{1-\phi-\overline{D}_{\rm{hw,r}}}{r-\phi}.
   \label{eqn:exp_sr_critical_angle}
\end{equation}
In figure \ref{fig:exp_critical_angle_hemiwicking}(a) we show the variation in $\theta_{\rm{hw,R}}$ with the pillar area fraction and the $\phi_{\rm{sr}}$ values for DMSO, DMF, ACN and heptanol, assuming $\overline{D}_{\rm{hw,r}}$ to be zero. Note that equations (\ref{eqn:exp_sr_critical_angle}) and (\ref{eqn:exp_hemiwicking}) for split-advancing have the same form if dissipation is neglected. We observe that DI water does not exhibit split-receding at any value of pillar area fraction.

\subsubsection{Permanently pinned TPCL}
\label{sec:pinned TPCL}

During the receding motion of the interface, it is possible to obtain a permanently pinned TPCL. When this happens, the TPCL remains at the same location while the macroscopic contact angle decreases, eventually approaching 0\textdegree. Hence, a condition for the permanently pinned TPCL to be preferred over homogeneous wetting can be obtained following equation (\ref{eqn:exp_meb_non_dim}) in a manner similar to the transition from homogeneous to heterogeneous wetting state discussed in \S\ref{sec:wenzel-to-cassie} that is,
\begin{equation}
    \cos\theta_{\rm{r}} = r \cos\theta_{\rm{R}} + \overline{D}_{\rm{r}} > 1,
\end{equation}
or assuming unit aspect ratio pillars,
\begin{equation}
    \overline{D}_{\rm{r}} > 1 - (1+4 \phi) \cos\theta_{\rm{R}}.
    \label{eqn:D_r_pihi_pin_r}
\end{equation}
The area fraction that represents the transition from homogeneous (Wenzel) to a permanently pinned TPCL wetting state is given by
\begin{equation}
 \phi_{\rm{pin,r}} = \frac{{1-\cos\theta_{\rm{R}}-\overline{D}_{\rm{r}}}}{4 \cos\theta_{\rm{R}}}.  
 \label{eqn:phi_pin_r_D_r}
\end{equation}
Assuming $\overline{D}_{\rm{r}}$ is zero in equation (\ref{eqn:phi_pin_r_D_r}), allows the area fraction at which the TPCL gets permanently pinned to be obtained (neglecting energy dissipation) as
\begin{equation}
    \phi_{\rm{pin,r}} = \left( \frac{1 - \cos\theta_{\rm{R}}}{4 \cos \theta_{\rm{R}}} \right).
 \label{eqn:exp_Dpin_zero}
\end{equation}
Instead, if the energy dissipation is not neglected, the area fraction at which the TPCL becomes permanently pinned ($\phi_{\rm{pin,r}}$) can be obtained by solving equations (\ref{eqn:D_r_pihi_pin_r}) and (\ref{eqn:exp_diss_rec}). The method for calculating $\phi_{\rm{pin,r}}$ is similar the $\phi_{\rm{tr,a}}$ calculation discussed in \S\ref{sec:wenzel-to-cassie}, that is
\begin{equation}
    \phi_{\rm{pin,r}} =  \frac{1 - \cos\theta_{\rm{R}}}{A_{\rm{r}} W \left( \frac{(1 - \cos\theta_{\rm{R}})e^{\left({\frac{4\cos\theta_{\rm{R}}}{A_{\rm{r}}} + \frac{B_{\rm{r}}}{A_{\rm{r}}}} \right)} } {A_{\rm{r}}} \right)}, 
\end{equation}
or
\begin{equation}
    \phi_{\rm{pin,r}} = \left( \frac{A_{\rm{r}} e^{-(1 + B_{\rm{r}}/A_{\rm{r}})} + (1-\cos\theta_{\rm{R}})}{4 \cos\theta_{\rm{R}}} \right),
\end{equation}
if the maximum dissipation, that is the dissipation corresponding to the area fraction ($\phi$) at which the first derivative of $\overline{D}_{\rm{r}}$ with respect to $\phi$ is zero, that is $\overline{D}_{\rm{max,r}}$ is used. In a general manner, the area fraction above which a pinned TPCL is favoured over a homogeneous wetting state can be written as
\begin{equation}
\begin{split}
    \phi_{\rm{pin,r}} = \text{min}\Biggl[ \frac{1 - \cos\theta_{\rm{R}}}{A_{\rm{r}} W \left( \frac{(1 - \cos\theta_{\rm{R}})e^{\left({\frac{4\cos\theta_{\rm{R}}}{A_{\rm{r}}} + \frac{B_{\rm{r}}}{A_{\rm{r}}}} \right)} } {A_{\rm{r}}} \right)},\\
    \left( \frac{A_{\rm{r}} e^{-(1 + B_{\rm{r}}/A_{\rm{r}})} + (1-\cos\theta_{\rm{R}})}{4 \cos\theta_{\rm{R}}} \right) \Biggr].
    \end{split}
    \label{eqn:phi_pin_r_combined}
\end{equation}
where $A_{\rm{r}}=-2.08\cos\theta_{\rm{R}}-0.99$ and $B_{\rm{r}}=-6.97\cos\theta_{\rm{R}}-0.04$, and $W$ is the Lambert $W$-function \cite{Mathematica}. 

\subsubsection{Wetting states summary}
\label{sec:exp_wetting_state_summary}

In the preceding sections, we discussed a number of different wetting scenarios which can be observed during the advancing and receding motion of an interface on a rough surface over a range of $\theta_{\rm{A}}, \theta_{\rm{R}}$ and $\phi$ values. In figure \ref{fig:exp_wetting_regimes} we summarise these six wetting states and the corresponding limits on the pillar area fractions ($\phi$) for the entire range of $\theta_{\rm{A}}$/$\theta_{\rm{R}}$ values.
%
\begin{figure}
    \centering
    \includegraphics[width=0.46\textwidth]{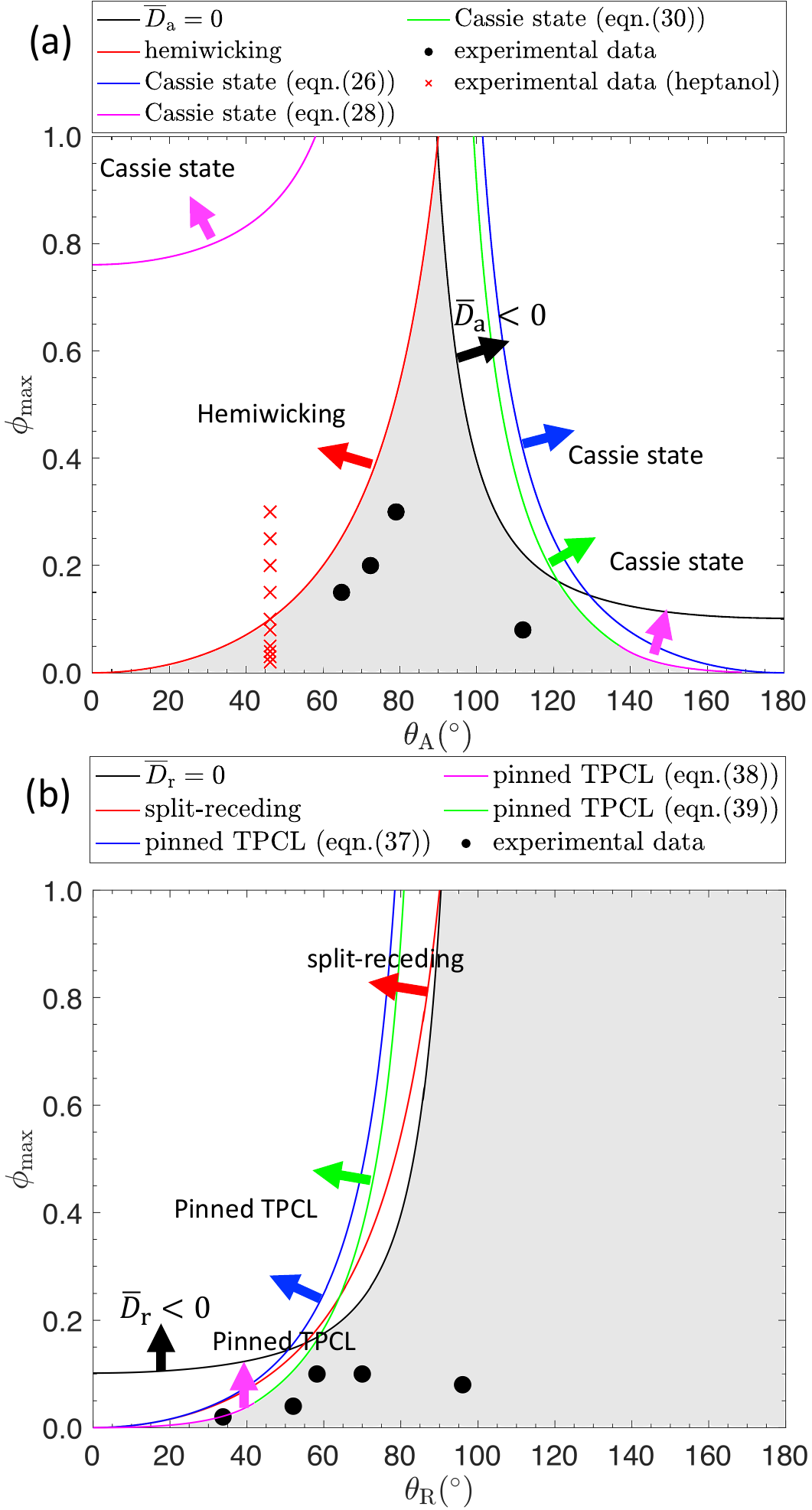}
    \caption{Range of validity of the proposed model (shaded region) represented by the variation in maximum pillar area fraction ($\phi_{\rm{max}}$) up to which the model is valid over a range of (a) inherent advancing ($\theta_{\rm{A}}$) and (b) inherent receding ($\theta_{\rm{R}}$) angles. For the receding interface, it is assumed that the interface has advanced in a homogeneous wetting state, which may not be true for high $\theta_{\rm{A}}$ values. This puts an upper limit in terms of $\theta_{\rm{R}}$ on the model's validity for predicting the receding motion. The maximum pillar area fraction tested ($\phi_{\rm{max}}$) and the corresponding inherent advancing/receding contact angles for the experimental data (DI water, DMSO, DMF, ACN and heptanol (receding)) are plotted as black circles. For the advancing heptanol case, all area fractions are shown as red crosses. We observe that only up to an area fraction of 0.1 the advancing heptanol case lies within the validity range. }
    \label{fig:model_validity}
\end{figure}
In figures \ref{fig:model_validity}(a) and \ref{fig:model_validity}(b) we graphically represent the regions in the $\theta_{\rm{A}}$ vs. $\phi_{\rm{max}}$ and $\theta_{\rm{R}}$ vs. $\phi_{\rm{max}}$ plots respectively, in which the proposed equation (\ref{eqn:exp_proposed_eqn}) is valid for predicting the macroscopic advancing ($\theta_{\rm{a}}$) and receding ($\theta_{\rm{r}}$) contact angles. Here, $\phi_{\rm{max}}$ represents the maximum pillar area fraction up to which the proposed equations for predicting CAH are valid. The maximum area fraction ($\phi_{\rm{max}}$) and the inherent advancing/receding contact angles for the experimental data obtained in this study are also plotted as scatter points. We observe that all the experimental data lies inside the validity range as the maximum pillar area fractions corresponding to the respective inherent advancing/receding contact angles (except for the heptanol advancing and receding case). It should be noted that in figure \ref{fig:model_validity}(b) it may appear that the model is valid for a wide range of inherent receding angles ($\theta_{\rm{R}}>90$\textdegree). However, the validity of the model for predicting the receding motion is based on the assumption that the interface has previously advanced in a homogeneous wetting state. Therefore, the transition from a homogeneous to a heterogeneous wetting state during the advancing motion (characterised by $\phi_{\rm{tr,a}}$) puts up an upper limit on the model validity for $\theta_{\rm{R}}>90$\textdegree.

Another notable point from figure \ref{fig:model_validity}(b) is that the split-receding phenomenon is favourable over the pinned TPCL for a greater part of the $\theta_{\rm{R}}$ range. Since the split-receding behaviour is plotted by neglecting the energy dissipation ($\overline{D}_{\rm{hw,r}}$, see equation (\ref{eqn:phi_split_receding})), we may get a different variation between $\phi_{\rm{max}}$ and $\theta_{\rm{R}}$ if a finite energy dissipation in the film is included in the analysis. Also, upon a comparison between $\phi_{\rm{sr}}$ (equation (\ref{eqn:phi_split_receding})) and $\phi_{\rm{pin,r}}$ (equation (\ref{eqn:exp_Dpin_zero})), neglecting the energy dissipation in both cases, we observe that $\phi_{\rm{sr}}<\phi_{\rm{pin,r}}$ for all $\theta_{\rm{R}} \in (0,\pi/2)$. 

\subsection{Comparison with experiments}
\label{sec:model_comparison}

\begin{figure}
    \centering
    \includegraphics[width=0.48\textwidth]{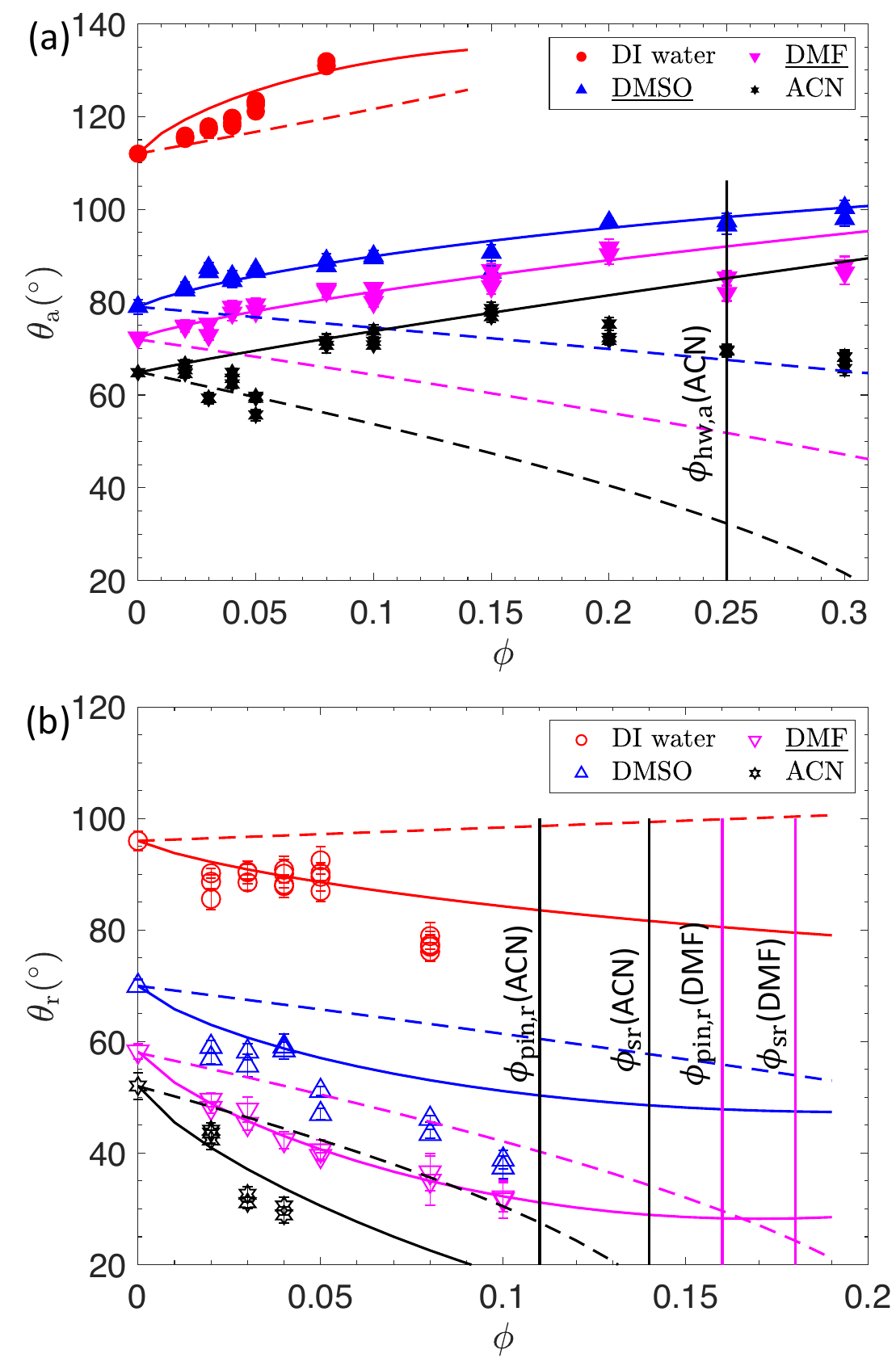}
    \caption{(a) Advancing ($\theta_{\rm{a}}$) and (b) receding ($\theta_{\rm{r}}$) contact angles measured on random surfaces for DI water, DMSO, DMF and ACN droplets.  Equation (\ref{eqn:exp_proposed_eqn}) and the Wenzel equation \cite{he2004contact} are shown by solid and dashed lines respectively. The limit on the maximum pillar area fraction as set forth by hemiwicking ($\phi_{\rm{hw,a}}$) for the advancing interface for ACN droplets; pinning of the TPCL ($\phi_{\rm{pin,r}}$) and the onset of split-receding motion ($\phi_{\rm{sr}}$) for the receding interface for ACN and DMF droplets are also shown.}
    \label{fig:exp_data}
\end{figure}

In figure \ref{fig:exp_data} we compare the advancing/receding contact angles predicted by the proposed equation (\ref{eqn:exp_proposed_eqn}) with the experimentally measured angles (on random surfaces) using DI water, dimethyl sulfoxide (DMSO), dimethylformamide (DMF) and acetonitrile (ACN) droplets - that is, the data that was used to produce equation (\ref{eqn:exp_proposed_eqn}), along with 5 other data sets that were independently measured.
We observe good agreement between the experimental data and the values predicted by equation (\ref{eqn:exp_proposed_eqn}), shown by solid lines, for both the advancing and receding contact angles and all liquids. Note that equation (\ref{eqn:exp_proposed_eqn}) was fitted based on advancing results for only DMSO, DMF and receding results for DMF only - other equation (\ref{eqn:exp_proposed_eqn}) results are predictions based on these fittings. Advancing and receding contact angles predicted by the Wenzel equation \cite{he2004contact} are also plotted as dashed lines for comparison - in general the Wenzel analysis does not describe the data. 

Some deviations between the experimental and predicted values are observed for DMF and ACN droplets, especially for the advancing interface. For example, at low area fractions for ACN droplets, equation (\ref{eqn:exp_proposed_eqn}) over-predicts the advancing contact angle. Also, in the experimental data, we observed a slight reduction in the advancing contact angle for pillar area fractions up to 0.05 compared to that on a flat surface. This may be due to the high volatility of acetonitrile (which has a vapour pressure of approximately 9.2 kPa at 20 \textdegree C \cite{ewing2004vapor}). Because of this high volatility, ACN droplets evaporate readily and the contact angle starts decreasing the moment the droplet is placed on the surface. The contact angle starts increasing again as the liquid is added to the droplet from the syringe. The presence of roughness pins the TPCL, resulting in an increase in the advancing contact angle. On the other hand, the presence of roughness also promotes the rate of evaporation from a droplet (a pinned TPCL results in a Constant Contact Radius (CCR) mode of evaporation, which has the highest evaporation rate compared to the other modes of evaporation \cite{stauber2014lifetimes, shaikeea2017universal}).  It may be due to these two competing behaviours that we observe a decrease in the advancing contact angle of an acetonitrile droplet for area fractions up to 0.05. However, a definite explanation for this behaviour can only be provided after a careful study of the variation in contact angles at different volumetric flow rates which is outside our scope. Another reason for the difference between the experimental and the predicted (equation (\ref{eqn:exp_proposed_eqn})) advancing contact angles, especially at higher area fractions is the onset of hemiwicking. In our experiments, we observed the onset of hemiwicking in DMF and ACN droplets at area fractions ($\phi=\phi_{\rm{hw,a}}$) of 0.25 and 0.20 respectively. Equation (\ref{eqn:phi_hw}), however, predicts $\phi_{\rm{hw,a}}$ for DMF and ACN droplets as 0.36 and 0.25 respectively (neglecting $\overline{D}_{\rm{hw,a}}$). We expect the reason for an early onset of hemiwicking on random surfaces is the presence of local regions (or clusters) with a higher area fraction as compared to the average area fraction for the whole surface. The liquid may undergo hemiwicking around these clusters even though the pillar area fraction ($\phi$) is smaller than $\phi_{\rm{hw,a}}$. Nevertheless, the decrease in advancing angles at high area fractions for these two liquids is qualitatively in line with our analysis of when hemiwicking occurs.

\subsection{Random vs. structured surfaces}
\label{sec:rand_vs_hex}

In \S \ref{sec:model_comparison} we presented the results for advancing and receding contact angles on surfaces with randomly distributed cylindrical pillars. We also proposed an equation for the advancing and receding contact angles on such surfaces (equation (\ref{eqn:exp_proposed_eqn})). Here, we present the experimental results for the advancing and receding contact angles on surfaces with cylindrical pillars of unit aspect ratio ($a=h=10$ $\mu$m) but arranged in a structured, hexagonal array. We measured the contact angles with DI water and dimethyl sulfoxide (DMSO) droplets on surfaces again coated with octadecanethiol (ODT).

\begin{figure}
    \centering
    \includegraphics[width=0.48\textwidth]{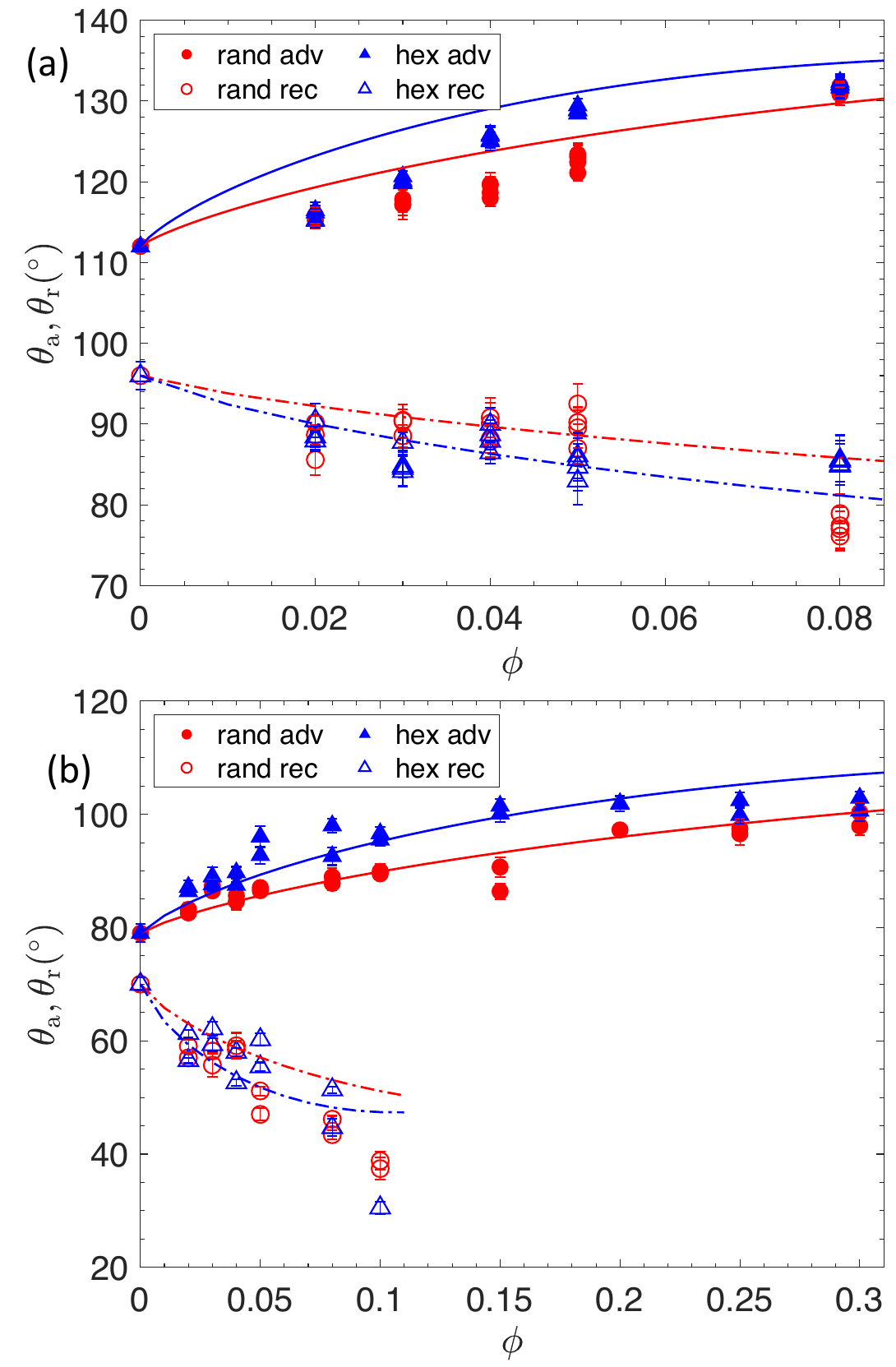}
    \caption{Advancing ($\theta_{\rm{a}}$) and receding ($\theta_{\rm{r}}$) contact angles on random (red circles) and hexagonal (blue triangles) surfaces for (a) DI water and (b) dimethyl sulfoxide (DMSO) droplets are shown. Equation (\ref{eqn:exp_proposed_eqn}) for the random surfaces is shown as solid (for advancing interface) and dashed-dotted (for receding interface) curves and equation (\ref{eqn:exp_ca_hex}) for hexagonal surfaces is shown as solid and dashed-dotted blue curves for advancing and receding interface respectively.}
    \label{fig:rand_vs_hex_exp}
\end{figure}

In figure \ref{fig:rand_vs_hex_exp} we show the variation in the advancing and receding contact angles with the pillar area fraction for random and hexagonal surfaces and droplets of DI water and DMSO. For both liquids, hexagonal surfaces exhibit a slightly higher advancing contact angle as compared to random surfaces over the entire range of area fractions. For the receding contact angle, the difference is less clear, probably due to higher uncertainties in measuring receding contact angles.

A possible reason for this behaviour could be the presence of local regions of closely spaced pillars on a random surface due to their random distribution. We refer to these local high-density regions as `clusters'. Our hypothesis is that the TPCL can pin/depin on more than one closely spaced pillar so that these `clusters of pillars' effectively act as one. This means that on a random surface, the effective number of pillars that act as pinning sites is less, which means that the effective distance between pinning sites ($d_{\rm{effective}}$) is larger. However, on a structured surface, all pillars are equally spaced so that we do not have these `clusters' and therefore the effective distance between pinning sites is the same as the inter-pillar distance ($d_{\rm{effective}}=d$, see figure \ref{fig:exp_surface_design}). Therefore, for a given pillar area fraction ($\phi$), a random surface has a larger $d_{\rm{effective}}$ and hence lower hysteresis. 

There is an alternative interpretation, i.e. that due to a larger effective inter-pillar distance, random surfaces have a slightly lower effective pillar area fraction ($\phi_{\rm{effective}}$) compared to their actual area fraction, $\phi$. 
Therefore, the proposed equation for the total non-dimensional energy dissipation during the advancing (and receding) motion of an interface on random surfaces (equations (\ref{eqn:exp_diss_adv}) and (\ref{eqn:exp_diss_rec})) 
actually represent the energy dissipation for a slightly lower pillar area fraction (that is $\phi_{\rm{effective}}$) instead of $\phi$. 

As an approximation to relate $\phi$ and $\phi_{\rm{effective}}$, let us consider a constant integer $a_{\rm{\phi}}(<1)$ such that $\phi_{\rm{effective}} = a_{\phi} \phi$. Noting that the contact angle models developed thus far for a random surface are actually in terms of $\phi_{\rm{effective}}$ (under this approximation), the total non-dimensional energy dissipation ($\overline{D}$) for an advancing/receding interface (using equation (\ref{eqn:exp_D_bar})) on a hexagonal surface can therefore be written as
\begin{equation}
\overline{D}_{\rm{hex}} = A \left(\frac{\phi}{a_{\rm{\phi}}}\right) \ln \left(\frac{\phi}{a_{\rm{\phi}}} \right) + B \left( \frac{\phi}{a_{\rm{\phi}}} \right),    
\label{eqn:exp_D_bar_hex}
\end{equation}
where $A$ and $B$ are the coefficients calculated for the interface dynamics on random surfaces (see equation (\ref{eqn:exp_A_and_B})). Equation (\ref{eqn:exp_D_bar_hex}) can be simplified further into,
\begin{equation}
\begin{split}
    \overline{D}_{\rm{hex}} &= \left( \frac{A}{a_{\rm{\phi}}}\right) \phi \ln \phi + \left[ \left( \frac{B}{a_{\phi}} \right) - \left( \frac{A}{a_{\phi}}\right) \ln a_{\phi}\right] \phi \\
    &= A_{\text{hex}} \phi \ln \phi + B_{\text{hex}} \phi,
    \end{split}
    \label{eqn:exp_D_bar_Ahex}
\end{equation}
where $A_{\rm{hex}}= A/a_{\phi}$ and $B_{\rm{hex}}=B/a_{\phi} - A/a_{\phi} \ln a_{\phi}$. Using equations (\ref{eqn:exp_D_bar_Ahex}) and (\ref{eqn:exp_meb_non_dim}) the advancing/receding contact angles on a hexagonal surface can be written as
\begin{equation}
\begin{split}
 \cos\theta_{\rm{a}} &= \left(1 + 4 \frac{\phi}{a_{\phi}} \right)\cos\theta_{\rm{e}} - \overline{D}_{\rm{hex,a}},  \\
  \cos\theta_{\rm{r}} &= \left(1 + 4  \frac{\phi}{a_{\phi}} \right)\cos\theta_{\rm{e}} + \overline{D}_{\rm{hex,r}},   
  \end{split}
  \label{eqn:exp_ca_hex}
\end{equation}
where $\overline{D}_{\rm{hex,a}}$ and $\overline{D}_{\rm{hex,r}}$ are respectively the total non-dimensional energy dissipation during the advancing and receding motion of the interface on a hexagonal surface. The dissipation terms $\overline{D}_{\rm{hex,a}}$/$\overline{D}_{\rm{hex,r}}$ can be expressed in terms of the inherent advancing/receding angles (that is $\theta_{\rm{A}}/\theta_{\rm{R}}$) using equations (\ref{eqn:exp_diss_adv}), (\ref{eqn:exp_diss_rec}) and (\ref{eqn:exp_D_bar_Ahex}). The parameter $a_{\phi}$ is calculated from the experimentally measured advancing angles for DMSO droplets (on hexagonal surfaces) using equations (\ref{eqn:exp_diss_adv}), (\ref{eqn:exp_D_bar_Ahex}) and (\ref{eqn:exp_ca_hex}).  
We obtained $a_{\phi}=0.54$ ($R^2=0.937$). The predicted advancing and receding contact angles on hexagonal surfaces for DI water and DMSO droplets based on equation (\ref{eqn:exp_ca_hex}) are shown in figure \ref{fig:rand_vs_hex_exp} as solid and dashed-dotted blue curves respectively. We observe good agreement between the experimental $\theta_{\rm{a}}/\theta_{\rm{r}}$ values and the respective predictive equations ((\ref{eqn:exp_proposed_eqn}) and (\ref{eqn:exp_ca_hex}) for the random and hexagonal surfaces respectively) for all the data sets, noting that the value of $a_{\phi}$ was based on DMSO advancing data only. However, a better agreement was observed for the DMSO droplets as compared to DI water droplets, which may be due to a limited number of data points for the DI water droplets. Based on our approximate model we expect the parameter $a_{\phi}$ to be independent of the inherent advancing/receding contact angle and depend only upon the surface roughness. However, a conclusive remark would require more experimental data covering a large range of area fractions. While such study is left to future work the reasonable agreement between advancing and receding angle for random and periodic surfaces displayed in figure \ref{fig:rand_vs_hex_exp} does support the concept of `clusters' and a lower area fraction applying to randomly distributed surface textures.

\begin{figure}
    \centering
    \includegraphics[width=0.46\textwidth]{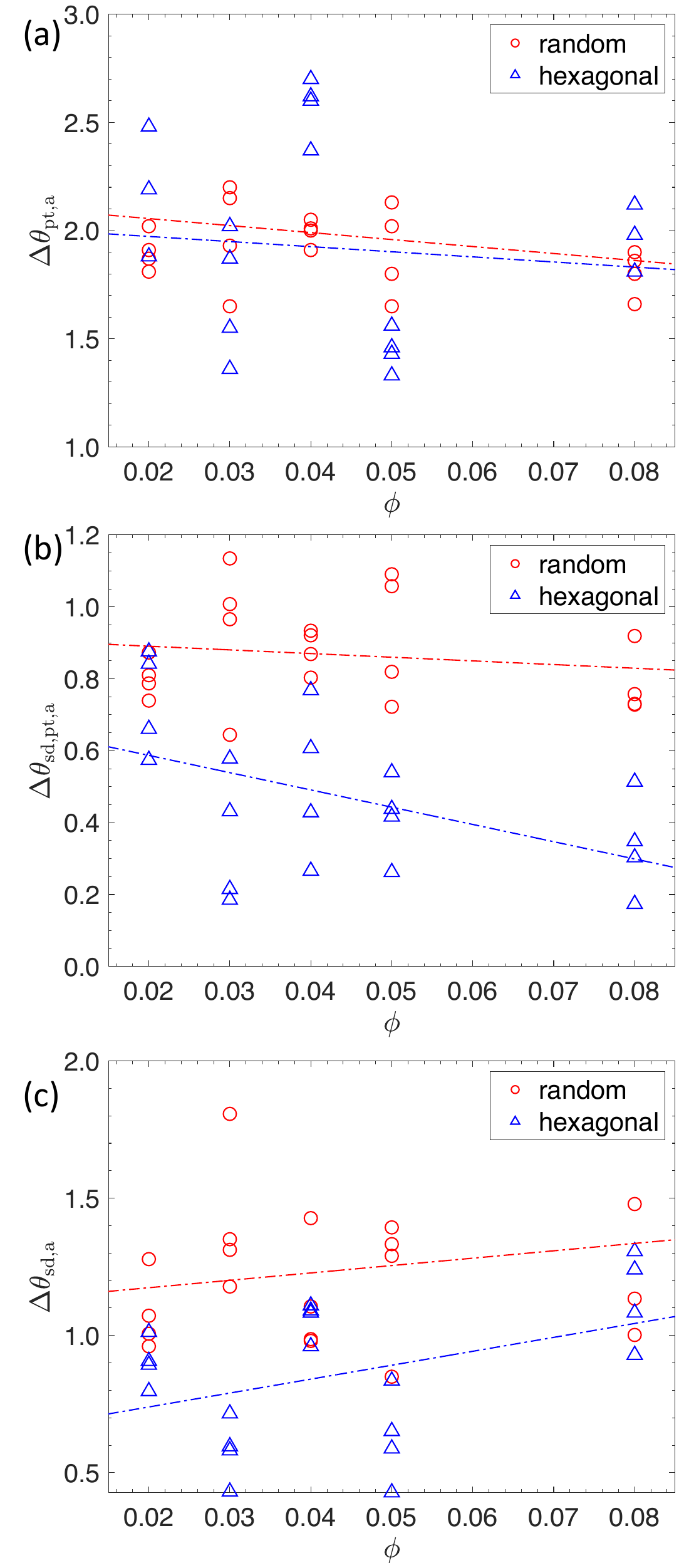}
    \caption{Variation in the average peak-to-trough angle difference ($\Delta \theta_{\rm{pt}}$) (a), standard deviation in the average peak-to-trough angle difference ($\Delta \theta_{\rm{sd,pt,a}}$) (b) and the standard deviation in the average advancing angle ($\Delta \theta_{\rm{sd,a}}$) (c), in degrees, with the pillar area fraction for an advancing interface on random and hexagonal surfaces (for DI water droplets).}
    \label{fig:average_p2t}
\end{figure}

To further understand the differences between the advancement of an interface on a random surface and a hexagonal surface, we plot the variation in average peak-to-trough angle difference (as discussed in \S\ref{sec:random_wetting}) for advancing interface ($\Delta \theta_{\rm{pt,a}}$) against the pillar area fraction ($\phi$). 
In figure \ref{fig:average_p2t}(a) we show the variation in $\Delta \theta_{\rm{pt},a}$ against the pillar area fraction for DI water droplets on random and hexagonal surfaces.
%
From figure \ref{fig:average_p2t}(a) we observe that $\Delta \theta_{\rm{pt,a}}$ decreases as the pillar area fraction increases for both the random as well as hexagonal surfaces. The reason for this pattern is that the sawtooth variation in the contact angle as the interface advances depends upon the average distance between the pillars. As the pillar area fraction increases, the average inter-pillar distance decreases ($d_{\rm{avg}} \propto 1/ \sqrt{\phi}$) and hence decreases the average peak-to-trough angle difference. We observe that $\Delta \theta_{\rm{pt,a}}$ is slightly higher for a random surface as compared to a hexagonal surface, for any given pillar area fraction. This is probably due to the `clustering effect' observed on random surfaces, where the closely spaced pillars in certain local regions of high pillar density act as a single pinning site. This results in an increase in the effective $d_{\rm{avg}}$ which results in a higher $\Delta \theta_{\rm{pt,a}}$ on random surfaces compared to hexagonal surfaces even at the same average pillar area fraction ($\phi$). However, the difference between the $\Delta \theta_{\rm{pt,a}}$ values for random and hexagonal surfaces is very small, probably due to the nature of the method used for averaging the peak-to-trough angle difference: The average peak-to-trough angle difference represents an average in time. As an interface advances on a random surface, the peak-to-trough angle difference can take a very small, moderate or very high value depending upon the local surface topology, such that the average peak-to-trough angle difference on a random surface is not very different from that on a hexagonal surface (or a structured surface in general). This is further illustrated in figure  \ref{fig:average_p2t}(b) where we plot the variation in the standard deviation in average peak-to-trough angle difference for an advancing interface ($\Delta \theta_{\rm{sd,pt,a}}$) on random and hexagonal surfaces against the pillar area fraction. We observe that $\Delta \theta_{\rm{sd,pt,a}}$ for a random surface is greater than that for a hexagonal surface at any pillar area fraction. Another difference between the interface dynamics on random and hexagonal surfaces is that on a random surface an advancing (or receding) interface pins/depins on unevenly spaced pillars as opposed to a structured surface where the pinning sites are evenly distributed. This results in a non-repeating sawtooth variation of the macroscopic contact angle ($\theta_{\rm{m}}$) (see figure S5 in the supplementary material) with greater variability in the advancing angle ($\theta_{\rm{a}}$), that is $\Delta \theta_{\rm{sd,a}}$ shown in figure \ref{fig:average_p2t}(c). We observe a high $\Delta \theta_{\rm{sd,a}}$ for a random surface as compared to a hexagonal surface at any given area fraction. This can also be explained in terms of the `clustering effect' - that is, that is as an interface advances on a random surface it comes across pillars which are unevenly distributed with the closely spaced pillars acting as a single pinning site. This results in a greater variation in the inter-pillar distance and hence a greater variation in the contact angle values.

\section{Conclusions}
\label{sec:conclusion}

We measured the advancing and receding contact angles on surfaces with randomly distributed cylindrical pillars of unit aspect ratio and a diameter of 10 $\mu$m using different liquids (DI water, dimethyl sulfoxide, dimethylformamide, acetonitrile and heptanol). We also measured the advancing and receding contact angles on surfaces with the same local pillar geometry but with the pillars arranged in a hexagonal pattern with DI water and dimethyl sulfoxide. All surfaces were coated with a layer of octadecanethiol to render them chemically homogeneous and immune to water adsorption from the surrounding atmosphere. The contact angle on a random surface shows a great variation during the interface movement, with many peaks of different magnitudes. This variability in the contact angle on random surfaces is due to the variation in the inter-pillar distance. We have proposed a set of robust parameters and the respective methodologies to calculate them to effectively characterise the wetting of random surfaces. 
The proposed parameters are: (a) average advancing/receding contact angle (and its standard deviation) which is most representative of wetting on a random surface and (b) average peak-to-trough variation in the macroscopic contact angle (and its standard deviation). Using the general mechanical energy balance equation \cite{dhcontact09} for a homogeneous wetting state, we calculated the dissipation in energy during interface movement due to the \textit{stick-slip} motion of the interface over each roughness element. We have proposed an equation, which is a logarithmic function of the average pillar area fraction and a linear function of the cosine of inherent advancing/receding contact angle, to describe the energy dissipation in homogeneous wetting of random surfaces. Using the experimental data (for dimethyl sulfoxide and dimethylformamide advancing angles and dimethylformamide receding angles) and the proposed form of the energy dissipation, we have presented an equation for predicting the advancing and receding contact angles on surfaces with randomly distributed micron-sized cylindrical pillars over a range of inherent advancing/receding contact angles. We have also defined the limits of inherent advancing/receding contact angle and pillar area fractions inside which the proposed equation for predicting hysteresis is valid. We have also derived (using the general mechanical energy balance equation \cite{dhcontact09}) expressions for the limiting values of pillar area fractions at which a liquid in a homogeneous wetting state may transition into hemiwicking mode, a heterogeneous wetting state, split-advancing mode, split-receding mode or pinned TPCL mode. We compared the advancing and receding contact angles predicted by the proposed equation with the experimental values obtained using different liquids and observed a good agreement between the two. We also compared the advancing and receding contact angles on random and hexagonal surfaces with a similar pillar geometry and observed higher advancing and lower receding contact angles on hexagonal surfaces compared to random ones. We have proposed a `cluster' theory to explain the lower hysteresis on random surfaces compared to structured surfaces at any given pillar area fraction. Under this theory, the effective distance between pinning sites is larger on a random surface as compared to a structured surface for any given pillar area fraction. We proposed that a larger effective pinning-sites distance on random surfaces results in a lower hysteresis as compared to structured surfaces. However, a detailed understanding of the proposed `cluster' theory would need a more detailed experimental or numerical investigation of the micro-scale interface dynamics on random and structured surfaces. As random surfaces are invariably part of the naturally occurring rough surfaces, this study will help to better understand such real surfaces as well as aid in designing new surfaces with functionalized wettability. 

\section*{Declaration of Competing Interest}

The authors declare that they have no known competing financial interests or personal relationships that could have appeared to influence the work reported in this paper.

\section*{Data availability}

Data will be made available on request.

\section*{Acknowledgements}

One of the authors (P.K.) acknowledges the financial support from the University of
Melbourne in the form of the Melbourne Research Scholarships program. P.K. also acknowledges the support from Melbourne India Postgraduate Program (MIPP). This work was performed in part at the Melbourne Centre for Nanofabrication (MCN), in the Victorian Node of the Australian National Fabrication Facility (ANFF).










\printcredits

\bibliographystyle{elsarticle-num}
\bibliography{cas-refs}



\end{document}


\maketitle

\setcounter{figure}{0}
\renewcommand{\figurename}{Figure}
\renewcommand{\thefigure}{S\arabic{figure}}

\setcounter{section}{0}
\renewcommand{\thesection}{S\arabic{section}}

\renewcommand{\theequation}{S.\arabic{equation}}

\section{Surface fabrication}
\label{app:sample_fabrication}

The surfaces were prepared using a photolithographic process on 0.5 mm thick silicon wafers. First, the silicon wafers were hard-baked at 180 \textdegree C for 15 minutes and then cooled to room temperature. The hard-baked silicon wafers were spin-coated with AZ1512 photoresist at 2000 rpm for 30 seconds using SUSS Delta 80 RC Spinner. A glass-chrome direct contact lithographic mask was used for exposing the photoresist to Ultra-Violet light using an ABM Flood Light Source. After exposure for 5 seconds, the exposed part of the photoresist was removed by AZ 726 M/F developer mixed with distilled water in a 2:3 ratio (by volume). The developed wafers were observed under an optical microscope to check for any defects in photoresist coating, UV exposure and developer application. After washing away the unwanted photoresist from wafers, cylindrical pillars on the silicon surface were created by etching away the uncoated portion of silicon. The wafers were etched up to a depth of 10 $\rm{\mu m}$ by {Deep Reactive Ion Etching (Oxford Plasma Lab 100)} followed by photoresist stripping using oxygen plasma. {The cleaned wafers were then coated with a 5 $\rm{nm}$ layer of chromium followed by a 25 $\rm{nm}$ layer of gold using the Intlvac Nanochrome AC/DC system.} The thickness of the gold layer deposited was checked by an {Ambios XP 200  profilometer}. The developed surface consists of 10 $\rm{\mu m}$ diameter cylindrical pillars of 10 $\rm{\mu m}$ height. We obtained a very good aspect ratio of the pillars with good dimensional accuracy. Scanning electron microscopy (SEM) was used to check the quality of the surfaces with a sample SEM image shown in figure \ref{fig:exp_sem}. A sacrificial 6 $\rm{\mu m}$ layer of AZ 4562 photoresist was put on the top of the wafer surface to protect the pillars while dicing the wafer into individual pieces - this layer was later removed.
\begin{figure}
    \centering
    \includegraphics[width=1.0\textwidth]{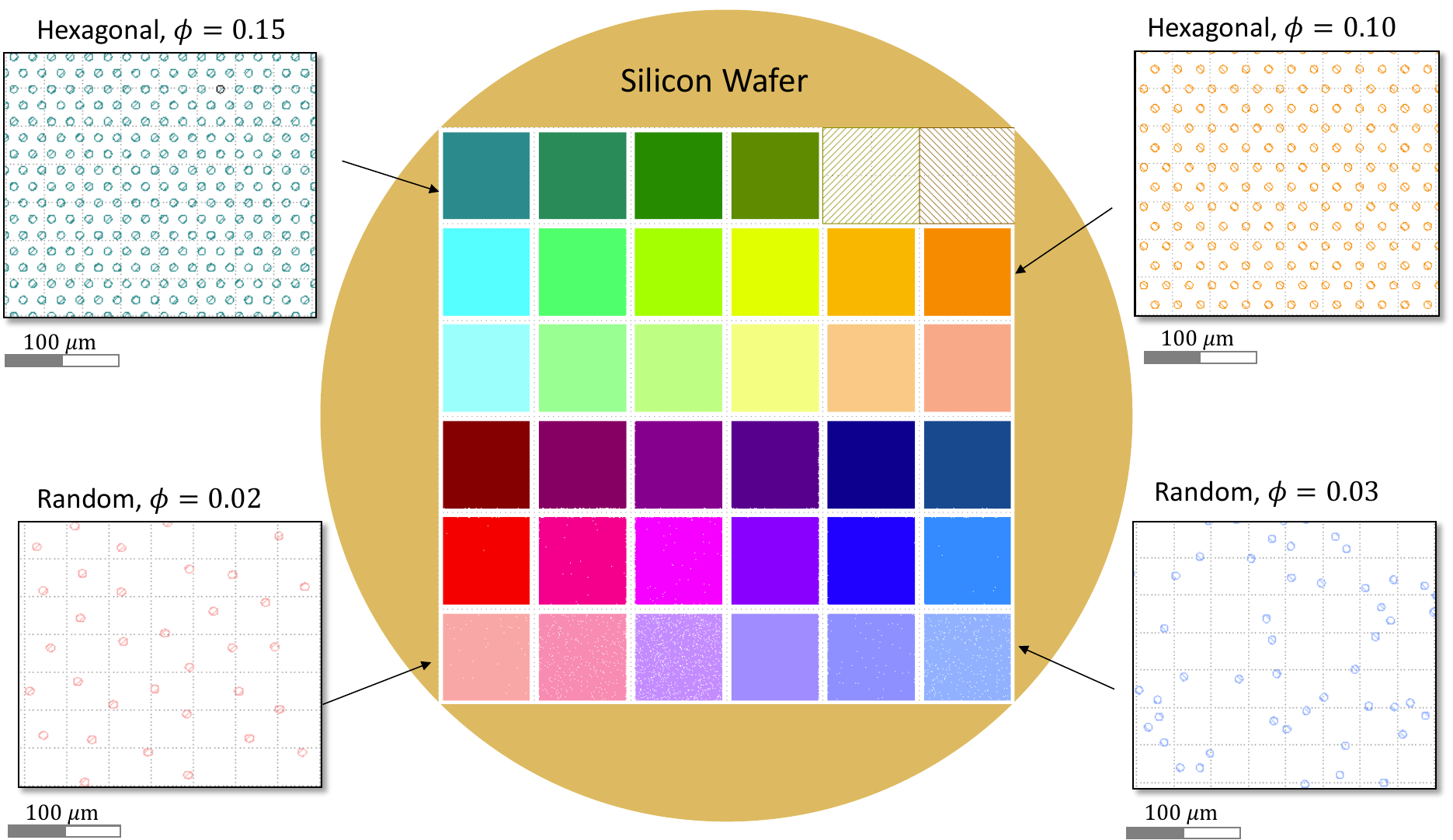}
    \caption[A representation of the fabricated silicon wafer with micron-sized pillars etched onto the surface.]{A representation of the fabricated silicon wafer with samples arranged in the increasing order of pillar area fraction. Each sample can be taken out of the wafer and tested separately for the contact angles.}
    \label{fig:app_silicon_wafer}
\end{figure}
\begin{figure}
    \centering
    \includegraphics[width=0.65\textwidth]{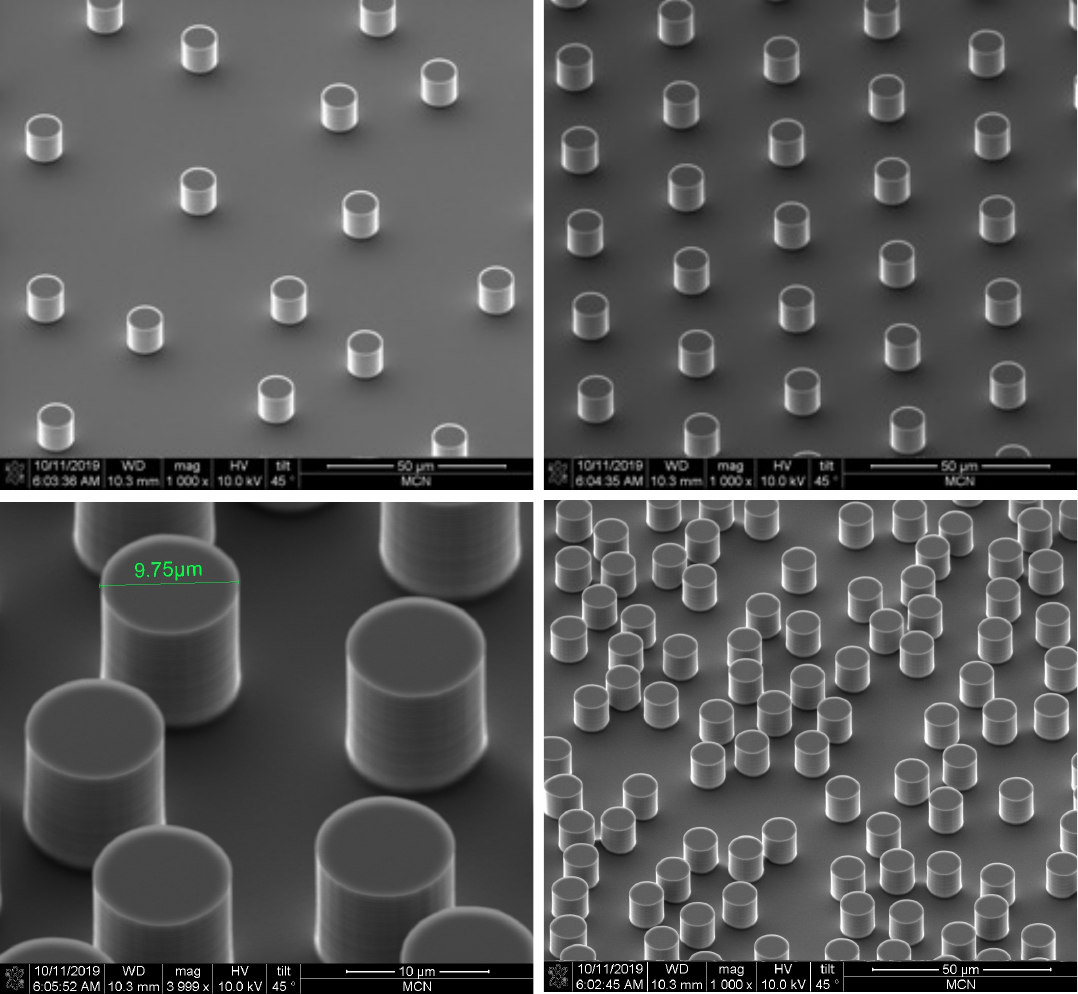}
    \caption{Scanning Electron Microscope (SEM) images of the fabricated surface.}
    \label{fig:exp_sem}
\end{figure}
\section{Surface cleaning and coating with thiol monolayer}
\label{app:surface_cleaning}

Before measuring the hysteresis on fabricated surfaces, care was taken to thoroughly clean them and cover them with a monolayer of appropriate thiol. During the fabrication, surfaces were coated with AZ 4562 photoresist to prevent any mechanical damage to the pillars. This coating also serves as a barrier against the adsorption of any chemical impurities onto the surface. There are different commonly used methods for removing a non-cross-linked photoresist layer and cleaning the silicon wafer. We tested some of the methods which are listed below, followed by our observations on their comparative ability to clean the surface.
\begin{enumerate}
    \item[(a)] \textbf{Acetone:} 
    Samples were dipped in acetone (Acros, purity $>99$\%) for 2 minutes followed by rinsing with isopropanol (ChemSupply) followed by drying in a stream of nitrogen gas.
   
    \item[(b)] \textbf{Dimethyl sulfoxide:} 
    Samples were dipped in  DMSO (ChemSupply) heated to 60 \textdegree C for 10 minutes followed by rinsing with DI water and drying in a stream of nitrogen gas.
    
    \item[(c)] \textbf{Ammonium peroxide solution:} Also known as base Piranha, is a 3:1 solution of ammonium hydroxide (\ce{NH4OH}) and hydrogen peroxide (\ce{H2O2}). To prepare the solution, we mixed two parts of DI water and one part of 30 wt.\% hydrogen peroxide (ChemSupply) by volume in a beaker. The solution was heated to 70 $^{\circ}\rm{C}$ using a water bath, and a drop of ammonium hydroxide (Sigma) was added to it. After a few seconds, the solution started fizzing and in this fizzing solution, the samples were dipped for approximately 20 seconds. After it, the samples were removed and washed thoroughly with DI water and ethanol (ChemSupply, undenatured) and then dried in a stream of nitrogen gas.
    
    \item[(d)] \textbf{Piranha solution:} Acid Piranha or Piranha solution is a mixture of 3 parts sulphuric acid (\ce{H2SO4}) and 1 part 30 wt.\% hydrogen peroxide. For preparing the Piranha solution we placed a beaker in an ice bath and slowly poured 3 parts of \ce{H2SO4} (RCI Labscan, purity 98\%) and 1 part of \ce{H2O2} (ChemSupply, 30\% w/w). The Piranha solution was then slowly deposited over the samples with the help of a glass pipette, to cover the entire surface area. Care was taken not to let the Piranha solution enter the edges of the samples, as it could lead to disruption of the gold coating. After 30 seconds, the samples were washed with copious amounts of DI water followed by drying under a nitrogen gas stream. 
    
    \item[(f)] \textbf{Oxygen plasma: } The oxygen plasma was generated inside the Harrick Plasma PDC-32G. Before using plasma, the samples were cleaned by dipping in acetone (Acros, purity $>99$\%) for 2 minutes followed by isopropanol (ChemSupply) for 1 minute and then rinsing with DI water. The samples were then dried in a nitrogen gas stream. The cleaned samples were then kept inside the plasma generator under a vacuum of 350-400 mTorr for approximately 10 minutes.
\end{enumerate}

To check the quality of cleaned surfaces, we deposited a small droplet of DI water. Pure gold has high surface energy and the water droplet should spread spontaneously on a perfectly clean sample \cite{smith1980hydrophilic}. Out of the methods listed above, the oxygen plasma cleaning method gave the best results judged by the ability of DI water drops to readily spread (DI water drops deposited on plasma-cleaned gold-coated samples readily spread out). Note that a pure gold surface absorbs moisture and other impurities from the air very rapidly causing it to lose its hydrophilicity \cite{troughton1988monolayer,bain1989formation}. We found that a good way to maintain the purity of the gold coating was to quickly coat the appropriate layer of thiol as soon as the surface was cleaned. We used octadecanethiol (\ce{C18H37SH}, ODT) for rendering the gold-coated samples hydrophobic. The ODT monolayer on gold coating has good stability and the surface maintains its hydrophobicity and chemical homogeneity for a long time \cite{bain1989formation}. We prepared a 1 mM solution of ODT  in toluene  by dissolving 7.165 mg of ODT (Aldrich) in 25 ml of toluene (ChemSupply) at room temperature. To coat the ODT layer, cleaned samples were dipped in the 1mM ODT solution in toluene for 30 minutes. After that, the samples were taken out and washed with copious amounts of ethanol and finally dried in a stream of nitrogen gas. 

\section{Contact angle measurements}

For measuring contact angles the inbuilt drop analysis software (SCA 20) was used. Each time a drop was deposited, a video of the drop while adding liquid to it as well as while removing liquid from it was taken. The maximum frame rate of the video depends upon the image size and was generally around 50 fps. This means that for a usual advancing and receding contact angle measurement, i.e. addition of 4 $\mu$L and removal of 6 $\mu$L of liquid at a rate of 0.06 $\mu$L$/s$, approximately 6,000 to 8,000 frames were analysed. Each frame of the video was analysed by using the \textit{ellipse-fitting} algorithm \cite{law2016surface} inbuilt into the software.

A typical goniometer setup is accurate within $\pm 2$\tc \cite{law2016surface}, however, this depends on the angle being measured and the fitting algorithm used for extracting data from the droplet image. We have used the \textit{ellipse fit} method \cite{sklodowska1999method} which offers fast and reliable measurements for moderate contact angle values \cite{lubarda2011analysis,xu2014algorithm,law2016surface}. 

\section{AFM scans of the sample}
\label{app:afm_scans}

We developed our predictive equation for the advancing and receding contact angles on the framework of mechanical energy balance and the assumption of zero inherent hysteresis on the flat surface. To make our equation useful for real surfaces, we used the inherent advancing/receding contact angle on the flat surface ($\theta_{\rm{A}}/\theta_{\rm{R}}$) instead of Young's angle ($\theta_{\rm{e}}$) in our analyses. The advancing and receding contact angle on the flat surface can be very different from Young's angle due to (for example) the presence of surface roughness at the nanometric scales. 

\begin{figure}
    \centering
    \includegraphics[width=\textwidth]{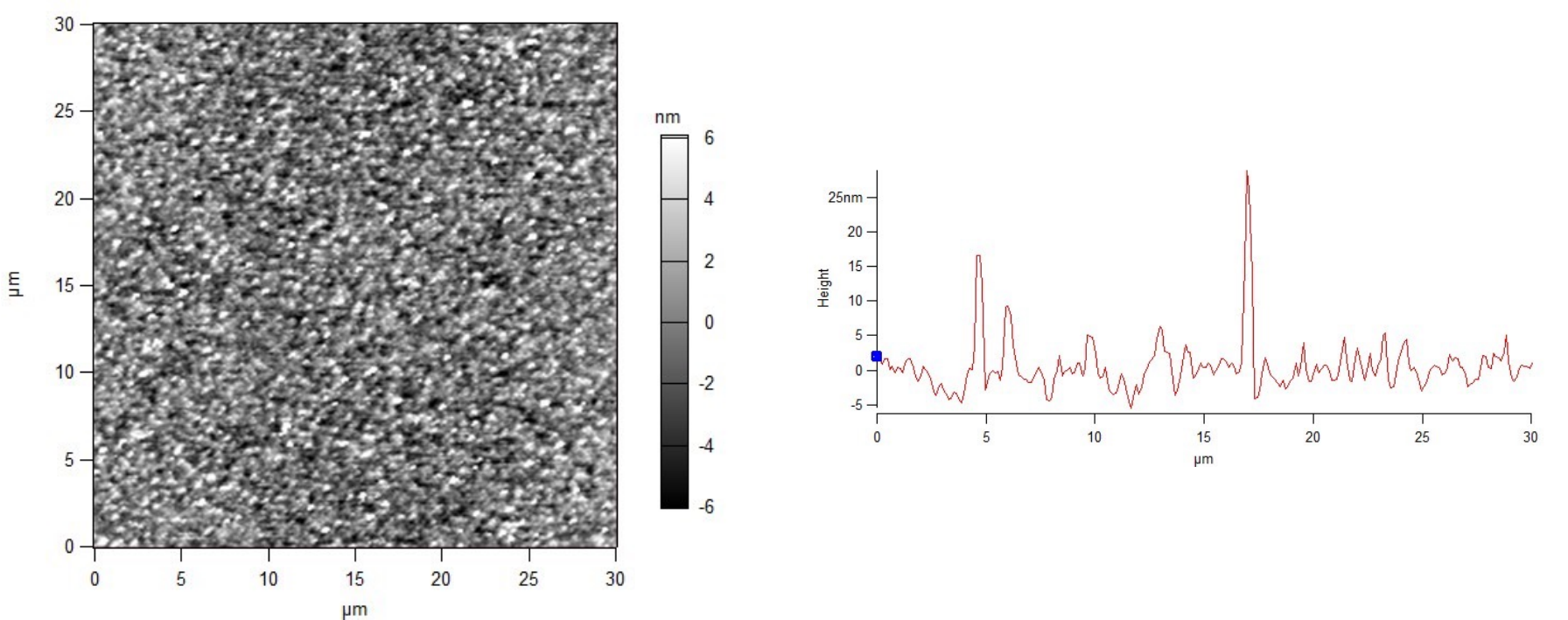}
    \caption{AFM image of the flat surface. This represents the area in between the pillars}
    \label{fig:app_afm_bottom}
\end{figure}

Figure \ref{fig:app_afm_bottom} shows the AFM scan of the flat surface which represents the area in between the pillars. We can see roughness at nano-scales present on the surface, we observed an average roughness of 2.5 nm and rms roughness of 5.0 nm.
%
The surface roughness on top of the pillars was also in the nanometric range with an average and rms roughness of 1.8 nm and 2.3 nm respectively. Importantly, the roughness on both surfaces is many orders of magnitude less than the pillar size. Further, independent inherent advancing angle measurements on the pillar top and base flat surfaces show negligible difference.
\begin{figure}
    \centering
    \includegraphics[width=\textwidth]{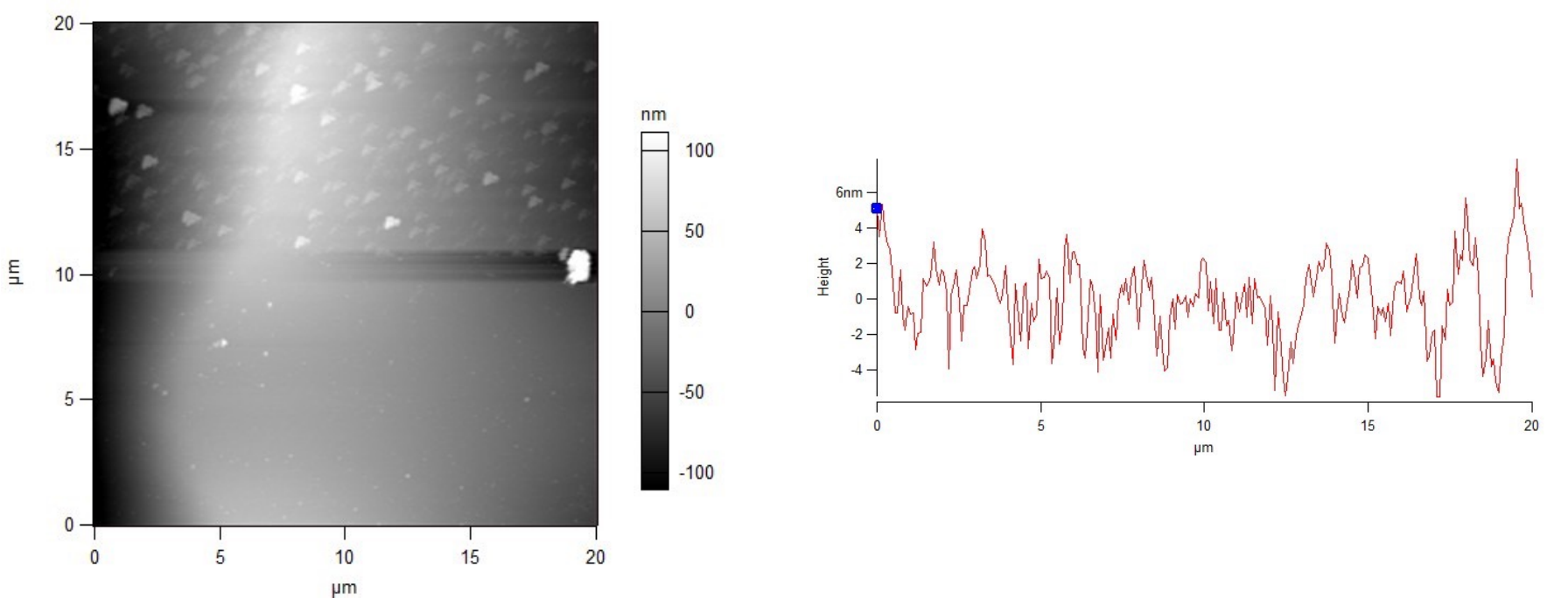}
    \caption{AFM image of the top surface of the pillars.}
    \label{fig:app_afm_top}
\end{figure}

\section{Typical contact angle variation on random and hexagonal surfaces}

\begin{figure}
    \centering
    \includegraphics[width=\textwidth]{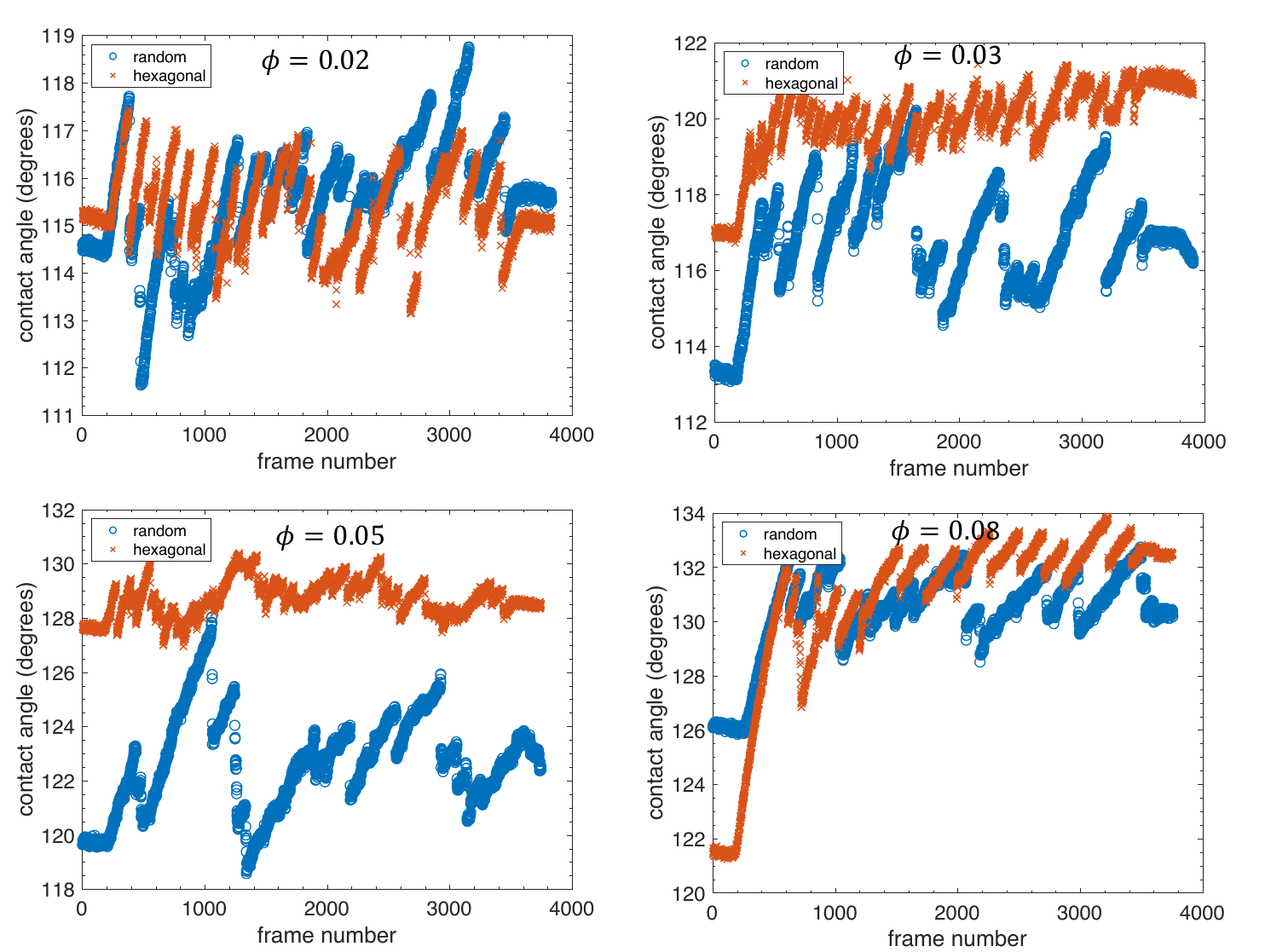}
    \caption{Typical variation in the macroscopic contact angle ($\theta_{\rm{m}}$) with respect to the image frame number during the advancing motion of DI water droplets (at a flow rate of 0.06 $\mu$l/s) on a random/hexagonal surface for different area fractions. }
    \label{fig:app_rand_hex_comp}
\end{figure}

Figure \ref{fig:app_rand_hex_comp} shows a typical contact angle vs image frame variation during the advancement of an interface on a random/hexagonal surface. We observed that fluctuations, i.e. the difference between the peaks and troughs are greater on random surfaces, even though the peak magnitudes were higher on hexagonal surfaces. This behaviour was observed over the entire range of pillar area fractions. 

\section{Calculating average advancing/receding contact angle and average peak-to-trough angle variation}
\label{app:avg_angle_method}

Here we present a methodology to choose a suitable starting point for averaging the contact angles. Suppose, $d_{\rm{initial}}$ is the base diameter of the droplet when initially deposited. According to the experimental protocol, more liquid is added to the droplet at a very small rate (0.06 or 0.10 $\mu \rm{L}/s$ in this study) and the motion of the TPCL is observed. The TPCL is said to be advancing when the drop base diameter ($d_{j}$) at an instant is greater than $d_{\rm{initial}}$ plus some constant multiple ($\zeta$) of $d_{\rm{avg}}$, that is
\begin{equation}
    d_{j} >  d_{\rm{initial}} + \zeta d_{\rm{avg}}.
    \label{eqn:advancing_definition}
\end{equation}
We start averaging contact angles once the interface has started advancing according to the equation (\ref{eqn:advancing_definition}). The selection of an appropriate $\zeta$ in equation (\ref{eqn:advancing_definition}) is discussed later in this section. 
Similarly, for finding the average receding contact angle, the averaging is started once we are sure that the entirety of the TPCL is receding. Defining $d_{\rm{final}}$ as the base diameter when the injection of liquid into the droplet is stopped, the TPCL is said to be receding when the drop base diameter ($d_{{j}}$) at that instant is smaller than $d_{\rm{final}}$ minus some constant multiple ($\zeta$) of $d_{\rm{avg}}$, that is
\begin{equation}
    d_{{j}} < d_{\rm{final}}  - \zeta d_{\rm{avg}}.
    \label{eqn:receding_definition}
\end{equation}
Therefore, the averaging of the advancing angle starts when the droplet diameter becomes greater than $d_{\rm{initial}} + \zeta d_{\rm{avg}}$ and is stopped when it reaches a value of $d_{\rm{final}}-\zeta d_{\rm{avg}}$. Similarly, the averaging of the receding angle starts when the droplet diameters become smaller $d_{\rm{final}}-\zeta d_{\rm{avg}}$ and continues as long as the droplet diameter is greater than $d_{\rm{final2}}+ \zeta d_{\rm{avg}}$, where $d_{\rm{final2}}$ represents the droplet diameter at the end of the experiment (that is when the removal of liquid from the droplet is stopped). Therefore, the average advancing (or receding) contact angle can be calculated by averaging the contact angles between the proper range, that is
%
\begin{equation}
\begin{split}
    \theta_{\rm{a}}&=\left ( \sum_{d_j>d_{\rm{initial}}+\zeta d_{\rm{avg}}}^{d_j<d_{\rm{final}}-\zeta d_{\rm{avg}}} \theta_{{j}} \right )/\sum j, \\
   \theta_{\rm{r}}&=\left ( \sum_{d_j<d_{\rm{final}}-\zeta d_{\rm{avg}}}^{d_j>d_{\rm{final2}}+\zeta d_{\rm{avg}}} \theta_{{j}} \right )/\sum j,
   \end{split}
   \label{eqn:exp_averaging}
\end{equation}
with the sums conducted over consecutive frame numbers. Note that, as frames are evenly distributed in time, equation (\ref{eqn:exp_averaging}) represents a temporal averaging of angles, with starting and stopping times constrained by displacement limits that recognize the structure of the surface. Note that in each test (advancing and receding) averaging is halted before the droplet motion stops to ensure that averages are truly representative of advancing/receding rather than equilibrium states.

To calculate the average peak-to-trough difference, we adopt the following methodology: For the advancing motion of the interface, we start with the contact angle corresponding to the first diameter in the base diameter data selected from the experimental values using equation (\ref{eqn:advancing_definition}). We represent this droplet diameter by $d_{\rm{initial,pt}}$ with 
\begin{equation}
   d_{\rm{initial,pt}} \approx d_{\rm{initial}} + \zeta d_{\rm{avg}}.
   \label{eqn:exp_d_initial_pt_def}
\end{equation}
We then search forward in time, measuring the change in droplet diameter relative to this starting value. When this change is greater than a certain value ($\Delta d$), that is
\begin{equation}
    d_{{j}}-d_{\rm{initial,pt}} > \Delta d,
    \label{eqn: delta_peak_trough}
\end{equation}
where $d_{j}$ is the droplet diameter at that instant of time, we assume a TPCL jumping event has occurred. Therefore, the peak and trough contact angles are contained in this $\Delta d$ range of base diameters. Within this contact angle range, we search the for maximum contact angle first, with this angle representing the peak contact angle ($\theta_{\rm{peak}}$). To find the trough contact angle ($\theta_{\rm{trough}}$) we search for the minimum contact angle in the contact angle range starting from the peak contact angle to the contact angle corresponding to $d_{j}$. The maximum ($\theta_{\rm{peak}}$) and minimum ($\theta_{\rm{trough}}$) contact angles and the difference $\theta_{\rm{peak}}-\theta_{\rm{trough}}$ represent the peak and trough contact angles and the peak-to-trough angle difference respectively, corresponding to the first TPCL jumping event. To find the peak/trough contact angles (and the peak-to-trough angle difference) for the next TPCL jump, we replace the $d_{\rm{initial,pt}}$ with $d_{j}$ and use equation (\ref{eqn: delta_peak_trough}) to find the new value of $d_j$, such that change in droplet base diameter from $d_{\rm{initial,pt}}$ to $d_{j}$, now captures the second TPCL jump.  
This process is repeated for all the TPCL jumping events until the droplet diameter reaches $d_{\rm{final}}-\zeta d_{\rm{avg}}$. The average of the difference between peak and trough contact angles (that is, $\theta_{\rm{peak}}-\theta_{\rm{trough}}$) corresponding to all the TPCL jumping events for the droplet diameter change from $d_{\rm{initial}}+\zeta d_{\rm{avg}}$ to $d_{\rm{final}}-\zeta d_{\rm{avg}}$ gives the average peak-to-trough angle difference ($\Delta \theta_{\rm{pt,a}}$) for an advancing interface. In addition to the average peak-to-trough angle difference ($\Delta \theta_{\rm{pt,a}}$), we also calculate the standard deviation in the peak-to-trough angle difference ($\Delta \theta_{\rm{sd,pt,a}}$).

\begin{sidewaysfigure}
    \centering
    \includegraphics[width=0.90\textwidth]{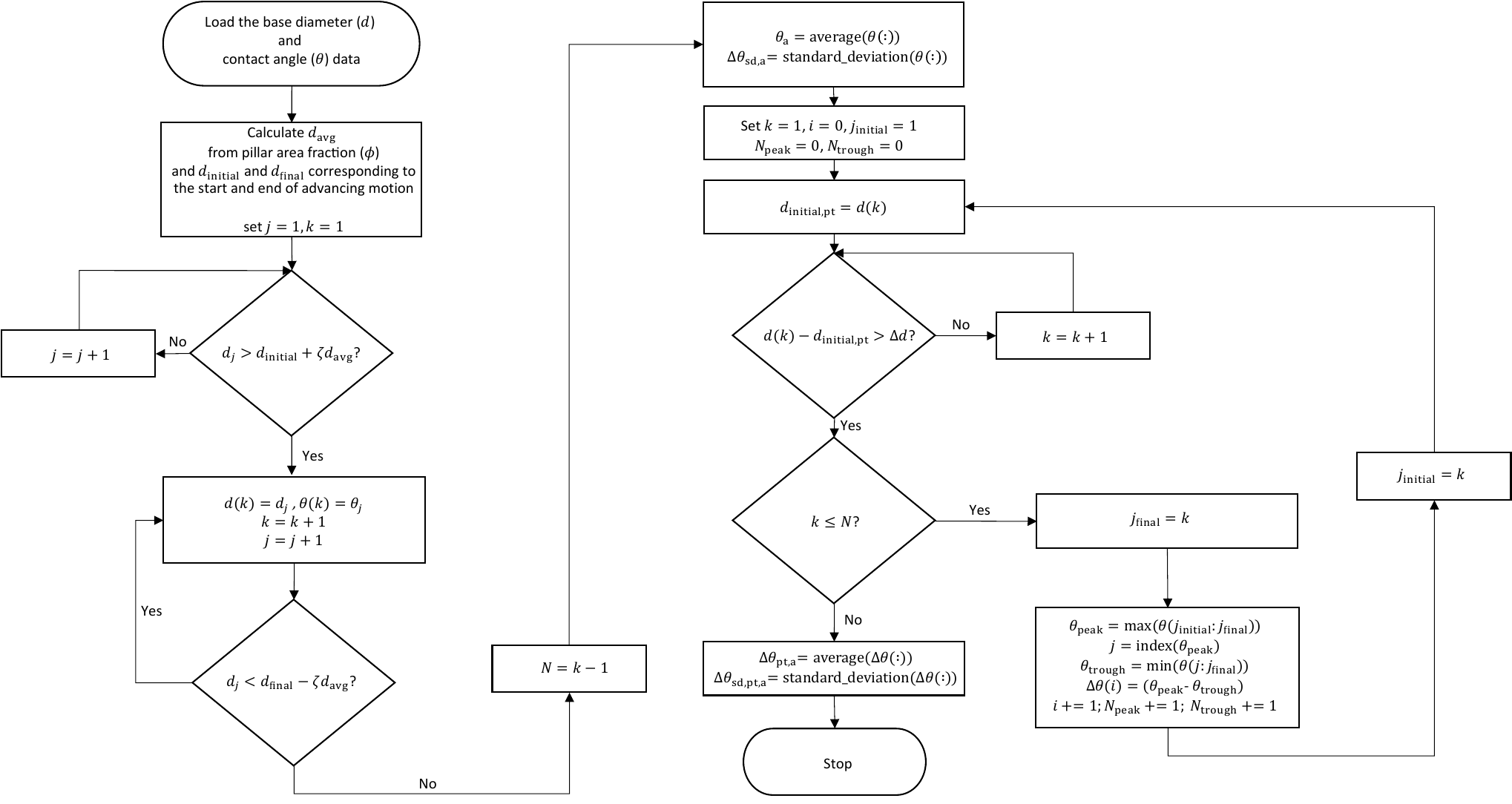}
    \caption{Flowchart of the algorithm used for calculating the average advancing/receding contact angle, its standard deviation, average peak-to-trough angle variation and its standard deviation.}
    \label{fig:app_algo}
\end{sidewaysfigure}

In figure \ref{fig:app_algo} we show a flowchart depicting the algorithm used for calculating the average advancing/receding contact angle, its standard deviation, average peak-to-trough angle variation and its standard deviation from the experimental data of macroscopic contact angle vs time (or frame number). Similar to the advancing interface, the receding motion of the interface is also analysed using an analogous procedure.

\begin{figure}
     \centering
     \includegraphics[width=0.65\textwidth]{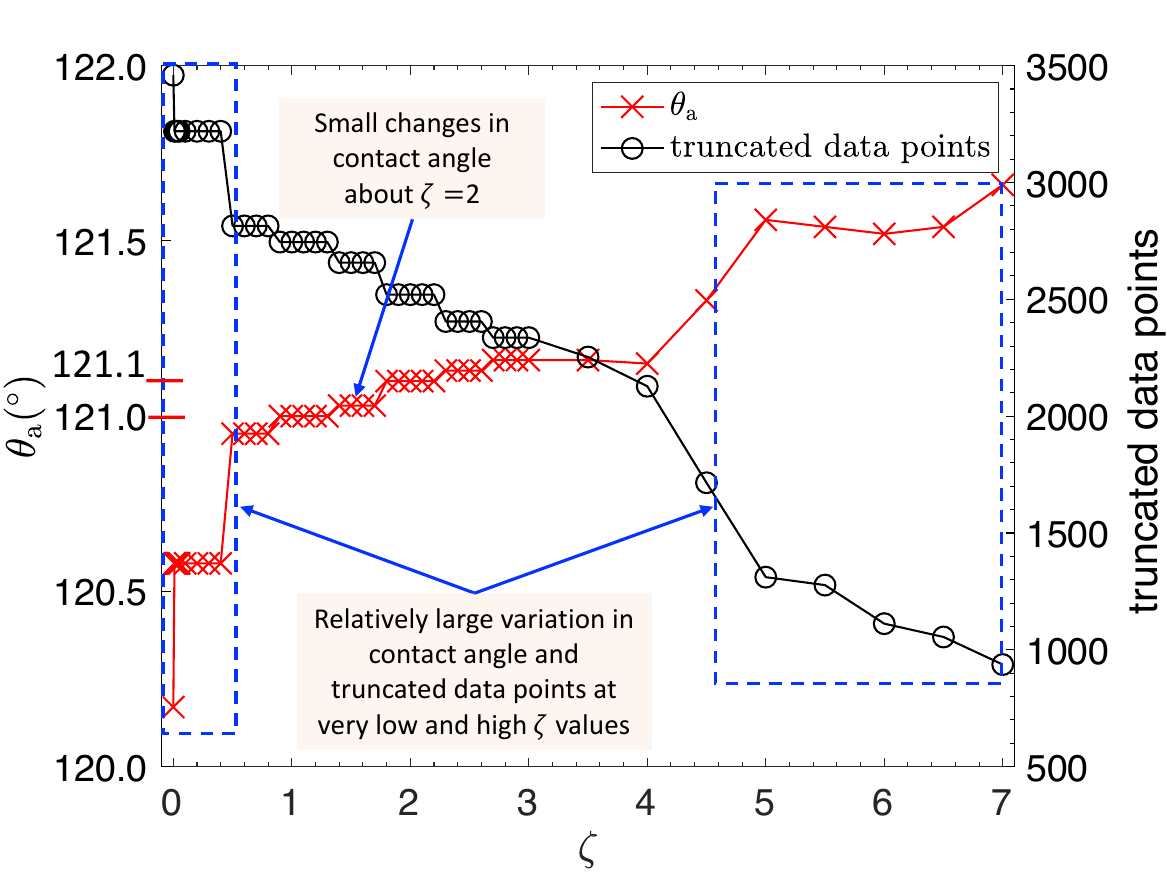}
     \caption{Variation in the averaged advancing contact angle ($\theta_{\rm{a}}$) and the number of truncated data points with $\zeta$ (see equations (\ref{eqn:advancing_definition}), (\ref{eqn:receding_definition})). At very low and high $\zeta$ values ($\zeta<0.5$ and $\zeta>4.0$), we see a comparatively large variation in $\theta_{\rm{a}}$ and the number of truncated data points. At moderate $\zeta$ values ($1.0<\zeta<3.0$), the variation in $\theta_{\rm{a}}$ becomes very small, showing a relatively constant averaged advancing contact angle. The loss in data points is also moderate in the moderate $\zeta$ limit. We choose $\zeta=2.0$ as it has a physical meaning to it, that is the contact line (of the full droplet) moves a distance equal to the average distance between pillar centres ($d_{\rm{avg}}$).
     The data presented represents the advancing angles measured with DI water on a random surface with an average pillar area fraction of 0.05.}
     \label{fig:avg_method}
 \end{figure}

We now discuss the particular choice of $\zeta$ which represents the actual advance/recede of interface with a minimum loss in experimental data. To illustrate, in figure \ref{fig:avg_method} we plot the variation in $\theta_{\rm{a}}$ (averaged)
and the number of truncated data points against $\zeta$ for an advancing interface (DI water droplet on a random surface with $\phi=0.05$). Most significant is the variation from $\zeta=0$ to $\zeta=0.5$ where we observe a significant variation in the truncated data points, as well as in $\theta_{\rm{a}}$. Then after the variation in $\theta_{\rm{a}}$ is gradual with $\zeta$, giving a relatively constant value of averaged advancing angle ($\theta_{\rm{a}}$). However, when $\zeta$ increases beyond a moderate value ($\zeta>4.0$), we observe a comparatively large variation in the $\theta_{\rm{a}}$. Based on the above observations, a moderate $\zeta$ ($1.0<\zeta<3.0$ in the example presented in figure \ref{fig:avg_method}) gives a good representation of the average angle. We chose $\zeta=2.0$ specifically because it represents an interface advancing $d_{\rm{avg}}$ everywhere around the circumference of a droplet, representing a TPCL that has advanced by a distance of $d_{\rm{avg}}$ in every direction prior to commencing angle averaging. 
Note that a high value of $\zeta$ results in a loss of data. 
For example, the loss in the data corresponding to $\zeta=2.0$ and $\zeta=4.0$ are 36.1\% and 46.1\% respectively (while the corresponding angles are 121.1\tc and 121.2\tc, respectively). The loss of data becomes more obvious at low area fractions as $d_{\rm{avg}}$ increases with a decrease in $\phi$. 
Note that for all test surfaces $d_{\rm{avg}}$ is larger than the resolution of the experimental droplet diameter measurements (10$\mu$m).

\section{Film propagation during hemiwicking}
\label{sec_app:hemiwicking}

In \S3.4.1 we discussed the phenomena of hemiwicking, in relation to the limits of validity of the proposed model (equation (24) in \S3.3). Here we discuss hemiwicking in detail, starting with the derivation of the relationship between the inherent advancing angle ($\theta_{\rm{A}}$) and the roughness area fraction for hemiwicking to be energetically favourable. 

Referring to Harvie \cite{dhcontact09}, an energy balance within a moving control volume around the TPCL can be written as
\begin{equation}
    T_0 (t_1,t_2) = \sum_{i=1}^{6}T_{i}(t_1,t_2),
    \label{eqn:T0_sum}
\end{equation}
where
\begin{align}
T_0 (t_1,t_2) & = \frac{1}{A_{\rm{CV}}} \left [ E(t=t_2)-E(t=t_1) \right ] \label{eqn:T0} \\
E(t) & = \int_{V_{\rm{CV}}} \left ( \frac{1}{2} \rho v^2 + \rho \hat{\Phi} \right ) \, dV + \sum_{i<j} \sigma_{ij} A_{ij} \label{eqn:E} \\
T_1 (t_1,t_2) & = \frac{1}{A_{\rm{CV}}} \int^{t_2}_{t_1} \int_{S_{\rm{cv}}} \sum_{i<j} \sigma _{ij} \delta_{\text{S},ij} \boldsymbol{n}_{\rm{CV}} \cdot \boldsymbol{v}_{\rm{CV}} \, dS dt \label{eq:T1} \\
T_2 (t_1,t_2) & = \frac{1}{A_{\rm{CV}}} \int^{t_2}_{t_1} \int_{S_{\rm{CV}}} \boldsymbol{n}_{\rm{CV}} \cdot \frac{1}{2}\rho v^2 ( \boldsymbol{v}_{\rm{CV}} - \boldsymbol{v} ) \, dS dt \label{eqn:T2} \\
T_3 (t_1,t_2) & = \frac{1}{A_{\rm{CV}}} \int^{t_2}_{t_1} \int_{S_{\rm{CV}}} \boldsymbol{n}_{\rm{CV}} \cdot \rho \hat{\Phi} ( \boldsymbol{v}_{\rm{CV}} - \boldsymbol{v} ) \, dS dt \label{eqn:T3} \\
T_4 (t_1,t_2) & = - \frac{1}{A_{\rm{CV}}} \int^{t_2}_{t_1} \int_{S_{\rm{CV}}} \sum_{i<j} \sigma_{ij} \delta_{\text{S},{ij}} \boldsymbol{n}_{\text{S},ij} \boldsymbol{n}_{\text{S},ij} : \boldsymbol{v} \boldsymbol{n}_{\rm{CV}} \, dS dt \label{eqn:T4} \\
T_5 (t_1,t_2) & = \frac{1}{A_{\rm{CV}}} \int^{t_2}_{t_1} \int_{S_{\rm{CV}}} \boldsymbol{\text{T}}_{\rm{M}}:\boldsymbol{v} \boldsymbol{n}_{\rm{CV}} \, dS dt \label{eqn:T5} \\
T_6 (t_1,t_2) & = - \frac{1}{A_{\rm{CV}}} \int^{t_2}_{t_1} \int_{V_{\rm{CV}}} \boldsymbol{\text{T}}_{\rm{M}}:\boldsymbol   {\nabla} \boldsymbol{v} \, dV dt \label{eqn:T6}
\end{align}
%
%
An order-of-magnitude analysis of all the terms from equations (\ref{eqn:T0}) to (\ref{eqn:T6}) for a control volume capturing the advancing motion of a thin film on a rough surface (see figure \ref{fig:app_hemiwicking_scenarios}) yields,
\begin{equation}
  0 = \overrightarrow{\sigma} - \overleftarrow{\sigma} - D_{\rm{hw,a}} -A\mu v_{\rm{CV}} \ln \left(\frac{h_{\rm{rough}}}{h_{\rm{mol}}} \right),
  \label{eqn:sigma_D}
\end{equation}
where $\overrightarrow{\sigma}$ and $\overleftarrow{\sigma}$ are the interfacial energies entering ($\sum_{i<j} \sigma_{ij} \overrightarrow{A}_{ij}$) and leaving ($\sum_{i<j} \sigma_{ij} \overleftarrow{A}_{ij}$) the control volume per unit area traversed by the TPCL ($A_{\rm{CV}}$), $D_{\rm{hw,a}}$ is the dissipation in energy within the control volume as it advances with the hemiwicking film, $A$ is a constant, $\mu$ is the viscosity of the fluid, $h_{\rm{rough}}$ is the roughness length scale and $h_{\rm{mol}}$ is the molecular length scale. The analysis that derives equation (\ref{eqn:sigma_D}) from $T_1$ to $T_4$ mirrors that in the work of Harvie \cite{dhcontact09}, with the exclusion of a control volume sized fluid-1/fluid-2 interface.

Using equation (\ref{eqn:sigma_D}), we now consider the three possible cases of spreading of a thin film on a surface with surface roughness in the form of cylindrical pillars. 

\subsection*{(1) The film spreads covering the pillar tops}

\begin{figure}
    \centering
    \includegraphics[width=0.50\textwidth]{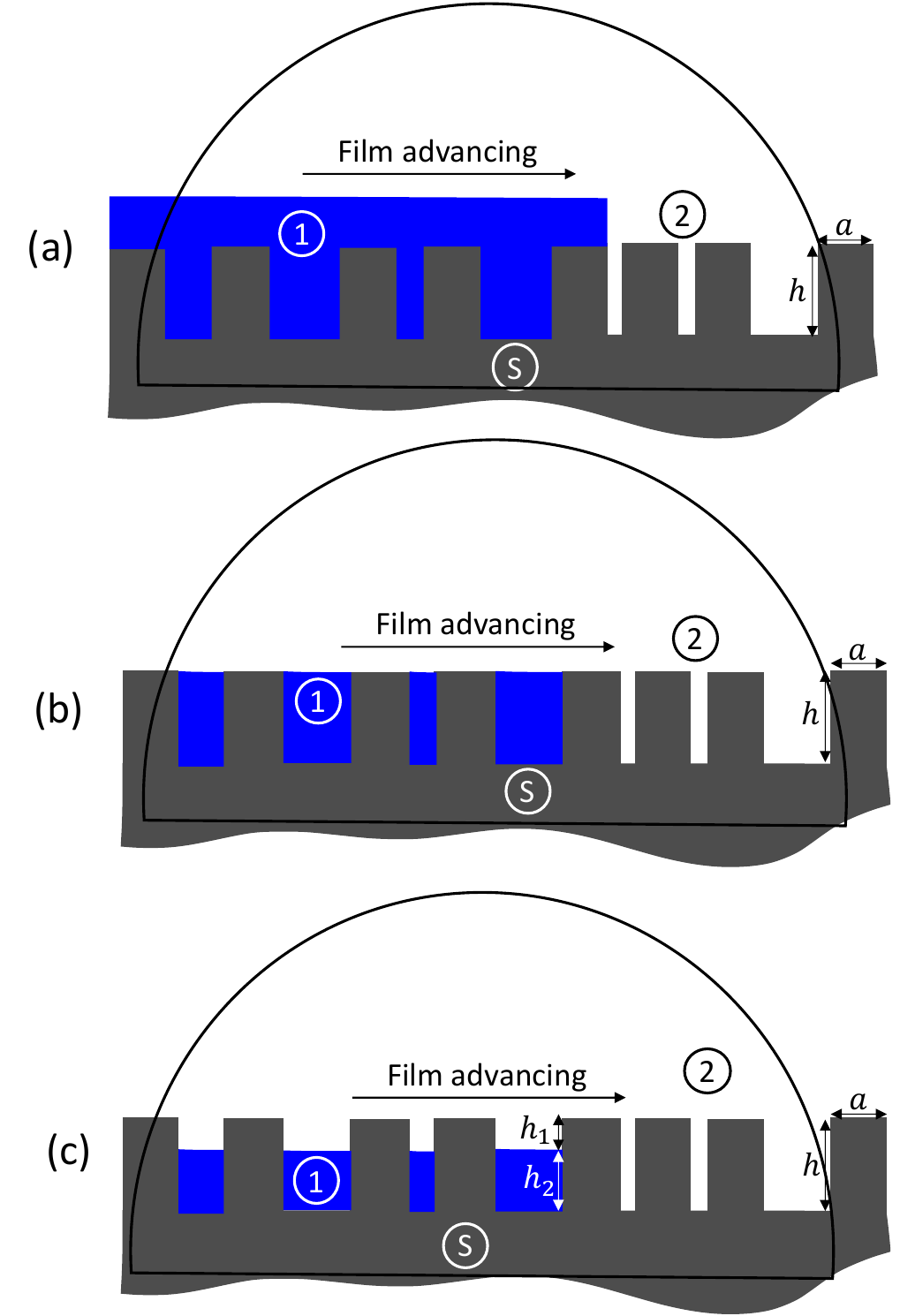}
    \caption{Schematic representing the different possible ways in which a liquid may undergo hemiwicking, and the control volumes used in the energy analysis, and the control volumes used in the energy analysis.}
    \label{fig:app_hemiwicking_scenarios}
\end{figure}

Figure \ref{fig:app_hemiwicking_scenarios}(a) shows a thin film spreading on a pillared surface with pillar tops submerged under the film. Various interfacial areas entering and leaving the control volume (CV) around the TPCL are
\begin{equation}
    \begin{split}
        &\overrightarrow{A}_{\rm{1S}} = 0, \quad \overrightarrow{A}_{\rm{2S}}=A_{\rm{CV}} + nA_{\rm{CV}}\pi a h, \quad \overrightarrow{A}_{12}=0, \\
        &\overleftarrow{A}_{\rm{1S}}=A_{\rm{CV}}+nA_{\rm{CV}}\pi a h, \quad \overleftarrow{A}_{\rm{2S}}=0, \quad \overleftarrow{A}_{12}=A_{\rm{CV}},
    \end{split}
    \label{eqn:top_wet_area}
\end{equation}
where $n$ is the number density of the pillars, $a$ and $h$ are respectively the diameter and height of the pillars. Substituting the areas from equation (\ref{eqn:top_wet_area}) into equation (\ref{eqn:sigma_D}) and using Young's equation \cite{young1805iii} for relating the interfacial tensions to the inherent advancing angle ($\theta_{\rm{A}}$), we get
\begin{equation}
    \sigma_{12}(r \cos \theta_{\rm{A}} - 1) - D_{\rm{hw,a}} - A\mu v_{\rm{CV,1}} \ln \left(\frac{h_{\rm{rough}}}{h_{\rm{mol}}} \right) = 0.
\end{equation}
Therefore, the control volume (or the TPCL) velocity ($v_{\rm{CV,1}}$) can be calculated as
\begin{equation}
    v_{\rm{CV,1}} = \sigma_{12} \left( \frac{r\cos\theta_{\rm{A}} -1 - \overline{D}_{\rm{hw,a}}}{A \mu \ln \left(\frac{h_{\rm{rough}}}{h_{\rm{mol}}} \right)} \right),
    \label{eqn:hemi_scene_1}
\end{equation}
where $\overline{D}_{\rm{hw,a}}={D}_{\rm{hw,a}}/\sigma_{12}$ is the non-dimensional energy dissipation within the CV. Since $\overline{D}_{\rm{hw,a}}$ is always positive, the velocity in equation (\ref{eqn:hemi_scene_1}) is negative for all values of $\theta_{\rm{A}}$, except when $r > \frac{1 + \overline{D}_{\rm{hw,a}}}{\cos\theta_{\rm{A}}}$. 
%

\subsection*{(2) The film spreads leaving the pillar tops dry}

Here we discuss the scenario when the film is advancing within the pillar forest and pillar tops are left untouched by the advancing film, as shown in figure \ref{fig:app_hemiwicking_scenarios}(b). In this case, various interfacial areas entering and leaving the control volume around the TPCL are
\begin{equation}
    \begin{split}
        &\overrightarrow{A}_{\rm{1S}} = 0, \quad \overrightarrow{A}_{\rm{2S}}=A_{\rm{CV}} + nA_{\rm{CV}}\pi a h, \quad \overrightarrow{A}_{12}=0, \\
        &\overleftarrow{A}_{\rm{1S}}=A_{\rm{CV}}+nA_{\rm{CV}}\pi a h - n A_{\rm{CV}} A_{\rm{top}}, \quad \overleftarrow{A}_{\rm{2S}}=n A_{\rm{CV}} A_{\rm{top}},\\ &\overleftarrow{A}_{12}=A_{\rm{CV}} - n A_{\rm{CV}} A_{\rm{top}},
    \end{split}
    \label{eqn:top_dry_area}
\end{equation}
where $A_{\rm{top}}$ is the area of the pillar tops.

Substituting the areas from equation (\ref{eqn:top_dry_area}) into equation (\ref{eqn:sigma_D}) and using Young's equation \cite{young1805iii} for relating the interfacial tensions to the inherent advancing angle ($\theta_{\rm{A}}$), we get
\begin{equation}
    \sigma_{12}\left[(r-\phi)\cos\theta_{\rm{A}} - (1-\phi) \right]- D_{\rm{hw,a}} - A\mu v_{\rm{CV,2}} \ln \left(\frac{h_{\rm{rough}}}{h_{\rm{mol}}} \right) = 0.
\end{equation}
Therefore,
\begin{equation}
    v_{\rm{CV,2}} = \sigma_{12}\left(\frac{(r-\phi) \cos \theta_{\rm{A}} - (1-\phi) - \overline{D}_{\rm{hw,a}}}{A \mu \ln \left(\frac{h_{\rm{rough}}}{h_{\rm{mol}}} \right)} \right).
        \label{eqn:hemi_scene_2}
\end{equation}
For the film to advance, $v_{\rm{CV,2}}>0$, therefore,
\begin{equation}
    \begin{split}
        (r-\phi) & \cos\theta_{\rm{A}} - (1-\phi) - \overline{D}_{\rm{hw,a}} > 0 \\
        &\cos\theta_{\rm{A}} > \left( \frac{1-\phi+\overline{D}_{\rm{hw,a}}}{r-\phi}\right)=\cos \theta_{\rm{hw,A}}.
    \end{split}
    \label{eqn:hemi_scene_2_angle}
\end{equation}
If we neglect the total non-dimensional energy dissipation within the CV ($\overline{D}_{\rm{hw,a}}$) during the film advancement, we get the classical condition for the onset of hemiwicking, i.e., 
\begin{equation}
    \theta_{\rm{A}} < \cos^{-1}\left( \frac{1-\phi}{r-\phi} \right).
    \label{eqn:app_hemi_critical}
\end{equation}
The presence of $\overline{D}_{\rm{hw,a}}$ in equation (\ref{eqn:hemi_scene_2_angle}) hampers the onset of hemiwicking, i.e., a liquid may not exhibit hemiwicking even if the inherent advancing angle ($\theta_{\rm{A}}$) is slightly lower than the critical angle ($\theta_{\rm{hw,A}}$) as given in equation (\ref{eqn:app_hemi_critical}). In other words, the area fraction at which, for a given $\theta_{\rm{A}}$, hemiwicking should be observed, i.e. $\phi_{\rm{hw,a}}$, increases with the increase in $\overline{D}_{\rm{hw,a}}$. Using equation (\ref{eqn:hemi_scene_2_angle}) we can write the following limiting condition on the pillar area fraction ($\phi$) for a hemiwicking film to advance, 
%
\begin{equation}
    \phi \geq \phi_{\rm{hw,a}} = \left( \frac{1-\cos\theta_{\rm{A}} + \overline{D}_{\rm{hw,a}}}{1 + 3\cos\theta_{\rm{A}}} \right).
    \label{eqn:hemi_critical_diss}
\end{equation}
%

Comparing the control volume velocities based on equations (\ref{eqn:hemi_scene_1}) and (\ref{eqn:hemi_scene_2}), that is
\begin{equation}
\begin{split}
v_{\rm{CV,2}} - v_{\rm{CV,1}} &=  \sigma_{12}\left(\frac{(r-\phi) \cos \theta_{\rm{A}} - (1-\phi) - \overline{D}_{\rm{hw,a}}}{A \mu \ln \left(\frac{h_{\rm{rough}}}{h_{\rm{mol}}} \right)} \right) - \sigma_{12} \left( \frac{r\cos\theta_{\rm{A}} -1 - \overline{D}_{\rm{hw,a}}}{A \mu \ln \left(\frac{h_{\rm{rough}}}{h_{\rm{mol}}} \right)} \right)\\
&= \sigma_{12} \left(  \frac{\phi(1-\cos\theta_{\rm{A}})}{A \mu \ln \left(\frac{h_{\rm{rough}}}{h_{\rm{mol}}}\right)} \right) \geq 0.
\end{split}
\label{eqn:exp_hemi_vc1_vc2}
\end{equation}
In equation (\ref{eqn:exp_hemi_vc1_vc2}) the dissipation within the CV in both cases is assumed to be the same. Clearly, $v_{\rm{CV,2}}>v_{\rm{CV,1}}$, therefore the hemiwicking film prefers to advance with pillar tops dry instead of completely submerging them. 

\subsection*{(3) The film spreads leaving the pillar tops and a portion of the  sides dry}

In this scenario, we assume that the liquid film is advancing in such a way that it fills up to a pillar height $h_2$ ($h_2<h$) and a portion of the pillar ($h_1$) remains untouched by the film as shown in figure \ref{fig:app_hemiwicking_scenarios}(c). The various interfacial areas entering and leaving a control volume around in this case are
\begin{equation}
    \begin{split}
        &\overrightarrow{A}_{\rm{1S}} = 0, \quad \overrightarrow{A}_{\rm{2S}}=A_{\rm{CV}} + nA_{\rm{CV}}\pi a h, \quad \overrightarrow{A}_{12}=0, \\
        &\overleftarrow{A}_{\rm{1S}}=A_{\rm{CV}}+nA_{\rm{CV}}\pi a h_2 - n A_{\rm{CV}} A_{\rm{top}}, \quad \overleftarrow{A}_{\rm{2S}}=n A_{\rm{CV}} A_{\rm{top}} + nA_{\rm{CV}} \pi a h_1,\\ &\overleftarrow{A}_{12}=A_{\rm{CV}} - n A_{\rm{CV}} A_{\rm{top}}.
    \end{split}
    \label{eqn:top_dry_area_half}
\end{equation}
Substituting the areas from equation (\ref{eqn:top_dry_area_half}) into equation (\ref{eqn:sigma_D}) and using Young's equation \cite{young1805iii} for relating the interfacial tensions to the equilibrium angle ($\theta_{\rm{A}}$), we get
\begin{equation}
    \sigma_{12}\left[(r^{'}-\phi)\cos\theta_{\rm{A}} - (1-\phi) \right]- D_{\rm{hw,a}} - A\mu v_{\rm{CV,3}} \ln \left(\frac{h_{\rm{rough}}}{h_{\rm{mol}}} \right) = 0,
\end{equation}
where $r^{'}=1+4 \phi h_2/a$. Therefore, the velocity of the control volume (or the TPCL), that is $v_{\rm{CV,3}}$ can be determined as,
\begin{equation}
    v_{\rm{CV,3}} = \sigma_{12}\left(\frac{(r^{'}-\phi) \cos \theta_{\rm{A}} - (1-\phi) - \overline{D}_{\rm{hw,a}}}{A \mu \ln \left(\frac{h_{\rm{rough}}}{h_{\rm{mol}}} \right)} \right).
        \label{eqn:hemi_scene_3}
\end{equation}
Since $r>r^{'}$, therefore the film propagation velocity, for a given inherent advancing angle and pillar area fraction, in the second scenario ($v_{\rm{CV,2}}$ in equation (\ref{eqn:hemi_scene_2})) is greater as compared to the third scenario ($v_{\rm{CV,3}}$  in equation (\ref{eqn:hemi_scene_3})). Hence, a hemiwicking film that completely occupies the space between the pillars will advance faster, and thus preferentially, compared to the case of partial film wetting (that is equation (\ref{eqn:hemi_scene_3})). Further, this analysis neglects pressure drops within the thin film due to finite advance velocities. While not affecting the existence and direction of the hemiwicking velocity, this neglect may cause the realised hemiwicking velocities to be lower than the prediction (noting however that the magnitude of $\overline{D}_{\rm{hw,a}}$ is unknown for hemiwicking). Thus, of the three hemiwicking cases considered, only the second case will occur in reality (equation (\ref{eqn:hemi_scene_2})). This conclusion does assume that the dissipation ($\overline{D}_{\rm{hw,a}}$) in the 2nd and 3rd cases are comparable.


\section{Formation of a thin film of liquid during receding motion of the interface}
\label{app_sec:receding_film}

In \ref{sec_app:hemiwicking}, we demonstrated the energetically most favourable morphology of an advancing film during a typical hemiwicking case is the one which advances within the pillar forest filling up to a height equal to that of the pillars while leaving the pillar tops untouched (figure \ref{fig:app_hemiwicking_scenarios}(b)). The interfacial advancement in such cases results in what we call a `split-advancing' motion. Similarly, during the receding motion of an interface, even if the interface advances in a homogeneous state we may observe a thin film left behind, while the bulk of the droplet recedes: We refer to this `split-receding' motion. It is not necessary for the droplet to exhibit hemiwicking to exhibit split-receding behaviour as the receding velocity of the film is governed by the inherent receding angle ($\theta_{\rm{R}}$), as opposed to the inherent advancing angle ($\theta_{\rm{A}}$) for the advancing case which must be equal or greater. Hence, there are a range of $\phi$ for which a thin film neither advances or recedes. In this section we find the inherent receding angle which causes the film to recede, thus defining the range of pillar area fractions for which a thin film is stationary.

Similar to the advancing case, for a receding film, the control volume velocity ($v_{\rm{CV}}$) for a case as shown in figure \ref{fig:app_hemiwicking_scenarios}(b) can be written as,
\begin{equation}
    v_{\rm{CV}} = \sigma_{12}\left(\frac{(1- \phi)-(r-\phi) \cos \theta_{\rm{R}} - \overline{D}_{\rm{hw,r}}}{A \mu \ln \left(\frac{h_{\rm{rough}}}{h_{\rm{mol}}} \right)} \right),
        \label{eqn:hemi_receding}
\end{equation}
where $\overline{D}_{\rm{hw,r}}$ is the total non-dimensional energy dissipation within the control volume as it recedes with the hemiwicking film. For the split-receding, $v_{\rm{CV}}$ based on $\theta_{\rm{R}}$ should be zero (or negative) noting that for split-receding case the thin film does not recede with the main droplet volume. Therefore, 
\begin{equation}
    (1-\phi) - (r-\phi) \cos\theta_{\rm{R}} - \overline{D}_{\rm{hw,r}} \leq 0.
    \label{eqn:split_receding1}
\end{equation}
Equation (\ref{eqn:split_receding1}) can be written as a condition on $\theta_{\rm{R}}$ that assumes split-receding, 
\begin{equation}
 \cos\theta_{\rm{R}} \geq \left( \frac{1-\phi-\overline{D}_{\rm{hw,r}}}{r - \phi} \right) =\cos \theta_{\rm{hw,R}}. 
 \label{eqn:exp_th_hw_R}
\end{equation}
If this equation isn't satisfied the receding will be homogeneous as any thin film present ahead of the TPCL will spontaneously move back towards the TPCL. Here $\theta_{\rm{hw,R}}$ is the critical angle such that a thin film prefers to recede (that is exhibit split-receding) if the inherent receding angle ($\theta_{\rm{R}}$) is smaller than the critical angle ($\theta_{\rm{hw,R}}$). 

Alternatively, equation (\ref{eqn:exp_th_hw_R}) can be written as a condition on $\phi$ that assures split-receding. Substituting ($r=1+4\phi$) in equation (\ref{eqn:exp_th_hw_R}) assuming the surface to be composed of unit aspect ratio pillars, the condition on the pillar area fraction ($\phi$) for the onset of split-receding can be written as
\begin{equation}
 \phi \geq \phi_{\rm{hw,r}} = \left( \frac{1-\cos\theta_{\rm{R}} - \overline{D}_{\rm{hw,r}}}{1+3 \cos\theta_{\rm{R}}} \right).   \label{eqn:split_receding2}
\end{equation}
The presence of energy dissipation term ($\overline{D}_{\rm{hw,r}}$) in equation (\ref{eqn:split_receding2}) promotes the onset of the split-receding phenomenon. That is, in the presence of a non-zero energy dissipation the split-receding phenomena can be observed at a lower pillar area fraction as compared to the zero dissipation case.

\section{Split-advancing and split-receding modes}
\label{app:split_advancing_receding}

In our experiments, hemiwicking was observed for the DMF, ACN and heptanol droplets, especially at high pillar area fractions. Ideally, hemiwicking is favoured if the inherent advancing angle ($\theta_{\rm{A}}$) is smaller than a certain critical angle ($\theta_{\rm{hw,A}}$) (see equation (\ref{eqn:hemi_scene_2_angle})). A low $\theta_{\rm{A}}$ and/or a high pillar area fraction ($\phi$) promotes hemiwicking. In Figure 8(c) inset we showed an advancing heptanol droplet (on a random surface with $\phi=0.20$) exhibiting split-advancing behaviour. In these cases, the bulk of the droplet is surrounded by a thin film of the liquid moving within the surface roughness. When such a droplet is made to recede, it results in what we call a split-receding motion. However, as discussed in \ref{app_sec:receding_film} this is not the only condition for split-receding and we may observe a split-receding behaviour even if the interface has advanced in a homogeneous state. 
\begin{figure}
    \centering
    \includegraphics[width=0.65\textwidth]{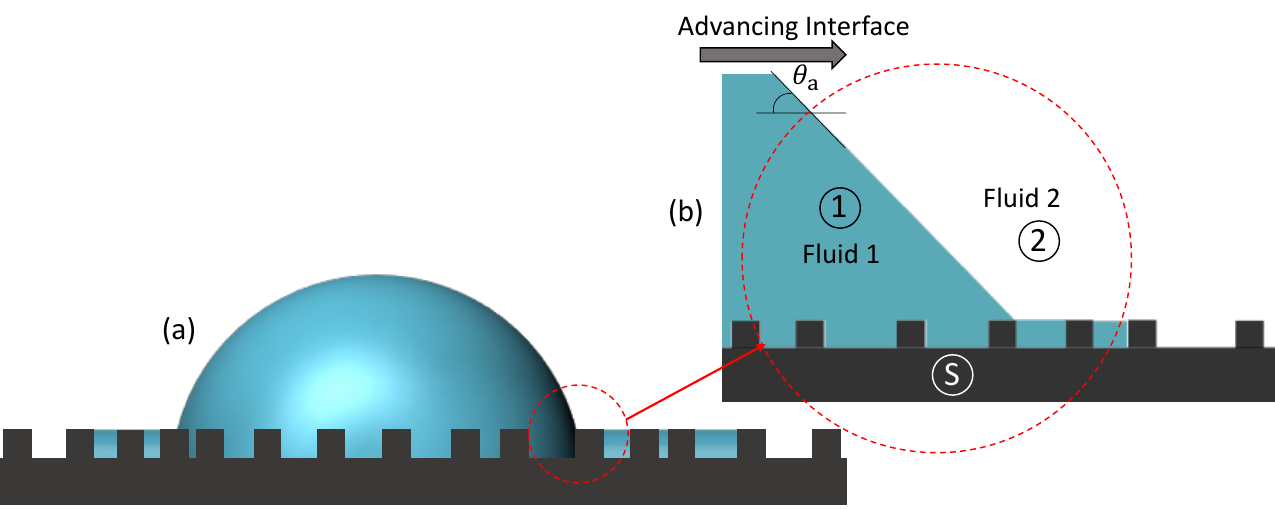}
    \caption[Schematic of a droplet exhibiting a hemiwicking wetting state.]{(a) Typical case of hemiwicking (not to scale). A control volume is chosen around the TPCL (shown by red dashed lines). (b) Shows a zoomed view of the interface morphology near the TPCL. A mechanical energy balance is performed for the interface between fluid-1 and fluid-2  while advancing on the pillared surface (S). The advancing contact angle ($\theta_{\rm{adv}}$) is measured at the top of the control volume.}
    \label{fig:app_schematic_meb_hemi}
\end{figure}

We now obtain an expression for the split-advance advancing contact angle (following hemiwicking) using the general mechanical energy balance equation \cite{dhcontact09}, that is
\begin{equation}
\sum_{i<j} \sigma_{ij}\frac{\overrightarrow A_{ij}-\overleftarrow A_{ij}}{\sigma_{12} A_{\rm{CV}}} -\cos\theta_{\rm{a}}-\overline{D}  = 0.
\label{eqn:general_meb}
\end{equation}
Here, $\overrightarrow A_{ij}$ and $\overleftarrow A_{jj}$ are $ij$ interface areas entering and leaving the control volume, $A_{\rm{CV}}$ is the area traversed by the TPCL as projected on a plane parallel to the surface. Figure \ref{fig:app_schematic_meb_hemi} shows a liquid/air (or two immiscible liquids) interface exhibiting hemiwicking on a surface decorated with cylindrical pillars (each pillar is of diameter $a$ and height $h$). The different interfacial areas entering and leaving the control volume as used in equation (\ref{eqn:general_meb}) are,
\begin{equation}
    \begin{split}
        \overrightarrow A_{1\rm{S}}&=A_{\rm{CV}}-nA_{\rm{CV}}\left(\frac{\pi a^2}{4} \right)+nA_{\rm{CV}}\pi a h, \quad \overrightarrow A_{2\rm{S}}=nA_{\rm{CV}}\left(\frac{\pi a^2}{4} \right),\\
        \overrightarrow A_{12}&=A_{\rm{CV}}-nA_{\rm{CV}}\left(\frac{\pi a^2}{4}\right), \quad \overleftarrow A_{1\rm{S}}=A_{\rm{CV}}+nA_{\rm{CV}} \pi a h,\\
        \overleftarrow A_{2\rm{S}}&=0,\quad \overleftarrow A_{12}=0.
    \end{split}
    \label{eqn:meb_hemi1}
\end{equation}

Using the areas from equation (\ref{eqn:meb_hemi1}) in equation (\ref{eqn:general_meb}) yields,
\begin{equation}
\begin{split}
      &\sigma_{1\rm{S}}\left( \frac{A_{\rm{CV}}-(nA_{\rm{CV}}\pi a^2 )/4 + nA_{\rm{CV}}\pi a h - A_{\rm{CV}} - nA_{\rm{CV}}\pi a h}{\sigma_{12} A_{\rm{CV}}} \right)+\\
      &\sigma_{2\rm{S}}\left( \frac{ (nA_{\rm{CV}} \pi a^2)/4 - 0}{\sigma_{12} A_{\rm{CV}}} \right) 
      +\left( \frac{A_{\rm{CV}}-nA_{\rm{CV}}(\pi a^2)/4}{A_{\rm{CV}}} \right)-\cos \theta_{\rm{sa}} - \overline{D}_{\rm{sa}}
      =0.
\end{split}
\label{eqn:meb_hemi2}
\end{equation}
Here, $\theta_{\rm{sa}}$ and $\overline{D}_{\rm{sa}}$ are the advancing contact angle and the total non-dimensional energy dissipation during the split-advancing mode respectively. Simplifying equation (\ref{eqn:meb_hemi2}) and using Young's equation \cite{young1805iii}, 
\begin{equation}
    \cos \theta_{\rm{sa}} = \phi(\cos\theta_{\rm{A}}-1) + 1 - \overline{D}_{\rm{sa}}.
    \label{eqn:app_meb_hemi_adv}
\end{equation}
%
%
%
Neglecting the energy dissipation in equation (\ref{eqn:app_meb_hemi_adv}) gives the classical equation for the equilibrium angle during hemiwicking \cite{bico2002wetting}. 

Similarly, for a receding interface, that is in split-receding mode the various interfacial areas entering and leaving the control volume as used in equation (\ref{eqn:general_meb}) are,
\begin{equation}
    \begin{split}
        \overrightarrow A_{1\rm{S}}&=A_{\rm{CV}} + nA_{\rm{CV}}\pi a h, \quad \overrightarrow A_{2\rm{S}}=0, \quad \overrightarrow A_{12}=0,\\ 
        \overleftarrow A_{1\rm{S}}&=A_{\rm{CV}}+nA_{\rm{CV}} \pi a h - nA_{\rm{CV}}\left(\frac{\pi a^2}{4}\right),
        \overleftarrow A_{2\rm{S}}=nA_{\rm{CV}}\left(\frac{\pi a^2}{4}\right),\\ \overleftarrow A_{12}&=A_{\rm{CV}}-nA_{\rm{CV}}\left(\frac{\pi a^2}{4}\right).
    \end{split}
    \label{eqn:meb_sr_areas}
\end{equation}
Using these area terms in equation (\ref{eqn:general_meb}) we get
\begin{equation}
\begin{split}
      &\sigma_{1\rm{S}}\left( \frac{A_{\rm{CV}}+ nA_{\rm{CV}}\pi a h - A_{\rm{CV}} - nA_{\rm{CV}}\pi a h + nA_{\rm{CV}} \pi a^2/4}{\sigma_{12} A_{\rm{CV}}} \right)+\\
      &\sigma_{2\rm{S}}\left( \frac{ 0 - nA_{\rm{CV}} \pi a^2/4}{\sigma_{12} A_{\rm{CV}}} \right) 
      +\left( \frac{0 -A_{\rm{CV}} + nA_{\rm{CV}}\pi a^2/4}{A_{\rm{CV}}} \right)-\cos (\pi -\theta_{\rm{sr}}) - \overline{D}_{\rm{sr}}
      =0.
\end{split}
\label{eqn:meb_sr_angle}
\end{equation}
Here, $\theta_{\rm{sr}}=\pi - \theta_{\rm{sa}}$ is the macroscopic contact angle and $\overline{D}_{\rm{sr}}$ is the total non-dimensional energy dissipation during the split-receding mode. Simplifying equation (\ref{eqn:meb_sr_angle}) and using Young's equation \cite{young1805iii}, 
\begin{equation}
    \cos \theta_{\rm{sr}} = \phi(\cos\theta_{\rm{R}}-1) + 1 + \overline{D}_{\rm{sr}}.
    \label{eqn:theta_sr}
\end{equation}

\section{Advancing angle in Cassie state}
\label{app_sec:cassie_advancing}

\begin{figure}
    \centering
    \includegraphics[width=0.65\textwidth]{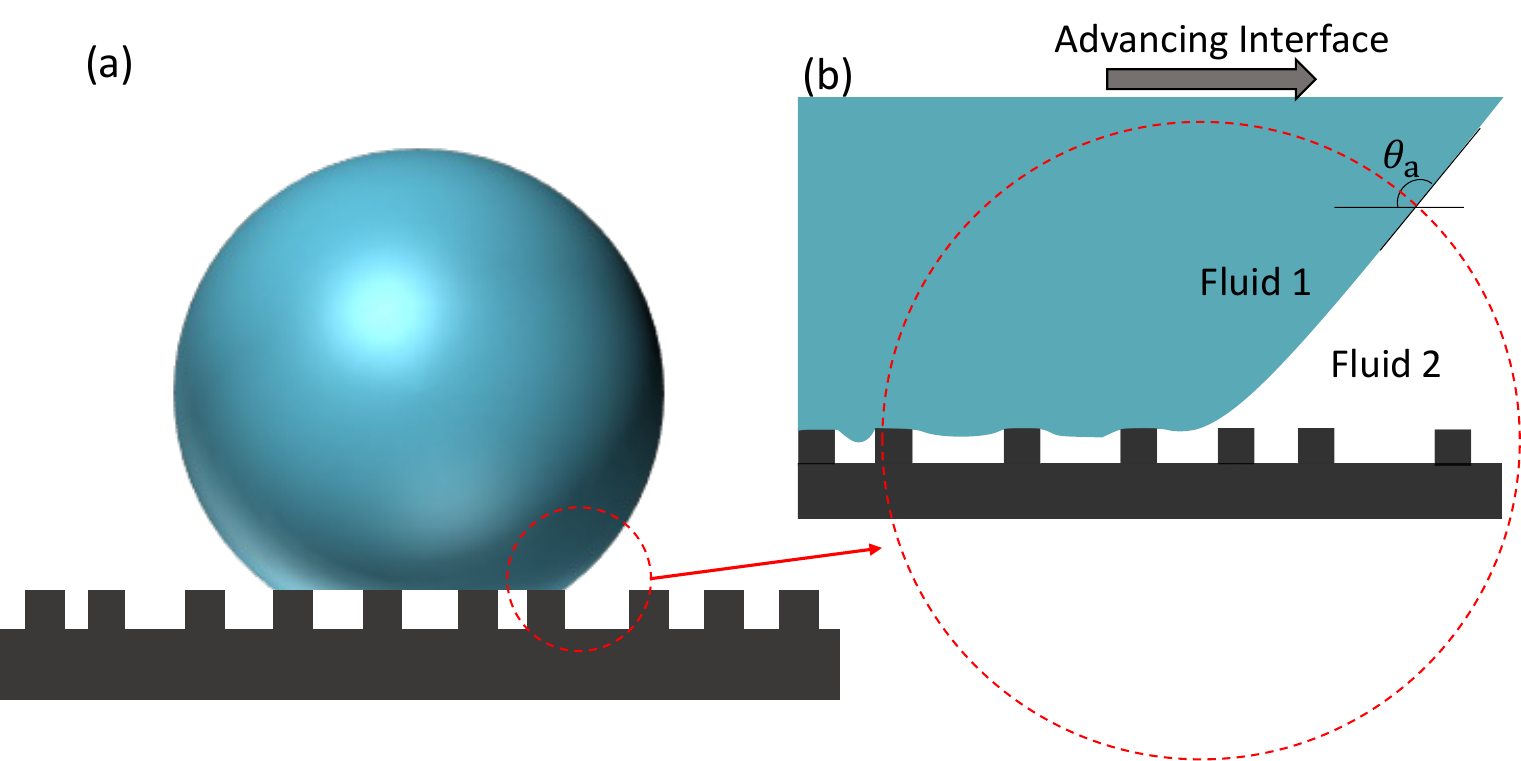}
    \caption{(a) A typical Cassie droplet on a pillared surface (not to scale). A control volume is chosen around the contact line (shown by red dashed lines). (b) Shows a zoomed view of the interface morphology near the contact line. A mechanical energy balance is performed for the interface between fluid-1 and fluid-2  while advancing on the pillared surface (S). The advancing contact angle ($\theta_{\rm{a}}$) is measured at the top of the control volume.}
    \label{fig_app:schematic_meb_cassie}
\end{figure}

We use the general mechanical energy balance equation (\ref{eqn:general_meb}) for deriving the advancing angle ($\theta_{\rm{a}}$) during the advancing motion of an interface in the Cassie wetting state. The various interfacial area terms in equation (\ref{eqn:general_meb}) for a typical control volume depicting the advancing motion in Cassie state as shown in figure \ref{fig_app:schematic_meb_cassie} can be written as
\begin{equation}
    \begin{split}
        \overrightarrow A_{1\rm{S}}&=0, \quad \overrightarrow A_{2\rm{S}}=A_{\rm{CV}}+nA_{\rm{CV}} A_{\rm{sides}}, \quad \overrightarrow A_{12}=0,\\
        \overleftarrow A_{1\rm{S}}&=nA_{\rm{CV}}A_{\rm{top}}, \quad \overleftarrow A_{2\rm{s}}=A_{\rm{CV}}-nA_{\rm{CV}}, A_{\rm{top}}+nA_{\rm{CV}}A_{\rm{sides}},\\
        \overleftarrow A_{12}&=\beta(A_{\rm{CV}}-nA_{\rm{CV}}A_{\rm{top}} ),
    \end{split}
    \label{eqn:meb_cassie1}
\end{equation}
where $A_{\rm{top}}=\pi a^2/h$ and $A_{\rm{sides}}=\pi a h$ are the pillar top and side areas and $\beta$ is a constant ($\beta \approx 1$) that accounts for the curvatures in the fluid-1/fluid-2 interface near the TPCL.
Substituting the areas from equation (\ref{eqn:meb_cassie1}) into equation (\ref{eqn:general_meb}), yields
\begin{equation}
\begin{split}
      \sigma_{1\rm{S}}\left( \frac{0-nA_{\rm{CV}}A_{\rm{top}}}{\sigma_{12} A_{\rm{CV}}} \right) &+ \sigma_{2\rm{S}}\left( \frac{A_{\rm{CV}}+nA_{\rm{CV}} A_{\rm{sides}} - A_{\rm{CV}} + nA_{\rm{CV}} A_{\rm{top}} - nA_{\rm{CV}} A_{\rm{sides}}}{\sigma_{12} A_{\rm{CV}}} \right) + \\
      &\sigma_{12}\left( \frac{0-\beta (A_{\rm{CV}}-nA_{\rm{CV}}A_{\rm{top}})}{\sigma_{12} A_{\rm{CV}}} \right) -\cos \theta_{\rm{a}} - \overline{D}_{\rm{Cassie}}=0,
\end{split}
\label{eqn:meb_cassie2}
\end{equation}
where $\overline{D}_{\rm{Cassie}}$ is the total non-dimensional energy dissipation in a Cassie wetting state. Upon further simplification by using the relationship between the three interfacial tension terms (Young's equation \cite{young1805iii}) and $\phi=nA_{\rm{top}}$, we get
\begin{equation}
    \cos \theta_{\rm{a}} = \phi (\beta + \cos \theta_{\rm{A}}) - \beta - \overline{D}_{\rm{Cassie}}.
    \label{eqn:meb_cassie4}
\end{equation}
If the energy dissipation and the interface's curvature near the TPCL are neglected, then equation (\ref{eqn:meb_cassie4}) reduces to the classic Cassie-Baxter equation, $\cos\theta_{\rm{a}}=\phi(1+\cos\theta_{\rm{A}})-1$.

\begin{figure}
    \centering
    \includegraphics[width=0.50\textwidth]{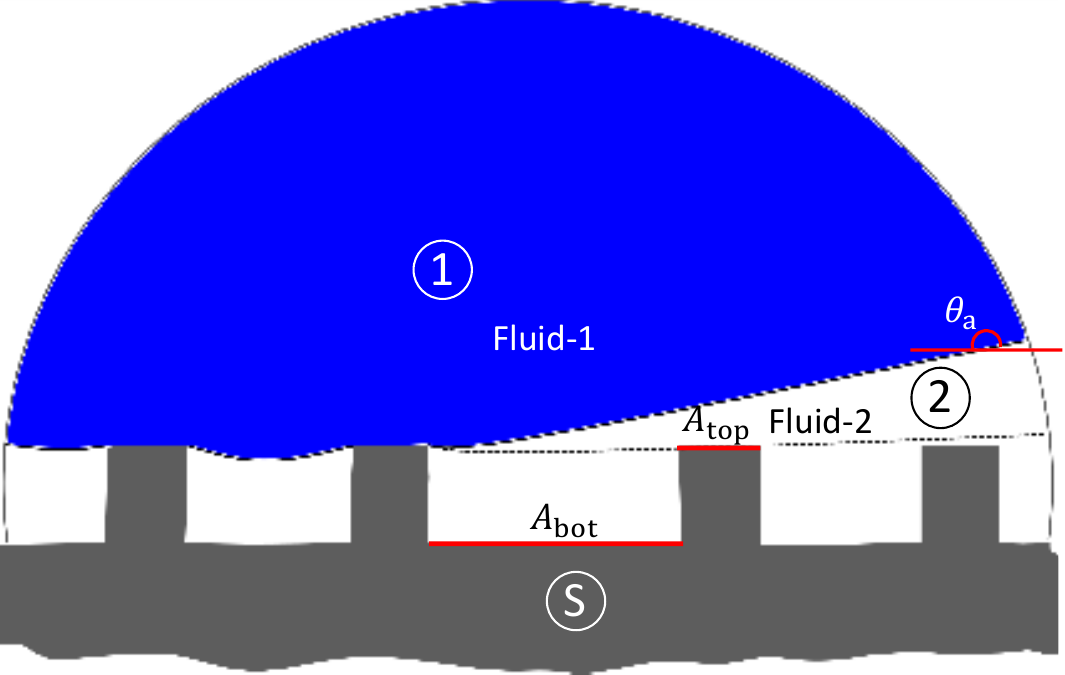}
    \caption{Control volume depicting the TPCL advancement during a typical Cassie wetting state. The dashed line represents the interface position when it just touches the next pillar in the advancement direction.}
    \label{fig_app:CV_Cassie_dissipation}
\end{figure}

\begin{figure}
    \centering
    \includegraphics[width=0.50\textwidth]{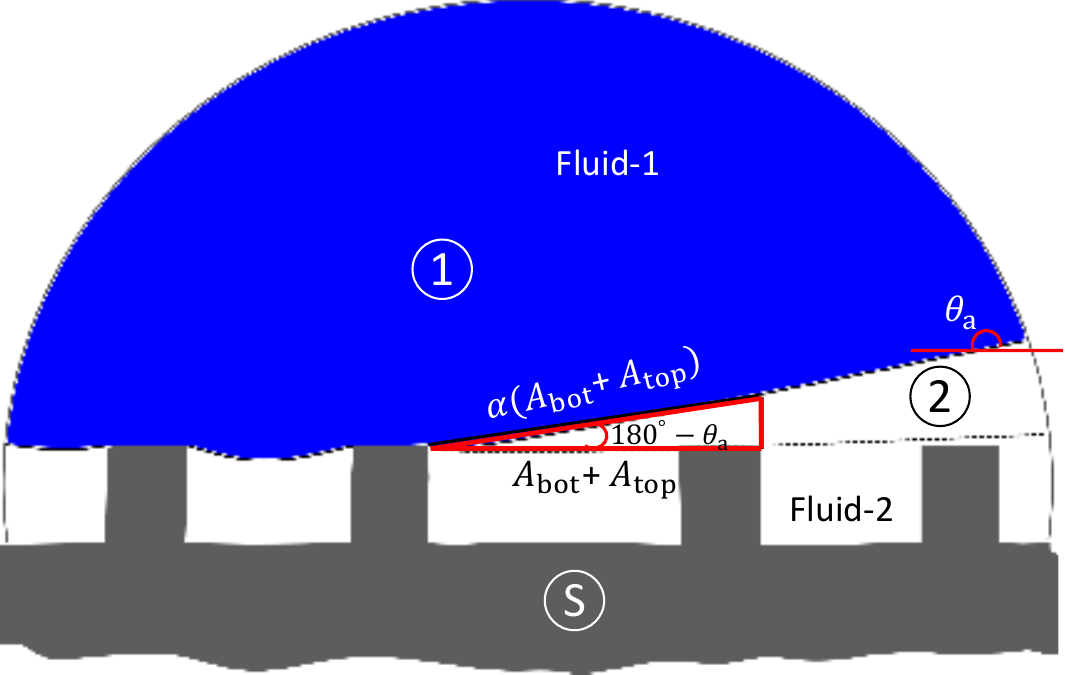}
    \caption{Schematic depicting the relationship between parameter $\alpha$ and the macroscopic contact angle capturing the state of the interface just before a TPCL jumping event.}
    \label{fig_app:alpha_Cassie}
\end{figure}

Following the work of Harvie \cite{dhcontact09}, the dissipation in energy (${D}_{\rm{Cassie}}$) can be obtained from the change in interfacial areas during a TPCL jumping event as discussed in \S3.2 (see equation (12)). In figure \ref{fig_app:CV_Cassie_dissipation} we show a schematic representing a single TPCL jumping event in a heterogeneous (Cassie-Baxter) wetting state such that the control volume only contains a single pillar in a direction normal to the TPCL advancement. 
Writing the change in different interfacial areas during a TPCL jump based on figure \ref{fig_app:CV_Cassie_dissipation},
\begin{equation}
\widehat{\Delta A_{12}} = \beta A_{\rm{bot}} - \alpha (A_{\rm{top}} + A_{\rm{bot}}), \quad \widehat{\Delta A_{\rm{1S}}} = A_{\rm{top}}, \quad \widehat{\Delta A_{\rm{2S}}} = -A_{\rm{top}},
\label{eqn:exp_hat_areas}
\end{equation}
where $A_{\rm{bot}}$ is the area between the pillars measured on the bottom surface, $\alpha$ is a constant which accounts for the fluid-1/fluid-2 interface curvature and angle near the TPCL ($\alpha \approx 1$) and $\beta$ is a constant that accounts for the curvature of the formed interface between pillars ($\beta \approx 1$). The total energy dissipation ($D_{\rm{Cassie}}$) can be expressed in terms of the interfacial areas given in equation (\ref{eqn:exp_hat_areas}), that is
\begin{equation}
\begin{split}
    D_{\rm{Cassie}} &= -nA_{\rm{CV}} \left[ \sigma_{12}\widehat{\Delta A_{12}} + \sigma_{1S}\widehat{\Delta A_{1S}} + \sigma_{2S}\widehat{\Delta A_{2S}} \right]\\
    &=-nA_{\rm{CV}} \sigma_{12} A_{\rm{top}} \left[ \frac{\beta A_{\rm{bot}}}{A_{\rm{top}}} - \alpha \left( 1 + \frac{A_{\rm{bot}}}{A_{\rm{top}}} \right) - \cos\theta_{\rm{A}}\right].
    \end{split}
    \label{eqn:exp_D_cassie_1}
\end{equation}
In equation (\ref{eqn:exp_D_cassie_1}) we have used the relationship $\cos \theta_{\rm{A}} = (\sigma_{\rm{2S}}-\sigma_{\rm{1S}})/\sigma_{12}$. Now the ratio $A_{\rm{bot}}/ A_{\rm{top}}$ can also be expressed in terms of pillar area fraction, that is
\begin{equation}
\begin{split}
 &\phi = \frac{A_{\rm{top}}}{A_{\rm{top}} + A_{\rm{bot}}}, \quad \text{or}\\
 &\frac{A_{\rm{bot}}}{A_{\rm{top}}} = \frac{1-\phi}{\phi}.
 \end{split}
 \label{eqn:exp_A_bot_A_top}
\end{equation}
Using the definition of $\phi$ from equation (\ref{eqn:exp_A_bot_A_top}) and also noting that $\phi=nA_{\rm{top}}$, equation (\ref{eqn:exp_D_cassie_1}) can be further simplified to
\begin{equation}
\begin{split}
 D_{\rm{Cassie}} &= -(\sigma_{12}A_{\rm{CV}}) \phi \left [ \beta \left( \frac{1-\phi}{\phi}\right) - \alpha \left(1 + \frac{1-\phi}{\phi} \right) - \cos\theta_{\rm{A}}\right]\\
 &= -(\sigma_{12}A_{\rm{CV}}) \left[\beta(1-\phi) - \alpha - \phi \cos\theta_{\rm{A}}\right].
 \end{split}
 \label{eqn:exp_D_cassie_2}
\end{equation}
We non-dimensionalize the energy dissipation by dividing it by $\sigma_{12}A_{\rm{CV}}$, that is $\overline{D}_{\rm{Cassie}} = D_{\rm{Cassie}}/ (\sigma_{12})A_{\rm{CV}}$, which when used with equation (\ref{eqn:exp_D_cassie_2}) gives,
\begin{equation}
 \overline{D}_{\rm{Cassie}} = \phi \cos\theta_{\rm{A}} + \alpha - \beta(1 - \phi).
 \label{eqn:exp_D_cassie_non-dim}
\end{equation}
Substituting $\overline{D}_{\rm{Cassie}}$ from equation (\ref{eqn:exp_D_cassie_non-dim}) into equation (\ref{eqn:meb_cassie4}), we get
\begin{equation}
    \cos\theta_{\rm{a}} = \phi(\beta + \cos\theta_{\rm{A}}) - \beta - \phi \cos\theta_{\rm{A}} - \alpha - \beta(1-\phi),
    \label{eqn:exp_tha_cassie}
\end{equation}
which after simplifications reduces to a very simple form
\begin{equation}
    \cos\theta_{\rm{a}} = - \alpha.
    \label{eqn:exp_tha_180}
\end{equation}
If the distortions in the interface near the TPCL are neglected, then referring to figure \ref{fig_app:alpha_Cassie} the increase in fluid-1/fluid-2 interface prior to the jump dictates that 
\begin{equation}
    \cos (180^{\circ} - \theta_{\rm{a}}) = \frac{1}{\alpha},
\end{equation}
or
\begin{equation}
\alpha = \frac{-1}{\cos \theta_{\rm{a}}}.    
\end{equation}
Substituting this into equation (\ref{eqn:exp_tha_180}) gives
\begin{equation}
    \cos^2 \theta_{\rm{a}} = 1,
\end{equation}
or
\begin{equation}
    \theta_{\rm{a}} = 180^{\circ}
\end{equation}
under our advancing scenario - that is, the advancing angle in a Cassie wetting state is 180\tc irrespective of the pillar geometry and area fraction (provided the pillars have flat tops as per figure \ref{fig_app:alpha_Cassie}).

\newpage
\bibliographystyle{plain} 
\bibliography{ref2.bib}